\documentclass[a4paper,11pt,onecolumn]{article}%
\usepackage{graphicx}
\usepackage{amsmath}
\usepackage{amsfonts}
\usepackage{amssymb}%
\begin{document}

\title{Do we really understand quantum mechanics?\\Strange correlations, paradoxes and theorems.}
\author{F.\ Lalo\"{e}\\Laboratoire de Physique de l'ENS,\\LKB, 24 rue Lhomond,\\F-75005 Paris, France}
\maketitle

\begin{abstract}
This article presents a general discussion of several aspects of our present
understanding of quantum mechanics.\ The emphasis is put on the very special
correlations that this theory makes possible: they are forbidden by very
general arguments based on realism and local causality.\ In fact, these
correlations are completely impossible in any circumstance, except the very
special situations designed by physicists especially to observe these purely
quantum effects. Another general point that is emphasized is the necessity for
the theory to predict the emergence of a single result in a single realization
of an experiment.\ For this purpose, orthodox quantum mechanics introduces a
special postulate: the reduction of the state vector, which comes in addition
to the Schr\"{o}dinger evolution postulate.\ Nevertheless, the presence in
parallel of two evolution processes of the same object (the state vector) may
be a potential source for conflicts; various attitudes that are possible to
avoid this problem are discussed in this text. After a brief historical
introduction, recalling how the very special status of the state vector has
emerged in quantum mechanics, various conceptual difficulties are introduced
and discussed.\ The Einstein Podolsky Rosen (EPR) theorem is presented with
the help of a botanical parable, in a way that emphasizes how deeply the EPR
reasoning is rooted into what is often called ``scientific method''.\ In
another section the GHZ argument, the Hardy impossibilities, as well as the
BKS theorem are introduced in simple terms.\ The final two sections attempt to
give a summary of the present situation: one section discusses non-locality
and entanglement as we see it presently, with brief mention of recent
experiments; the last section contains a (non-exhaustive) list of various
attitudes that are found among physicists, and that are helpful to alleviate
the conceptual difficulties of quantum mechanics.

\end{abstract}
\tableofcontents

\bigskip

\begin{center}
\_\_\_\_\_\_\_\_\_
\end{center}

\bigskip

Quantum mechanics describes physical systems through a mathematical object,
the state vector $\mid\Psi>$, which replaces positions and velocities of
classical mechanics.\ This is an enormous change, not only mathematically, but
also conceptually.\ The relations between $\mid\Psi>$ and physical properties
are much less direct than in classical mechanics; the distance between the
formalism and the experimental predictions leaves much more room for
discussions about the interpretation of the theory. Actually, many
difficulties encountered by those who tried (or are still trying) to ``really
understand'' quantum mechanics are related to questions pertaining to the
exact status of $\mid\Psi>$: for instance, does it describe the physical
reality itself, or only some partial knowledge that we might have of this
reality?\ Does it fully describe ensemble of systems only (statistical
description), or one single system as well (single events)?\ Assume that,
indeed, $\mid\Psi>$ is affected by an imperfect knowledge of the system; is it
then not natural to expect that a better description should exist, at least in
principle?\ If so, what would be this deeper and more precise description of
the reality?

Another confusing feature of $\mid\Psi>$ is that, for systems extended in
space (for instance, a system made of two particles at very different
locations), it gives an overall description of all its physical properties in
a single block from which the notion of space seems to have disappeared; in
some cases, the physical properties of the two remote particles seem to be
completely ``entangled'' (the word was introduced by Schr\"{o}dinger in the
early days of quantum mechanics) in a way where the usual notions of
space-time and local events seem to become dimmed.\ Of course, one could think
that this entanglement is just an innocent feature of the formalism with no
special consequence: for instance, in classical electromagnetism, it is often
convenient to introduce a choice of gauge for describing the fields in an
intermediate step, but we know very well that gauge invariance is actually
fully preserved at the end.\ But, and as we will see below, it turns out that
the situation is different in quantum mechanics: in fact, a mathematical
entanglement in $\mid\Psi>$ can indeed have important physical consequences on
the result of experiments, and even lead to predictions that are, in a sense,
contradictory with locality (we will see below in what sense).

Without any doubt, the state vector is a rather curious object to describe
reality; one purpose of this article is to describe some situations in which
its use in quantum mechanics leads to predictions that are particularly
unexpected.\ As an introduction, and in order to set the stage for this
discussion, we will start with a brief historical introduction, which will
remind us of the successive steps from which the present status of $\mid\Psi>$
emerged. Paying attention to history is not inappropriate in a field where the
same recurrent ideas are so often rediscovered; they appear again and again,
sometimes almost identical over the years, sometimes remodelled or rephrased
with new words, but in fact more or less unchanged.\ Therefore, a look at the
past is not necessarily a waste of time!

\section{Historical perspective}

\label{historical}

The founding fathers of quantum mechanics had already perceived the essence of
many aspects of the discussions on quantum mechanics; today, after almost a
century, the discussions are still lively and, if some very interesting new
aspects have emerged, at a deeper level the questions have not changed so
much. What is more recent, nevertheless, is a general change of attitude among
physicists: until about 20 years ago, probably as a result of the famous
discussions between Bohr, Einstein, Schr\"{o}dinger, Heisenberg, Pauli, de
Broglie and others (in particular at the famous Solvay meetings \cite{Bohr}),
most physicists seemed to consider that ``Bohr was right and proved his
opponents to be wrong'', even if this was expressed with more nuance.\ In
other words, the majority of physicists thought that the so called
``Copenhagen interpretation'' had clearly emerged from the infancy of quantum
mechanics as the only sensible attitude for good scientists.\ As we all know,
this interpretation introduced the idea that modern physics must contain
indeterminacy as an essential ingredient: it is fundamentally impossible to
predict the outcome of single microscopical events; it is impossible to go
beyond the formalism of the wave function (or its equivalent, the state vector
$\mid\Psi>$\footnote{In all this article, we will not make any distinction
between the words ``wave function'' and ``state vector''.}) and complete it;
for some physicists, the Copenhagen interpretation also includes the difficult
notion of ``complementarity''.... even if it is true that, depending on the
context, complementarity comes in many varieties and has been interpreted in
many different ways! By and large, the impression of the vast majority was
that Bohr had eventually won the debate with Einstein, so that discussing
again the foundations of quantum mechanics after these giants was pretentious,
useless, and maybe even bad taste.

Nowadays, the attitude of physicists is much more moderate concerning these
matters, probably partly because the community has better realized the
non-relevance of the ``impossibility theorems'' put forward by the defenders
of the Copenhagen orthodoxy, in particular by Von Neumann \cite{Von-Neumann}
(see \cite{Bell-1}, \cite{Bohm-Bub-1} and \cite{Bohm-Bub-2}, as well as the
discussion given in \cite{Mermin}); another reason is, of course, the great
impact of the discoveries and ideas of J. Bell \cite{Bell-livre}. At the turn
of the century, it is probably fair to say that we are no longer sure that the
Copenhagen interpretation is the only possible consistent attitude for
physicists - see for instance the doubts expressed in \cite{Shimony}%
.\ Alternative points of view are considered as perfectly consistent: theories
including additional variables (or ``hidden variables''\footnote{As we discuss
in more detail in \S \ \ref{additional}, we prefer to use the words
``additional variables'' since they are not hidden, but actually appear
directly in the results of measurements; what is actually hidden in these
theories is rather the wave function itself, since it evolves independently of
these variables and can never be measured directly.\label{note-hidden}})
\cite{Bohm} \cite{Wiener-Siegel}; modified dynamics of the state vector
\cite{Bohm-Bub-1} \cite{Pearle} \cite{Pearle-2} \cite{Ghirardi} (non-linear
and/or stochastic evolution) ; at the other extreme we have points of view
such as the so called ``many worlds interpretation'' (or multibranched
universe interpretation) \cite{DeWitt}, or more recently other interpretations
such as that of ``decoherent histories'' \cite{Griffiths} (the list is
non-exhaustive).\ All these interpretations will be discussed in
\S \ \ref{other}.\ For a recent review containing many references, see
\cite{Goldstein}, which emphasizes additional variables, but which is also
characteristic of the variety of positions among contemporary
scientists\footnote{It is amusing to contrast the titles of refs.
\cite{Shimony} and \cite{Goldstein}.}, as well as an older but very
interesting debate published in Physics Today \cite{debate}; another very
useful source of older references is the 1971 AJP ``Resource Letter''
\cite{AJP-RL}. But recognizing this variety of positions should not be the
source of misunderstandings! It should also be emphasized very clearly that,
until now, no new fact whatsoever (or no new reasoning) has appeared that has
made the Copenhagen interpretation obsolete in any sense.

\subsection{Three periods}

Three successive periods may be distinguished in the history of the
elaboration of the fundamental quantum concepts; they have resulted in the
point of view that we may call ``the orthodox interpretation'', with all
provisos that have just been made above. Here we give only a brief historical
summary, but we refer the reader who would like to know more about the history
of the conceptual development of quantum mechanics to the book of Jammer
\cite{Jammer}; see also \cite{Darrigol-1}; for detailed discussions of
fundamental problems in quantum mechanics, one could also look for references
such as \cite{d'Espagnat} \cite{d'Espagnat-2} \cite{Shimony} or those given in
\cite{AJP-RL}.

\subsubsection{Prehistory}

Planck's name is obviously the first that comes to mind when one thinks about
the birth of quantum mechanics: he is the one who introduced the famous
constant $h$, which now bears his name, even if his method was
phenomenological.\ His motivation was actually to explain the properties of
the radiation in thermal equilibrium (black body radiation) by introducing the
notion of finite grains of energy in the calculation of the entropy, later
interpreted by him as resulting from discontinuous exchange between radiation
and matter.\ It is Einstein who, later, took the idea more seriously and
really introduced the notion of quantum of light (which would be named
``photon'' much later), in order to explain the wavelength dependence of the
photoelectric effect- for a general discussion of the many contributions of
Einstein to quantum theory, see \cite{Pais}.

One should nevertheless realize that the most important and urgent question at
the time was not so much to explain fine details of the properties of
radiation-matter interaction, or the peculiarities of the blackbody radiation;
it was, rather, to understand the origin of the stability of atoms, that is of
all matter which surrounds us and of which we are made! Despite several
attempts, explaining why atoms do not collapse almost instantaneously was
still a complete challenge in physics.\ One had to wait a little bit more,
until Bohr introduced his celebrated atomic model, to see the appearance of
the first elements allowing to treat the question.\ He proposed the notion of
``quantized permitted orbits'' for electrons, as well as that of ``quantum
jumps'' to describe how they would go from one orbit to another, during
radiation emission processes for instance. To be fair, we must concede that
these notions have now almost disappeared from modern physics, at least in
their initial forms; quantum jumps are replaced by a much more precise theory
of spontaneous emission in quantum electrodynamics.\ But, on the other hand,
one may also see a resurgence of the old quantum jumps in the modern use of
the postulate of the wave packet reduction. After Bohr, came Heisenberg who
introduced the theory that is now known as ``matrix mechanics'', an abstract
intellectual construction with a strong philosophical component, sometimes
close to positivism; the classical physical quantities are replaced by
``observables'', mathematically matrices, defined by suitable postulates
without much help of the intuition. Nevertheless, matrix mechanics contained
many elements which turned out to be building blocks of modern quantum mechanics!

In retrospect, one can be struck by the very abstract and somewhat mysterious
character of atomic theory at this period of history; why should electrons
obey such rules which forbid them to leave a given class of orbits, as if they
were miraculously guided on simple trajectories? What was the origin of these
quantum jumps, which were supposed to have no duration at all, so that it
would make no sense to ask what were the intermediate states of the electrons
during such a jump? Why should matrices appear in physics in such an abstract
way, with no apparent relation with the classical description of the motion of
a particle? One can guess how relieved many physicists felt when another point
of view emerged, a point of view which looked at the same time much simpler
and in the tradition of the physics of the 19th century: the undulatory (or
wave) theory.

\subsubsection{The undulatory period}

\label{undulatory}

It is well known that de Broglie was the first who introduced the idea of
associating a wave with every material particle; this was soon proven to be
correct by Davisson and Germer in their famous electron diffraction
experiment. Nevertheless, for some reason, at that time de Broglie did not
proceed much further in the mathematical study of this wave, so that only part
of the veil of mystery was raised by him (see for instance the discussion in
\cite{Darrigol-2}).\ It is sometimes said that Debye was the first who, after
hearing about de Broglie's ideas, remarked that in physics a wave generally
has a wave equation: the next step would then be to try and propose an
equation for this new wave.\ The story adds that the remark was made in the
presence of Schr\"{o}dinger, who soon started to work on this program; he
successfully and rapidly completed it by proposing the equation which now
bears his name, one of the most basic equations of all physics.\ Amusingly,
Debye himself does not seem to have remembered the event. The anecdote may not
be accurate; in fact, different reports about the discovery of this equation
have been given and we will probably never know exactly what happened.\ What
remains clear anyway is that the introduction of the Schr\"{o}dinger equation
is one of the essential milestones in the history of physics.\ Initially, it
allowed one to understand the energy spectrum of the hydrogen atom, but we now
know that it also gives successful predictions for all other atoms, molecules
and ions, solids (the theory of bands for instance), etc. It is presently the
major basic tool of many branches of modern physics and chemistry.

Conceptually, at the time of its introduction, the undulatory theory was
welcomed as an enormous simplification of the new mechanics; this is
particularly true because Schr\"{o}dinger and others (Dirac, Heisenberg)
promptly showed how it allowed one to recover the predictions of the
complicated matrix mechanics from more intuitive considerations on the
properties of the newly introduced ``wave function'' - the solution of the
Schr\"{o}dinger equation. The natural hope was then to be able to extend this
success, and to simplify all problems raised by the mechanics of atomic
particles: one would replace it by a mechanics of waves, which would be
analogous to electromagnetic or sound waves.\ For instance, Schr\"{o}dinger
thought initially that all particles in the universe looked to us like point
particles just because we observe them at a scale which is too large; in fact,
they are tiny ``wave packets'' which remain localized in small regions of
space.\ He had even shown that these wave packets remain small (they do not
spread in space) when the system under study is a harmonic oscillator... alas,
we now know that this is only one of the very few special cases where this is
true; in general, they do constantly spread in space!

\subsubsection{Emergence of the Copenhagen interpretation}

\label{Cop}

It did not take a long time before it became clear that the undulatory theory
of matter also suffers from very serious difficulties, actually so serious
that physicists were soon led to abandon it. A first example of difficulty is
provided by a collision between particles, where the Schr\"{o}dinger wave
spreads in all directions, exactly as the water wave stirred in a pond by a
stone thrown into it; but, in all collision experiments, particles are
observed to follow well-defined trajectories which remain perfectly localized,
going in some precise direction. For instance, every photograph taken in the
collision chamber of a particle accelerator shows very clearly that particles
never get ``diluted'' in all space! This remark stimulated the introduction,
by Born, of the probabilistic interpretation of the wave function.\ Another
difficulty, even more serious, arises as soon as one considers systems made of
more than one single particle: then, the Schr\"{o}dinger wave is no longer an
ordinary wave since, instead of propagating in normal space, it propagates in
the so called ``configuration space'' of the system, a space which has $3N$
dimensions for a system made of $N$ particles! For instance, already for the
simplest of all atoms, the hydrogen atom, the wave which propagates in $6$
dimensions (if spins are taken into account, four such waves propagate in 6
dimensions); for a macroscopic collection of atoms, the dimension quickly
becomes an astronomical number.\ Clearly the new wave was not at all similar
to classical waves, which propagate in ordinary space; this deep difference
will be a sort of Leitmotiv in this text\footnote{For instance, the
non-locality effects occurring with two correlated particles can be seen as a
consequence of the fact that the wave function propagates locally, but in a
$6$ dimension space, while the usual definition of locality refers to ordinary
space which has $3$ dimensions.}, reappearing under various aspects here and
there\footnote{One should probably mention at this point that quantum
mechanics can indeed be formulated in a way which does not involve the
configuration space, but just the ordinary space: the formalism of field
operators (sometimes called second quantization for historical reasons). The
price to pay, however, is that the wave function (a complex number) is then
replaced by an operator, so that any analogy with a classical field is even
less valid.}.

In passing, and as a side remark, it is amusing to notice that the recent
observation of the phenomenon of Bose-Einstein condensation in dilute gases
\cite{BEC} can be seen, in a sense, as a sort of realization of the initial
hope of Schr\"{o}dinger: this condensation provides a case where the
many-particle matter wave does propagate in ordinary space.\ Before
condensation takes place, we have the usual situation: the atoms belong to a
degenerate quantum gas, which has to be described by wave functions defined in
a huge configuration space.\ But, when they are completely condensed, they are
restricted to a much simpler many-particle state that can be described by the
same wave function, exactly as a single particle.\ In other words, the matter
wave becomes similar to a classical field with two components (the real part
and the imaginary part of the wave function), resembling an ordinary sound
wave for instance.\ This illustrates why, somewhat paradoxically, the
``exciting new states of matter'' provided by Bose-Einstein condensates are
not an example of an extreme quantum situation; they are actually more
classical than the gases from which they originate (in terms of quantum
description, interparticle correlations, etc.). Conceptually, of course, this
remains a very special case and does not solve the general problem associated
with a naive view of the Schr\"{o}dinger waves as real waves.

The purely undulatory description of particles has now disappeared from modern
quantum mechanics. In addition to Born and Bohr, Heisenberg \cite{Heisenberg},
Jordan, Dirac \cite{Dirac} and others played an essential role in the
appearance of a new formulation of quantum mechanics \cite{Darrigol-1}, where
probabilistic and undulatory notions are incorporated in a single complex
logical edifice.\ The now classical Copenhagen interpretation of quantum
mechanics (often also called ``orthodox interpretation'') incorporates both a
progressive, deterministic, evolution of the wave function/state vector
according to the Schr\"{o}dinger equation, as well as a second postulate of
evolution that is often called the ``wave packet reduction'' (or also ``wave
function collapse'').\ The Schr\"{o}dinger equation in itself does not select
precise experimental results, but keeps all of them as potentialities in a
coherent way; forcing the emergence of a single result in a single experiment
is precisely the role of the postulate of the wave packet reduction.\ In this
scheme, separate postulates and equations are therefore introduced, one for
the ``natural'' evolution of the system, another for measurements performed on it.

\subsection{The status of the state vector}

\label{status}

With two kinds of evolution, it is no surprise if the state vector should get,
in orthodox quantum theory, a non-trivial status - actually it has no
equivalent in all the rest of physics.

\subsubsection{Two extremes and the orthodox solution}

Two opposite mistakes should be avoided, since both ``miss the target'' on
different sides. The first is to endorse the initial hopes of Schr\"{o}dinger
and to decide that the (many-dimension) wave function directly describes the
physical properties of the system.\ In such a purely undulatory view, the
position and velocities of particles are replaced by the amplitude of a
complex wave, and the very notion of point particle becomes diluted; but the
difficulties introduced by this view are now so well known - see discussion in
the preceding section - that few physicists seem to be tempted to support it.
Now, by contrast, it is surprising to hear relatively often colleagues falling
to the other extreme, and endorsing the point of view where the wave function
does not attempt to describe the physical properties of the system itself, but
just the information that we have on it - in other words, the wave function
should get a relative (or contextual) status, and become analogous to a
classical probability distribution in usual probability theory. Of course, at
first sight, this would bring a really elementary solution to all fundamental
problems of quantum mechanics: we all know that classical probabilities
undergo sudden jumps, and nobody considers this as a special problem.\ For
instance, as soon as a new information becomes available on any system to us,
the probability distribution that we associate with it changes suddenly; is
this not the obvious way to explain the sudden wave packet reduction?

One first problem with this point of view is that it would naturally lead to a
relative character of the wave function: if two observers had different
information on the same system, should they use different wave functions to
describe the same system\footnote{Here we just give a simplified discussion;
in a more elaborate context, one would introduce for instance the notion of
intersubjectivity, etc. \cite{Shimony} \cite{d'Espagnat}.}? In classical
probability theory, there would be no problem at all with
``observer-dependent'' distribution probabilities, but standard quantum
mechanics clearly rejects this possibility: it certainly does not attribute
such a character to the wave function\footnote{We implicitly assume that the
two observers use the same space-time referential; otherwise, one should apply
simple mathematical transformations to go from one state vector to the
other.\ But this has no more conceptual impact than the transformations which
allow us, in classical mechanics, to transform positions and conjugate
momenta.
\par
We should add that there is also room in quantum mechanics for classical
uncertainties arising from an imperfect knowledge of the system; the formalism
of the density operator is a convenient way to treat these
uncertainties.\ Here, we intentionally limit ourselves to the discussion of
wave functions (pure states).}. Moreover, when in ordinary probability theory
a distribution undergoes a sudden ``jump'' to a more precise distribution, the
reason is simply that more precise values of the variables already exist -
they actually existed before the jump.\ In other words, the very fact that the
probability distribution reflected our imperfect knowledge implies the
possibility for a more precise description, closer to the reality of the
system itself.\ But this is in complete opposition with orthodox quantum
mechanics, which negates the very idea of a better description of the reality
than the wave function.\ In fact, introducing the notion of pre-existing
values is precisely the basis of unorthodox theories with additional variables
(hidden variables)! So the advocates of this ``information
interpretation''\footnote{Normally, in physics, information (or probabilities)
is about something! (meaning about something which has an independent reality,
see for instance \S \ VII of \cite{Stapp-2}).} are often advocates of
additional variables (often called hidden variables - see
\S \ \ref{additional} and note \ref{note-hidden}), without being aware of it!
It is therefore important to keep in mind that, in the classical
interpretation of quantum mechanics, the wave function (or state vector) gives
THE ultimate physical description of the system, with all its physical
properties; it is neither contextual, nor observer dependent; if it gives
probabilistic predictions on the result of future measurements, it
nevertheless remains inherently completely different from an ordinary
classical distribution of probabilities.

If none of these extremes is correct, how should we combine them?\ To what
extent should we consider that the wave function describes a physical system
itself (realistic interpretation), or rather that it contains only the
information that we may have on it (positivistic interpretation), presumably
in some sense that is more subtle than a classical distribution
function?\ This is not an easy question, and various authors answer the
question with different nuances; we will come back to it question in
\S \ \ref{Wigner}, in particular in the discussion of the ``Schr\"{o}dinger
cat paradox''.\ Even if it not so easy to be sure about what the perfectly
orthodox interpretation is, we could probably express it by quoting Peres
\cite{Peres-2}: ``a state vector is not a property of a physical system, but
rather represents an experimental procedure for preparing or testing one or
more physical systems''; we could then add another quotation from the same
article, as a general comment: ``quantum theory is incompatible with the
proposition that measurements are processes by which we discover some unknown
and preexisting property''. In this context, a wave function is an absolute
representation, but of a preparation procedure rather than of the isolated
physical system itself; nevertheless, but, since this procedure may also imply
some information on the system itself (for instance, in the case of repeated
measurements of the same physical quantity), we have a sort of intermediate
situation where none of the answers above is completely correct, but where
they are combined in a way that emphasizes the role of the whole experimental setup.

\subsubsection{An illustration}

\label{illustration}

Just as an illustration of the fact that the debate is not closed, we take a
quotation from a recent article\ \cite{mental} which, even if taken out of its
context, provides an interesting illustration of the variety of nuances that
can exist within the Copenhagen interpretation (from the context, is seems
clear that the authors adhere to this interpretation); after criticizing
erroneous claims of colleagues concerning the proper use of quantum concepts,
they write: ``(One) is led astray by regarding state reductions as physical
processes, rather than accepting that they are nothing but mental
processes''.\ The authors do not expand much more on this sentence, which they
relate on a ``minimalistic interpretation of quantum mechanics''; actually
they even give a general warning that it is dangerous to go beyond it (``Van
Kampen's caveat'').\ Nevertheless, let us try to be bold and to cross this
dangerous line for a minute; what is the situation then? We then see that two
different attitudes become possible, depending on the properties that we
attribute to the Schr\"{o}dinger evolution itself: is it also a ``purely
mental process'', or is it of completely different nature and associated more
closely with an external reality?\ Implicitly, the authors of \cite{mental}
seem to favor the second possibility - otherwise, they would probably have
made a more general statement about all evolutions of the state vector - but
let us examine both possibilities anyway.\ In the first case, the relation of
the wave function to physical reality is completely lost and we meet all the
difficulties mentioned in the preceding paragraph as well as some of the next
section; we have to accept the idea that quantum mechanics has nothing to say
about reality through the wave function (if the word reality even refers to
any well-defined notion!).\ In the second case, we meet the conceptual
difficulties related to the co-existence of two processes of completely
different nature for the evolution of the state vector, as discussed in the
next section. What is interesting is to note that Peres's point of view (end
of the preceding subsection), while also orthodox, corresponds to neither
possibilities: it never refers to mental process, but just to preparation and
tests on physical systems, which is clearly different; this illustrates the
flexibility of the Copenhagen interpretation and the variety of ways that
different physicists use to describe it.

Another illustration of the possible nuances is provided by a recent note
published by the same author together with Fuchs \cite{Fuchs-Peres} entitled
``Quantum theory needs no `interpretation' ''.\ These authors take explicitly
a point of view where the wave function is not absolute, but observer
dependent: ``it is only a mathematical expression for evaluating probabilities
and depends on the knowledge of whoever is doing the computing''.\ The wave
function becomes similar to a classical probability distribution which,
obviously, depends on the knowledge of the experimenter, so that several
different distributions can be associated with the same physical system (if
there are several observers).\ On the other hand, as mentioned above,
associating several different wave functions with one single system is not
part of what is usually called the orthodox interpretation (except, of course,
for a trivial phase factor).

To summarize, the orthodox status of the wave function is indeed a subtle
mixture between different, if not opposite, concepts concerning reality and
the knowledge that we have of this reality. Bohr is generally considered more
as a realist than a positivist or an operationalist \cite{Jammer}; he would
probably have said that the wave function is indeed a useful tool, but that
the concept of reality can not properly be defined at a microscopic level
only; it has to include all macroscopic measurement apparatuses that are used
to have access to microscopic information (we come back to this point in more
detail in \S \ \ref{transpo}).\ In this context, it is understandable why he
once even stated that ``there is no quantum concept'' \cite{Chevalley}!

\section{Difficulties, paradoxes}

\label{diff}

We have seen that, in most cases, the wave function evolves gently, in a
perfectly predictable and continuous way, according to the Schr\"{o}dinger
equation; in some cases only (as soon as a measurement is performed),
unpredictable changes take place, according to the postulate of wave packet
reduction. Obviously, having two different postulates for the evolution of the
same mathematical object is unusual in physics; the notion was a complete
novelty when it was introduced, and still remains unique in physics, as well
as the source of difficulties.\ Why are two separate postulates
necessary?\ Where exactly does the range of application of the first stop in
favor of the second?\ More precisely, among all the interactions - or
perturbations- that a physical system can undergo, which ones should be
considered as normal (Schr\"{o}dinger evolution), which ones are a measurement
(wave packet reduction)?\ Logically, we are faced with a problem that did not
exist before, when nobody thought that measurements should be treated as
special processes in physics.\ We learn from Bohr that we should not try to
transpose our experience of everyday's world to microscopic systems; this is
fine, but where exactly is the limit between the two worlds?\ Is it sufficient
to reply that there is so much room between macroscopic and microscopic sizes
that the exact position of the border does not matter\footnote{Proponents of
the orthodox interpretation often remark that one is led to the same
experimental predictions, independently of the exact position of this border,
so that any conflict with the experiments can be avoided.}?

Moreover, can we accept that, in modern physics, the ``observer'' should play
such a central role, giving to the theory an unexpected anthropocentric
foundation, as in astronomy in the middle ages? Should we really refuse as
unscientific to consider isolated (unobserved) systems, because we are not
observing them? These questions are difficult, almost philosophical, and we
will not attempt to answer them here.\ Rather, we will give a few
characteristic quotations, which illustrate\footnote{With, of course, the
usual proviso: short quotations taken out of their context may, sometimes,
give a superficial view on the position of their authors.} various positions:

(i) Bohr (second ref. \cite{Jammer}, page 204): \ ``There is no quantum
world......it is wrong to think that the task of physics is to find out how
Nature is.\ Physics concerns what we can say about Nature''.

(ii) Heisenberg (same ref. page 205): ``But the atoms or the elementary
particles are not real; they form a world of potentialities or possibilities
rather than one of things and facts''\footnote{Later, Heisenberg took a more
moderate attitude and no longer completely rejected the idea of a wave
functions describing some physical reality.}.

(iii) Jordan (as quoted by Bell in \cite{Bertlmann}): ``observations not only
disturb what has to be measured, they \textit{produce} it.\ In a measurement
of position, the electron is forced to a decision.\ We compel it to assume a
definite position; previously it was neither here nor there, it had not yet
made its decision for a definite position.. ''.

(iv) Mermin \cite{Mermin}, summarizing the ``fundamental quantum doctrine''
(orthodox interpretation): ``the outcome of a measurement is brought into
being by the act of measurement itself, a joint manifestation of the state of
the probed system and the probing apparatus. Precisely how the particular
result of an individual measurement is obtained - Heisenberg's transition from
the possible to the actual - is inherently unknowable''.

(v) Bell \cite{Bell-speakable}, speaking of ``modern'' quantum theory
(Copenhagen interpretation): ``it never speaks of events in the system, but
only of outcomes of observations upon the system, implying the existence of
external equipment\footnote{One could add ``and of external observers''.}''
(how, then, do we describe the whole universe, since there can be no external
equipment in this case?).

(vi) Shimony \cite{Shimony}: ``According to the interpretation proposed by
Bohr, the change of state is a consequence of the fundamental assumption that
the description of any physical phenomenon requires reference to the
experimental arrangement''.

(vii) Rosenfeld \cite{Rosenfeld}: ``the human observer, whom we have been at
pains to keep out of the picture, seems irresistibly to intrude into it,...''.

(viii) Stapp \cite{Stapp-2} ``The interpretation of quantum theory is clouded
by the following points: (1) Invalid classical concepts are ascribed
fundamental status; (2) The process of measurement is not describable within
the framework of the theory; (3) The subject-object distinction is blurred;
(4) The observed system is required to be isolated in order to be defined, yet
interacting to be observed''.

\subsection{Von Neumann's infinite regress}

\label{Neumann}

In this section, we introduce the notion of the Von Neumann regress, or Von
Neumann chain, a process that is at the source of phenomenon of
decoherence.\ Both actually correspond to the same basic physical process, but
the word decoherence usually refers to its initial stage, when the number of
degrees of freedom involved in the process is still relatively limited.\ The
Von Neumann chain, on the other hand, is more general since it includes this
initial stage as well as its continuation, which goes on until it reaches the
other extreme where it really becomes paradoxical: the Schr\"{o}dinger cat,
the symbol of a macroscopic system, with an enormous number of degrees of
freedom, in a impossible state (Schr\"{o}dinger uses the word ``ridiculous''
to describe it). Decoherence in itself is an interesting physical phenomenon
that is contained in the Schr\"{o}dinger equation and introduces no particular
conceptual problems; the word is relatively recent, and so is the observation
of the process in beautiful experiments in atomic physics \cite{Brune} - for
more details on decoherence, see \S \ref{decoherence}.\ Since for the moment
we are at the stage of an historical introduction of the difficulties of
quantum mechanics, we will not discuss microscopic decoherence further, but
focus the interest on macroscopic systems, where serious conceptual
difficulties do appear.

Assume that we take a simple system, such as a spin 1/2 atom, which enters
into a Stern-Gerlach spin analyzer. If the initial direction of the spin is
transverse (with respect to the magnetic field which defines the eigenstates
associated with the apparatus), the wave function of the atom will split into
two different wave packets, one which is pulled upwards, the other pushed
downwards; this is an elementary consequence of the linearity of the
Schr\"{o}dinger equation. Propagating further, each of the two wave packets
may strike a detector, with which they interact by modifying its state as well
as theirs; for instance, the incoming spin 1/2 atoms are ionized and produce
electrons; as a consequence, the initial coherent superposition now
encompasses new particles.\ Moreover, when a whole cascade of electrons is
produced in photomultipliers, all these additional electrons also become part
of the superposition.\ In fact, there is no intrinsic limit in what soon
becomes an almost infinite chain: rapidly, the linearity of the
Schr\"{o}dinger equation leads to a state vector which is the coherent
superposition of states including a macroscopic number of particles,
macroscopic currents and, maybe, pointers or recorders which have already
printed zeros or ones on a piece of paper! If we stick to the Schr\"{o}dinger
equation, there is nothing to stop this ``infinite Von Neumann regress'',
which has its seed in the microscopic world but rapidly develops into a
macroscopic consequence.\ Can we for instance accept the idea that, at the
end, it is the brain of the experimenter (who becomes aware of the results)
and therefore a human being, which enters into such a superposition?

Needless to say, no-one has ever observed two contradictory results at the
same time, and the very notion is not even very clear: it would presumably
correspond to an experimental result printed on paper looking more or less
like two superimposed slides, or a double exposure of a photograph.\ But in
practice we know that we always observe only one single result in a single
experiment; linear superpositions somehow resolve themselves before they
become sufficiently macroscopic to involve measurement apparatuses and
ourselves.\ It therefore seems obvious\footnote{Maybe not so obvious after
all?\ There is an interpretation of quantum mechanics that precisely rests on
the idea of never breaking this chain: the Everett interpretation, which will
be discussed in \S \ \ref{ever}.} that a proper theory should break the Von
Neumann chain, and stop the regress when (or maybe before) it reaches the
macroscopic world.\ But when exactly and how precisely?

\subsection{Wigner's friend}

\label{Wigner}

The question can also be asked differently: in a theory where the observer
plays such an essential role, who is entitled to play it?\ Wigner discusses
the role of a friend, who has been asked to perform an experiment, a
Stern-Gerlach measurement for instance \cite{Wigner-friend}; the friend may be
working in a closed laboratory so that an outside observer will not be aware
of the result before he/she opens the door. What happens just after the
particle has emerged from the analyzer and when its position has been observed
inside the laboratory, but is not yet known outside?\ For the outside
observer, it is natural to consider the whole ensemble of the closed
laboratory, containing the experiment as well as his friend, as the ``system''
to be described by a big wave function.\ As long as the door of the laboratory
remains closed and the result of the measurement unknown, this wave function
will continue to contain a superposition of the two possible results; it is
only later, when the result becomes known outside, that the wave packet
reduction should be applied.\ But, clearly, for Wigner's friend who is inside
the laboratory, this reasoning is just absurd! He/she will much prefer to
consider that the wave function is reduced as soon as the result of the
experiment is observed inside the laboratory. We are then back to a point that
we already discussed, the absolute/relative character of the wave function:
does this contradiction mean that we should consider two state vectors, one
reduced, one not reduced, during the intermediate period of the experiment?
For a discussion by Wigner of the problem of the measurement, see
\cite{Wigner1}.

An unconventional interpretation, sometimes associated with Wigner's
name\footnote{The title of ref \cite{Wigner-friend} is indeed suggestive of
this sort of interpretation; moreover, Wigner writes in this reference that
``it follows (from the Wigner friend argument) that the quantum description of
objects is influenced by impressions entering my consciousness''.\ At the end
of the article, he also discusses the influence of non-linearities which would
put a limit on the validity of the Schr\"{o}dinger equation, and be
indications of life.}, assumes that the reduction of the wave packet is a real
effect which takes place when a human mind interacts with the surrounding
physical world and acquires some consciousness of its state; in other words,
the electrical currents in the human brain may be associated with a reduction
of the state vector of measured objects, by some yet unknown physical process.
Of course, in this view, the reduction takes place under the influence of the
experimentalist inside the laboratory and the question of the preceding
paragraph is settled.\ But, even if one accepts the somewhat provocative idea
of possible action of the mind (or consciousness) on the environment, this
point of view does not suppress all logical difficulties: what is a human
mind, what level of consciousness is necessary to reduce the wave packet, etc.?

\subsection{Schr\"{o}dinger's cat}

\label{cat}

The famous story of the Schr\"{o}dinger cat \cite{SC} \cite{SC2} illustrates
the problem in a different way; it is probably too well known to be described
once more in detail here.\ Let us then just summarize it: the story
illustrates how a living creature could be put into a very strange state,
containing life and death, by correlation with a decaying radioactive atom,
and through a Von Neumann chain; the chain includes a gamma ray detector,
electronic amplification, and finally a mechanical system that automatically
opens a bottle of poisonous gas if the atomic decay takes place.\ The
resulting paradox may be seen as an illustration of the question: does an
animal such a cat have the intellectual abilities that are necessary to
perform a measurement and resolve all Von Neumann branches into one?\ Can it
perceive its own state, projecting itself onto one of the alive or dead
states? Or do humans only have access to a sufficient level of introspection
to become conscious of their own observations, and to reduce the wave
function?\ In that case, when the wave function includes a cat component, the
animal could remain simultaneously dead and alive for an arbitrarily long
period of time, a paradoxical situation indeed.

Another view on the paradox is obtained if one just considers the cat as a
symbol of any macroscopic object; such objects can obviously never be in a
``blurred'' state containing possibilities that are obviously contradictory
(open and closed bottle, dead and alive cat, etc.). Schr\"{o}dinger considers
this as a ``quite ridiculous case'', which emerges from the linearity of his
equation, but should clearly be excluded from any reasonable theory - or at
best considered as the result of some incomplete physical description.\ In
Schr\"{o}dinger's words: ''an indeterminacy originally restricted to the
atomic domain becomes transformed into a macroscopic indeterminacy''.\ The
message is simple: standard quantum mechanics is not only incapable of
avoiding these ridiculous cases, it actually provides a recipe for creating
them; one obviously needs some additional ingredients in the theory in order
to resolve the Von Neumann regress, select one of its branches, and avoid
stupid macroscopic superpositions.\ It is amusing to note in passing that
Schr\"{o}dinger's name is associated to two contradictory concepts that are
actually mutually exclusive, a continuous equation of evolution and the
symbolic cat, a limit that the equation should never reach! Needless to say,
the limit of validity of the linear equation does not have to be related to
the cat itself: the branch selection process may perfectly take place before
the linear superposition reaches the animal. But the real question is that the
reduction process has to take place somewhere, and where exactly?

Is this paradox related to decoherence?\ Not really.\ Coherence is completely
irrelevant for Schr\"{o}dinger, since the cat is actually just a symbol of a
macroscopic object that is in an impossible blurred state, encompassing two
possibilities that are incompatible in ordinary life; the state in question is
not necessarily a pure state (only pure states are sensitive to decoherence)
but can also be a statistical mixture.\ Actually, in the story, the cat is
never in a coherent superposition, since its blurred state is precisely
created by correlation with some parts of the environment (the bottle of
poison for instance); the cat is just another part of the environment of the
radioactive atom.\ In other words, the cat is not the seed of a Von Neumann
chain; it is actually trapped into two (or more) of its branches, in a tree
that has already expanded into the macroscopic world after decoherence has
already taken place at a microscopic level (radioactive atom and radiation
detector), and will continue to expand after it has captured the cat.
Decoherence is irrelevant for Schr\"{o}dinger since his point is not to
discuss the features of the Von Neumann chain, but to emphasize the necessity
to break it: the question is not to have a coherent or a statistical
superposition of macroscopically different states, it is to have no
superposition at all\footnote{This is for instance the purpose of theories
with a modified non-linear Schr\"{o}dinger dynamics: providing equations of
motion where during measurements all probabilities dynamically go to zero,
except one that goes to 1.}!

So the cat is the symbol of an impossibility, an animal that can never exist
(a Schr\"{o}dinger gargoyle?), and a tool for a ``reductio ad absurdum''
reasoning that puts into light the limitations of the description of a
physical system by a Schr\"{o}dinger wave function only. Nevertheless, in the
recent literature in quantum electronics, it has become more and more frequent
to weaken the concept, and to call ``Schr\"{o}dinger cat (SC)'' any coherent
superposition of states that can be distinguished macroscopically,
independently of the numbers of degree of freedom of the system.\ SC states
can then be observed (for instance an ion located in two different places in a
trap), but often undergo rapid decoherence through correlation to the
environment.\ Moreover, the Schr\"{o}dinger equation can be used to calculate
how the initial stages of the Von Neumann chain take place, and how rapidly
the solution of the equation tends to ramify into branches containing the
environment. Since this use of the words SC has now become rather common in a
subfield of physics, one has to accept it; it is, after all, just a matter of
convention to associate them with microscopic systems - any convention is
acceptable as long as it does not create confusion. But it would be an insult
to Schr\"{o}dinger to believe that decoherence can be invoked as the solution
of his initial cat paradox: Schr\"{o}dinger was indeed aware of the properties
of entanglement in quantum mechanics, a word that he introduced (and uses
explicitly in the article on the cat), and he was not sufficiently naive to
believe that standard quantum mechanics would predict possible interferences
between dead and alive cats!

To summarize, the crux of most of our difficulties with quantum mechanics is
the question: what is exactly the process that forces Nature to break the
regress and to make its choice among the various possibilities for the results
of experiments? Indeed, the emergence of a single result in a single
experiment, in other words the disappearance of macroscopic superpositions, is
a major issue; the fact that such superpositions cannot be resolved at any
stage within the linear Schr\"{o}dinger equation may be seen as the major
difficulty of quantum mechanics. As Pearle nicely expresses it \cite{Pearle-2}%
, the problem is to ``explain why events occur''!

\subsection{Unconvincing arguments}

We have already emphasized that the invention of the Copenhagen interpretation
of quantum mechanics has been, and remains, one of the big achievements of
physics.\ One can admire even more, in retrospect, how early its founders
conceived it, at a time when experimental data were relatively scarce.\ Since
that time, numerous ingenious experiments have been performed, precisely with
the hope of seeing the limits of this interpretation but, until now, not a
single fact has disproved the theory.\ It is really a wonder of pure logic
that has allowed the early emergence of such an intellectual construction.

This being said, one has to admit that, in some cases, the brilliant authors
of this construction may sometimes have gone too far, pushed by their great
desire to convince.\ For instance, authoritative statements have been made
concerning the absolute necessity of the orthodox interpretation which now, in
retrospect, seem exaggerated - to say the least. According to these
statements, the orthodox interpretation would give the only ultimate
description of physical reality; no finer description would ever become
possible.\ In this line of thought, the fundamental probabilistic character of
microscopic phenomena should be considered as a proven fact, a rule that
should be carved into marble and accepted forever by scientists. But, now, we
know that this is not proven to be true: yes, one may prefer the orthodox
interpretation if one wishes, but this is only a matter of taste; other
interpretations are still perfectly possible; determinism in itself is not
disproved at all.\ As discussed for instance in \cite{Mermin}, and initially
clarified by Bell \cite{Bell-1} \cite{Bell-livre} and Bohm \cite{Bohm-Bub-1}
\cite{Bohm-Bub-2} , the ``impossibility proofs'' put forward by the proponents
of the Copenhagen interpretation are logically unsatisfactory for a simple
reason: they arbitrarily impose conditions that may be relevant to quantum
mechanics (linearity), but not to the theories that they aim to dismiss - any
theory with additional variables such as the Bohm theory, for instance.
Because of the exceptional stature of the authors of the impossibility
theorems, it took a long time to the physics community to realize that they
were irrelevant; now, this is more widely recognized so that the plurality of
interpretations is more easily accepted.

\section{Einstein, Podolsky and Rosen}

It is sometimes said that the article by Einstein, Podolsky and Rosen (EPR)
\cite{EPR} is, by far, that which has collected the largest number of
quotations in the literature; the statement sounds very likely to be true.
There is some irony in this situation since, so often, the EPR\ reasoning has
been misinterpreted, even by prominent physicists!\ A striking example is
given in the Einstein-Born correspondence \cite{Einstein-Born} where Born,
even in comments that he wrote after Einstein's death, still clearly shows
that he never really understood the nature of the objections raised by
EPR.\ Born went on thinking that the point of Einstein was an a priori
rejection of indeterminism (``look, Einstein, indeterminism is not so bad''),
while actually the major concern of EPR\ was locality and/or separability (we
come back later to these terms, which are related to the notion of
space-time).\ If giants like Born could be misled in this way, no surprise
that, later on, many others made similar mistakes! This is why, in what
follows, we will take an approach that may look elementary, but at least has
the advantage of putting the emphasis on the logical structure of the arguments.

\subsection{A theorem}

One often speaks of the ``EPR paradox'', but the word ``paradox'' is not
really appropriate in this case. For Einstein, the basic motivation was not to
invent paradoxes or to entertain colleagues inclined to philosophy; it was to
build a strong logical reasoning which, starting from well defined assumptions
(roughly speaking: locality and some form of realism), would lead ineluctably
to a clear conclusion (quantum mechanics is incomplete, and even: physics is
deterministic\footnote{Born's mistake, therefore, was to confuse assumptions
and conclusions.}). To emphasize this logical structure, we will speak here of
the ``EPR theorem'', which formally could be stated as follows:

Theorem: \textit{If the predictions of quantum mechanics are correct ( even
for systems made of remote correlated particles) and if physical reality can
be described in a local (or separable) way, then quantum mechanics is
necessarily incomplete: some ``elements of reality\footnote{These words are
carefully defined by the authors of the theorem; see the beginning of
\S \ \ref{transpo}.}'' exist in Nature that are ignored by this theory.}

The theorem is valid, and has been scrutinized by many scientists who have
found no flaw in its derivation; indeed, the logic which leads from the
assumptions to the conclusion is perfectly sound.\ It would therefore be an
error to repeat (a classical mistake!) ``the theorem was shown by Bohr to be
incorrect'' or, even worse, ``the theorem is incorrect since experimental
results are in contradiction with it\footnote{The contradiction in question
occurs through the Bell theorem (which is therefore sometimes criticized for
the same reason), which was introduced as a continuation of the EPR
theorem.}''.\ Bohr himself, of course, did not make the error: in his reply to
EPR\ \cite{Bohr-EPR}, he explains why he thinks that the assumptions on which
the theorem is based are not relevant to the quantum world, which makes it
inapplicable to a discussion on quantum mechanics; more precisely, he uses the
word ``ambiguous'' to characterize these assumptions, but he never claims that
the reasoning is faulty (for more details, see \S \ \ref{transpo}).\ A theorem
which is not applicable in a particular case is not necessarily incorrect:
theorems of Euclidean geometry are not wrong, or even useless, because one can
also build a consistent non-Euclidean geometry! Concerning possible
contradictions with experimental results we will see that, in a sense, they
make a theorem even more interesting, mostly because it can then be used
within a ``reductio ad absurdum'' reasoning.

Goods texts on the EPR argument are abundant; for instance, a classic is the
wonderful little article by Bell \cite{Bertlmann}; another excellent
introductory text is, for instance, ref. \cite{Mermin-moon}, which contains a
complete description of the scheme (in the particular case where two settings
only are used) and provides an excellent general discussion of many aspects of
the problem; for a detailed source of references, see for instance
\cite{Home-Selleri}.\ Most readers are probably already familiar with the
basic scheme considered, which is summarized in figure 1: a source $S$ emits
two correlated particles, which propagate towards two remote regions of space
where they undergo measurements; the type of these measurements are defined by
``settings'', or ``parameters''\footnote{Here we will use the words
``settings'' and ``parameters'' indifferently.} (typically orientations of
Stern-Gerlach analyzers, often noted $a$ and $b$), which are at the choice of
the experimentalists; in each region, a result is obtained, which can take
only two values symbolized by $\pm1$ in the usual notation; finally, we will
assume that, every time both settings are chosen to be the same value, the
results of both measurements are always the same.

\begin{figure}[ptb]
\begin{center}
\includegraphics[width=10cm]{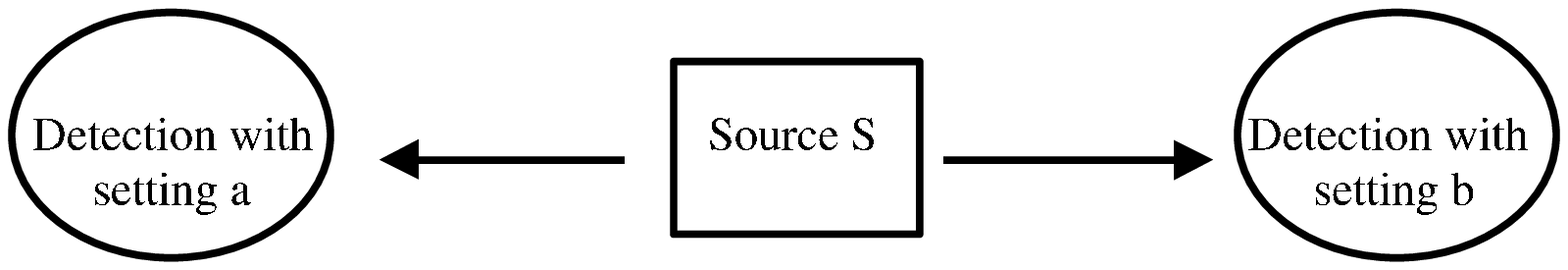}\newline
\end{center}
\par
\label{fig 1}\end{figure}

\ Here, rather than trying to paraphrase the good texts on EPR\ with more or
less success, we will purposefully take a different presentation, based on a
comparison, a sort of a parable.\ Our purpose is to emphasize a feature of the
reasoning: the essence of the EPR reasoning is actually nothing but what is
usually called ``the scientific method'' in the sense discussed by Francis
Bacon and Claude Bernard.\ For this purpose, we will leave pure physics for
botany! Indeed, in both disciplines, one needs rigorous scientific procedures
in order to prove the existence of relations and causes, which is precisely
what we want to do.

\subsection{Of peas, pods and genes}

\label{peas}

When a physicist attempts to infer the properties of microscopic objects from
macroscopic observations, ingenuity (in order to design meaningful
experiments) must be combined with a good deal of logic (in order to deduce
these microscopic properties from the macroscopic results).\ Obviously, some
abstract reasoning is indispensable, merely because it is impossible to
observe with the naked eye, or to take in one's hand, an electron or even a
macromolecule for instance.\ The scientist of past centuries who, like Mendel,
was trying to determine \ the genetic properties of plants, had exactly the
same problem: he did not have access to any direct observation of the DNA
molecules, so that he had to base his reasoning on adequate experiments and on
the observation of their macroscopic outcome.\ In our parable, the scientist
will observe the color of flowers (the ``result'' of the measurement, $+1$ for
red, $-1$ for blue) as a function of the condition in which the peas are grown
(these conditions are the ``experimental settings'' $a$ and $b$, which
determine the nature of the measurement).\ The basic purpose is to infer the
intrinsic properties of the peas (the EPR ``element of reality'') from these observations.

\subsubsection{Simple experiments; no conclusion yet.}

It is clear that many external parameters such as temperature, humidity,
amount of light, etc. may influence the growth of vegetables and, therefore,
the color of a flower; it seems very difficult in a practical experiment to be
sure that all the relevant parameters have been identified and controlled with
a sufficient accuracy.\ Consequently, if one observes that the flowers which
grow in a series of experiments are sometimes blue, sometimes red, it is
impossible to identify the reason behind these fluctuation; it may reflect
some trivial irreproducibility of the conditions of the experiment, or
something more fundamental. In more abstract terms, a completely random
character of the result of the experiments may originate either from the
fluctuations of uncontrolled external perturbations, or from some intrinsic
property that the measured system (the pea) initially possesses, or even from
the fact that the growth of a flower (or, more generally, life?) is
fundamentally an indeterministic process - needless to say, all three reasons
can be combined in any complicated way.\ Transposing the issue to quantum
physics leads to the following formulation of the question: are the results of
the experiments random because of the fluctuation of some uncontrolled
influence taking place in the macroscopic apparatus, of some microscopic
property of the measured particles, or of some more fundamental process?

The scientist may repeat the ``experiment'' a thousand times and even more: if
the results are always totally random, there is no way to decide which
interpretation should be selected; it is just a matter of personal taste.\ Of
course, philosophical arguments might be built to favor or reject one of them,
but from a pure scientific point of view, at this stage, there is no
compelling argument for a choice or another. Such was the situation of quantum
physics before the EPR argument.

\subsubsection{Correlations; causes unveiled.}

\label{causes}

The stroke of genius of EPR was to realize that correlations could allow a big
step further in the discussion.\ They exploit the fact that, when the choice
of the settings are the same, the observed results turn out to be always
identical; in our botanical analogy, we will assume that our botanist observes
correlations between colors of flowers. Peas come together in pods, so that it
is possible to grow peas taken from the same pod and observe their flowers in
remote places. It is then natural to expect that, when no special care is
taken to give equal values to the experimental parameters (temperature, etc.),
nothing special is observed in this new experiment.\ But assume that, every
time the parameters are chosen to the same values, the colors are
systematically the same; what can we then conclude? Since the peas grow in
remote places, there is no way that they can be influenced by the any single
uncontrolled fluctuating phenomenon, or that they can somehow influence each
other in the determination of the colors.\ If we believe that causes always
act locally, we are led to the following conclusion: the only possible
explanation of the common color is the existence of some common property of
both peas, which determines the color; the property in question may be very
difficult to detect directly, since it is presumably encoded inside some tiny
part of a biological molecule, but it is sufficient to determine the results
of the experiments.

Since this is the essence of the argument, let us make every step of the
EPR\ reasoning completely explicit, when transposed to botany.\ The key idea
is that the nature and the number of ``elements of reality'' associated with
each pea can not vary under the influence of some remote experiment, performed
on the other pea. For clarity, let us first assume that the two experiments
are performed at different times: one week, the experimenter grows a pea, then
only next week another pea from the same pod; we assume that perfect
correlations of the colors are always observed, without any special influence
of the delay between the experiments.\ Just after completion of the first
experiment (observation of the first color), but still before the second
experiment, the result of that future experiment has a perfectly determined
value; therefore, there must already exist one element of reality attached to
the second pea that corresponds to this fact - clearly, it can not be attached
to any other object than the pea, for instance one of the measurement
apparatuses, since the observation of perfect correlations only arises when
making measurements with peas taken from the same pod.\ Symmetrically, the
first pod also had an element of reality attached to it which ensured that its
measurement would always provide a result that coincides with that of the
future measurement. The simplest idea that comes to mind is to assume that the
elements of reality associated with both peas are coded in some genetic
information, and that the values of the codes are exactly the same for all
peas coming from the same pod; but other possibilities exist and the precise
nature and mechanism involved in the elements of reality does not really
matter here.\ The important point is that, since these elements of reality can
not appear by any action at a distance, they necessarily also existed before
any measurement was performed - presumably even before the two peas were separated.

Finally, let us consider any pair of peas, when they are already spatially
separated, but before the experimentalist decides what type of measurements
they will undergo (values of the parameters, delay or not, etc.).\ We know
that, if the decision turns out to favor time separated measurements with
exactly the same parameter, perfect correlations will always be
observed.\ Since elements of reality can not appear, or change their values,
depending of experiments that are performed in a remote place, the two peas
necessarily carry some elements of reality with them which completely
determine the color of the flowers; any theory which ignores these elements of
reality is incomplete. This completes the proof.

It seems difficult not to agree that the method which led to these conclusions
is indeed the scientific method; no tribunal or detective would believe that,
in any circumstance, perfect correlations could be observed in remote places
without being the consequence of some common characteristics shared by both
objects. Such perfect correlations can then only reveal the initial common
value of some variable attached to them, which is in turn a consequence of
some fluctuating common cause in the past (a random choice of pods in a bag
for instance).\ To express things in technical terms, let us for instance
assume that we use the most elaborate technology available to build elaborate
automata, containing powerful modern computers\footnote{We are assuming here
that the computers are not quantum computers (if quantum computers can ever be
built, which is another question).} if necessary, for the purpose of
reproducing the results of the remote experiments: whatever we do, we must
ensure that, somehow, the memory of each computer contains the encoded
information concerning all the results that it might have to provide in the
future (for any type of measurement that might be made).

To summarize this section, we have shown that each result of a measurement may
be a function of two kinds of variables\footnote{In Bell's notation, the $A$
functions\ depend on the settings $a$ and $b$ as well as on $\lambda$.}:

(i) intrinsic properties of the peas, which they carry along with them.

(ii) the local setting of the experiment (temperature, humidity, etc.);
clearly, a given pair that turned out to provide two blue flowers could have
provided red flowers in other experimental conditions.

We may also add that:

(iii) the results are well-defined functions, in other words that no
fundamentally indeterministic process takes place in the experiments.

(iv) when taken from its pod, a pea cannot ``know in advance'' to which sort
of experiment it will be submitted, since the decision may not yet have been
made by the experimenters; when separated, the two peas therefore have to take
with them all the information necessary to determine the color of flowers for
any kind of experimental conditions.\ What we have shown actually is that each
pea carries with it as many elements of reality as necessary to provide ``the
correct answer''\footnote{Schr\"{o}dinger used to remark that, if all students
of a group always give the right answer to a question chosen randomly by the
professor among two, they all necessarily knew the answer to both questions
(and not only the one they actually answer).} to all possible questions it
might be submitted to.

\subsubsection{Transposition to physics}

\label{transpo}

The starting point of EPR is to assume that quantum mechanics provides correct
predictions for all results of experiments; this is why we have built the
parable of the peas in a way that exactly mimics the quantum predictions for
measurements performed on two spin 1/2 particles for some initial quantum
state: the red/blue color is obviously the analogue to the result that can be
obtained for a spin in a Stern-Gerlach apparatus, the parameters (or settings)
are the analogous to the orientation of these apparatuses (rotation around the
axis of propagation of the particles).\ Quantum mechanics predicts that the
distance and times at which the spin measurements are performed is completely
irrelevant, so that the correlations will remain the same if they take place
in very remote places.

Another ingredient of the EPR\ reasoning is the notion of ``elements of
reality''; EPR first remark that these elements cannot be found by a priori
philosophical considerations, but must be found by an appeal to results of
experiments and measurements.\ They then propose the following criterion:
``if, without in any way disturbing a system, we can predict with certainty
the value of a physical quantity, then there exists an element of physical
reality corresponding to this physical quantity''. In other words, certainty
can not emerge from nothing: an experimental result that is known in advance
is necessarily the consequence of some pre-existing physical property.\ In our
botanical analogy, we implicitly made use of this idea in the reasoning of
\S \ \ref{causes}.

A last, but essential, ingredient of the EPR reasoning is the notion of
space-time and locality: the elements of reality in question are attached to
the region of space where the experiment takes place, and they cannot vary
suddenly (or even less appear) under the influence of events taking place in
very distant region of space.\ The peas of the parable were in fact not so
much the symbol of some microscopic object, electrons or spin 1/2 atoms for
instance.\ Rather, they symbolize regions of space where we just know that
``something is propagating''; it can be a particle, a field, or anything else,
with absolutely no assumption on its structure or physical
description.\ Actually, in the EPR\ quotation of the preceding paragraph, one
may replace the word ``system'' by ``region of space'', without altering the
rest of the reasoning.\ One may summarize the situation by saying that the
basic belief of EPR is that regions of space can contain elements of reality
attached to them (attaching distinct elements of reality to separate regions
of space is sometimes called ``separability'') and that they evolve locally.
From these assumptions, EPR prove that the results of the measurements are
functions of:

(i) intrinsic properties of the spins that they carry with them (the EPR
elements of reality)

(ii) of course, also of the orientations of the Stern-Gerlach analyzers

In addition, they show that:

(iii) the functions giving the results are well-defined functions, which
implies that no indeterministic process is taking place; in other words, a
particle with spin carries along with it all the information necessary to
provide the result to any possible measurement.

(iv) since it is possible to envisage future measurements of observables that
are called ``incompatible'' in quantum mechanics, as a matter of fact,
incompatible observables can simultaneously have a perfectly well defined value.

Item (i) may be called the EPR-1 result: quantum mechanics is incomplete (EPR
require from a complete theory that ``every element of physical reality must
have a counterpart in the physical theory''); in other words, the state vector
may be a sufficient description for a statistical ensemble of pairs, but for
one single pair of spins, it should be completed by some additional
information; in still other words, inside the ensemble of all pairs, one can
distinguish between sub-ensembles with different physical properties.\ Item
(iii) may be called EPR-2, and establishes the validity of determinism from a
locality assumption.\ Item (iv), EPR-3 result, shows that the notion of
incompatible observables is not fundamental, but just a consequence of the
incomplete character of the theory; it actually provides a reason to reject
complementarity. Curiously, EPR-3 is often presented as the major EPR\ result,
sometimes even with no mention of the two others; actually, the rejection of
complementarity is almost marginal or, at least, less important for EPR than
the proof of incompleteness.\ In fact, in all that follows in this article, we
will only need EPR-1,2.

Niels Bohr, in his reply to the EPR article \cite{Bohr-EPR}, stated that their
criterion for physical reality contains an essential ambiguity when it is
applied to quantum phenomena. A more extensive quotation of Bohr's reply is
the following:

``The wording of the above mentioned criterion (the EPR criterion for elements
of reality)... contains an ambiguity as regards the expression 'without in any
way disturbing a system'.\ Of course there is in a case like that considered
(by EPR) no question of a mechanical disturbance of the system under
investigation during the last critical stage of the measuring procedure.\ But
even at this stage there is essentially the question of an influence of the
very conditions which define the possible types of predictions regarding the
future behavior of the system.... the quantum description may be characterized
as a rational utilization of all possibilities of unambiguous interpretation
of measurements, compatible with the finite and uncontrollable interactions
between the objects and the measuring instruments in the field of quantum theory''.

{}Indeed, in Bohr's view, physical reality cannot be properly defined without
reference to a complete and well-defined experiment.\ This includes, not only
the systems to be measured (the microscopic particles), but also all the
measurement apparatuses: ``these (experimental) conditions must be considered
as an inherent element of any phenomenon to which the term physical reality
can be unambiguously applied''. Therefore EPR's attempt to assign elements of
reality to one of the spins only, or to a region of space containing it, is
incompatible with orthodox quantum mechanics\footnote{One could add that the
EPR disproval of the notion of incompatible observables implies that, at
least, two different settings are considered for one of the measurement
apparatuses; this should correspond, in Bohr's view, to two different physical
realities (every different couple $a$,$b$ actually corresponds to a different
physical reality), and not to a single one as assumed in the EPR reasoning.} -
even if the region in question is very large and isolated from the rest of the
world. Expressed differently, a physical system that is extended over a large
region of space is to be considered as a single entity, within which no
attempt should be made to distinguish physical subsystems or any substructure;
trying to attach physical reality to regions of space is then automatically
bound to failure. In terms of our Leitmotiv of \S \ \ref{Cop}, the difference
between ordinary space and configuration space, we could say the following:
the system has a single wave function for both particles that propagates in a
configuration space with more than 3 dimensions, and this should be taken very
seriously; no attempt should be made to come back to three dimensions and
implement locality arguments in a smaller space.

Bohr's point of view is, of course, not contradictory with relativity, but
since it minimizes the impact of basic notions such as space-time, or events
(a measurement process in quantum mechanics is not local; therefore it is not
an event stricto sensu), it does not fit very well with it. One could add that
Bohr's article is difficult to understand; many physicists admit that a
precise characterization of his attitude, in terms for instance of exactly
what traditional principles of physics should be given up, is delicate (see
for example the discussion of ref. \cite{Shimony}).\ In Pearle's words:
``Bohr's rebuttal was essentially that Einstein's opinion disagreed with his
own'' \ \cite{Pearle-3}.\ It is true that, when scrutinizing Bohr's texts, one
never gets completely sure to what extent he fully realized all the
consequences of his position.\ Actually, in most of his reply to EPR
\cite{Bohr-EPR} in Physical Review, he just repeats the orthodox point of view
in the case of a single particle submitted to incompatible measurements, and
even goes through considerations that are not obviously related to the EPR
argument, as if he did not appreciate how interesting the discussion becomes
for two remote correlated particles; the relation to locality is not
explicitly discussed, as if this was an unimportant issue (while it was the
starting point of further important work, the Bell theorem for
instance\footnote{If Bohr had known the Bell theorem, he could merely have
replied to EPR\ that their logical system was inconsistent (see
\S \ \ \ref{contradictions})!}).\ The precise reply to EPR is actually
contained in only a short paragraph of this article, from which the quotations
given above have been taken.\ Even Bell confessed that he had strong
difficulties understanding Bohr (``I have very little idea what this means..''
- see the appendix of ref. \cite{Bertlmann})!

\section{Quantitative theorems: Bell, GHZ, Hardy, BKS}

\label{quantit}

The Bell theorem \cite{Bell-2} may be seen in many different ways.\ In fact,
Bell initially invented it as a logical continuation of the EPR\ theorem: the
idea is to take completely seriously the existence of the EPR\ elements of
reality, and introduce them into the mathematics with the notation $\lambda$;
one then proceeds to study all possible kinds of correlations that can be
obtained from the fluctuations of the $\lambda$'s, making the condition of
locality explicit in the mathematics (locality was already useful in the EPR
theorem, but not used in equations). As \ a continuation of EPR, the reasoning
necessarily develops from a deterministic framework and deals with classical
probabilities; it studies in a completely general way all kinds of correlation
that can be predicted from the fluctuations in the past of some classical
common cause - if one prefers, from some uncertainty concerning the initial
state of the system. This leads to the famous inequalities. But subsequent
studies have shown that the scope of the Bell theorem is not limited to
determinism; for instance, the $\lambda$'s may influence the results of future
experiments by fixing the values of probabilities of the results, instead of
these results themselves (see appendix I). We postpone the discussion of the
various possible generalizations to \S \ \ref{general} and, for the moment, we
just emphasize that the essential condition for the validity of the Bell
theorem is locality: all kinds of fluctuations can be assumed, but their
effect must affect physics only locally.\ If we assume that throwing dice in
Paris may influence physical events taking place in Tokyo, or even in other
galaxies, the proof of the theorem is no longer possible.\ For non-specialized
discussions of the Bell theorem, see for instance \cite{Bertlmann}
\cite{Mermin-moon} \cite{La-Recherche} \cite{FL}.

\subsection{Bell inequalities}

The Bell inequalities are relations satisfied by the average values of product
of random variables that are correlated classically (their correlations arise
from the fluctuations of some common cause in the past, as above for the
peas). As we will see, the inequalities are especially interesting in cases
where they are contradictory with quantum mechanics; one of these situations
occurs in the EPRB (B for Bohm \cite{Bohm-2}) version of the EPR argument,
where two spin 1/2 particles undergo measurements.\ This is why we begin this
section by briefly recalling the predictions of quantum mechanics for such a
physical system - but the only ingredient we need from quantum mechanics at
this stage is the predictions concerning the probabilities of results.\ Then
we leave again standard quantum mechanics and come back to the EPR-Bell
argument, discuss its contradictions with quantum mechanics, and finally
emphasize the generality of the theorem.

\subsubsection{Two spins in a quantum singlet state}

We assume that two spin 1/2 particles propagate in opposite directions after
leaving a source which has emitted them in a singlet spin state.\ Their spin
state is then described by:%
\begin{equation}
\mid\Psi>=\frac{1}{\sqrt{2}}\left[  \mid+,->-\mid-,+>\right]  \label{0}%
\end{equation}
When they reach distant locations, they are then submitted to spin
measurements, with Stern-Gerlach apparatuses oriented along angles $a$ and $b$
around the direction of propagation.\ If $\theta$ is the angle between $a$ and
$b$, quantum mechanics predicts that the probability for a double detection of
results $+1$, $+1$ (or of $-1$, $-1$) is:
\begin{equation}
\mathcal{P}_{+,+}=\mathcal{P}_{-,-}=\frac{1}{2}\sin^{2}\frac{\theta}{2}
\label{1}%
\end{equation}
while the probability of two opposite results is:
\begin{equation}
\mathcal{P}_{+,-}=\mathcal{P}_{-,+}=\frac{1}{2}\cos^{2}\frac{\theta}{2}
\label{2}%
\end{equation}
This is all that we want to know, for the moment, of quantum mechanics:
probability of the results of measurements. We note in passing that, if
$\theta=0$, (when the orientations of the measurements apparatuses are
parallel) the formulas predict that one of the probabilities vanishes, while
the other is equal to one; therefore the condition of perfect correlations
required by the EPR reasoning is fulfilled (in fact, the results of the
experiments are always opposed, instead of equal, but it is easy to convince
oneself that this does not have any impact on the reasoning).

\subsubsection{Proof}

\label{proof}

We now come back to the line of the EPR\ theorem.\ In the framework of strict
deterministic theories, the proof of the Bell theorem is the matter of a few
lines; the longest part is actually the definition of the notation. Following
Bell, we assume that $\lambda$ represents all ``elements of reality''
associated to the spins; it should be understood that $\lambda$ is only a
concise notation which may summarize a vector with many components, so that we
are not introducing any limitation here. In fact, one can even include in
$\lambda$ components which play no special role in the problem; the only thing
which matters it that $\lambda$ does contain all the information concerning
the results of possible measurements performed on the spins. We use another
classical notation, $A$ and $B$, for these results, and small letters $a$ and
$b$ for the settings (parameters) of the corresponding apparatuses.\ Clearly
$A$ and $B$ may depend, not only on $\lambda$, but also on the settings $a$
and $b$; nevertheless locality requests that $b$ has no influence on the
result $A$ (since the distance between the locations of the measurements can
be arbitrarily large); conversely, $a$ has no influence on result $B$.\ We
therefore call $A(a,\lambda)$ and $B(b,\lambda)$ the corresponding functions
(their values are either $+1$ or $-1$).

In what follows, it is sufficient to consider two directions only for each
separate measurement; we then use the simpler notation:
\begin{equation}
A(a,\lambda)=A\;\;\;\;\;\;;\;\;\;\;\;\;\;\;A(a^{^{\prime}},\lambda
)=A^{^{\prime}} \label{3}%
\end{equation}
and:
\begin{equation}
B(b,\lambda)=B\;\;\;\;\;\;;\;\;\;\;\;\;\;\;B(b^{^{\prime}},\lambda
)=B^{^{\prime}} \label{4}%
\end{equation}
For each pair of particles, $\lambda$ is fixed, and the four numbers have
well-defined values (which can only be $\pm1$).\ With Eberhard \cite{Eberhard}
we notice that the product:
\begin{equation}
M=AB+AB^{^{\prime}}-A^{^{\prime}}B+A^{^{\prime}}B^{^{\prime}}=(A-A^{^{\prime}%
})B+(A+A^{^{\prime}})B^{^{\prime}} \label{5}%
\end{equation}
is always equal to either $+2$, or to $-2$; this is because one of the
brackets in the right hand side of this equation always vanishes, while the
other is $\pm2$. Now, if we take the average value of $M$ over a large number
of emitted pairs (average over $\lambda$), since each instance of $M$ is
limited to these two values, we necessarily have:
\begin{equation}
-2\leq\,\,\,\,\,\,\,<M>\,\,\,\,\,\,\,\leq+2 \label{6}%
\end{equation}
This is the so called BCHSH\ form \cite{BCHSH} of the Bell theorem: the
average values of all possible kinds of measurements that provide random
results, whatever the mechanism behind them may be (as long as the randomness
is local and arises from the effect of some common fluctuating cause in the
past), necessarily obey this strict inequality.

\subsubsection{Contradiction with quantum mechanics and with experiments}

\label{contradictions}

The generality of the proof is such that one could reasonably expect that any
sensible physical theory will automatically give predictions that also obey
this inequality; the big surprise was to realize that quantum mechanics does
not: it turns out that, for some appropriate choices of the four directions
$a$, $a^{^{\prime}}$, $b$, $b^{^{\prime}}$ (the precise values do not matter
for the discussion here), the inequality is violated by a factor $\sqrt{2}$,
which is more than 40\%. Therefore, the EPR-Bell reasoning leads to a
quantitative contradiction with quantum mechanics; indeed, the latter is not a
local realistic theory in the EPR\ sense. How is this contradiction possible,
and how can a reasoning that is so simple be incorrect within quantum
mechanics? The answer is the following: what is wrong, if we believe quantum
mechanics, is to attribute well-defined values $A$, $A^{^{\prime}}$, $B$,
$B^{^{\prime}}$ to each emitted pair; because only two of them at maximum can
be measured in any experiment, we can not speak of these four quantities, or
reason on them, even as unknown quantities.\ As nicely emphasized by Peres in
an excellent short article \cite{Peres}, ``unperformed experiments have no
result'', that is all!

Wheeler expresses a similar idea when he writes: \textquotedblleft No
elementary quantum phenomenon is a phenomenon until it is a recorded
phenomenon\textquotedblright\ \cite{Wheeler}. As for Wigner, he emphasizes in
\cite{Wigner-AJP} that the proof of the Bell inequalities relies on a very
simple notion: the number of categories into which one can classify all pairs
of particles\footnote{In this reference, Wigner actually reasons explicitly in
terms of hidden variables; he considers domains for these variables, which
correspond to given results for several possible choices of the settings.\ But
these domains also correspond to categories of pairs of particles, which is
why, here, we use the notion of categories.}.\ Each category is associated
with well-defined results of measurements, for the various choices of the
settings $a$ and $b$ that are considered; in any sequence of repeated
experiments, each category will contribute with some given weight, its
probability of occurrence, which has to positive or zero.\ Wigner then notes
that, if one introduces the notion of locality, each category becomes the
intersection of a sub-ensemble that depends on $a$ only, by another
sub-ensemble that depends on $b$ only.\ This operation immediately reduces the
number of categories: in a specific example (involving three possible values
of each setting), he shows that their number reduces from $4^{9}$ to
$(2^{3})^{2}=2^{6}$; with two values only for each setting, the reduction
would be from $4^{4}$ to $(2^{2})^{2}=2^{4}$.\ The mathematical origin of the
Bell inequalities lies precisely in the possibility of distributing all pairs
into this smaller number of categories, with positive probabilities.

A general way to express the Bell theorem in logical terms is to state that
the following system of three assumptions (which could be called the EPR
assumptions) is self-contradictory:

1. validity of their notion of ``elements of reality''

2. locality

3. the predictions of quantum mechanics are always correct.\newline The Bell
theorem then becomes a useful tool to build a ``reductio ad absurdum''
reasoning: it shows that, among all three assumptions, one (at least) has to
be given up.\ The motivation of the experimental tests of the Bell
inequalities was precisely to check if it was not the third which should be
abandoned. Maybe, after all, the Bell theorem is nothing but an accurate
pointer towards exotic situations where the predictions of quantum mechanics
are so paradoxical that they are actually wrong?\ Such was the hope of some
theorists, as well as the exciting challenge to experimentalists.

Experiments were performed in the seventies, initially with photons
\cite{Clauser} \cite{Fry} where they already gave very clear results, as well
as with protons \cite{Lamehi}; in the eighties, they were made more and more
precise and convincing \cite{Aspect} - see also \cite{Kleinpoppen}; ever
since, they have been constantly improved (see for instance \cite{parametric},
but the list of references is too long to be given here); all these results
have clearly shown that, in this conflict between local realism and quantum
mechanics, the latter wins completely. A fair summary of the situation is
that, even in these most intricate situations invented and tested by the
experimentalists, no one has been able to disprove quantum mechanics.\ In this
sense, we can say that Nature obeys laws which are non-local, or non-realist,
or both. It goes without saying that no experiment in physics is perfect, and
it is always possible to invent ad hoc scenarios where some physical
processes, for the moment totally unknown, ``conspire'' in order to give us
the illusion of correct predictions of quantum mechanics - we come back to
this point in \S \ \ref{loopholes} - but the quality and the number of the
experimental results does not make this attitude very attractive intellectually.

\subsubsection{Generality of the theorem}

\label{general}

We have already mentioned that several generalizations of the Bell theorem are
possible; they are at the same time mathematically simple and conceptually
interesting.\ For instance, in some of these generalizations, it is assumed
that the result of an experiment becomes a function of several fluctuating
causes: the fluctuations taking place in the source as usual, but also
fluctuations taking place in the measuring apparatuses \cite{Bell-Varenna},
or/and perturbations acting on the particles during their motion towards the
apparatuses; actually, even fundamentally indeterministic (but local)
processes may influence the results.\ The two former cases are almost trivial
since they just require the addition of more dimensions to the vector variable
$\lambda$; the latter requires replacing the deterministic functions $A$ and
$B$ by probabilities, but this is also relatively straightforward \cite{FL}
(see also footnote 10 in \cite{Bell-Varenna} and appendix I of this article).
Moreover, one should realize that the role of the $A$ and $B$ functions is
just to relate the conditions of production of a pair of particles (or of
their propagation) to their behavior when they reach the measurements
apparatuses (and to the effects that they produce on them); they are, so to
say, solutions of the equation of motion whatever these are.\ The important
point is that they may perfectly include, in a condensed notation, a large
variety of physical phenomena: propagation of point particles, propagation of
one or several fields from the source to the detectors (see for instance the
discussion in \S 4 of \cite{Bertlmann}), particles and fields in interaction,
or whatever process one may have in mind (even random propagations can be
included) - as long as they do not depend on the other setting ($A$ is
supposed to be a function of $a$, not of $b$).\ The exact mathematical form of
the equations of propagation is irrelevant; the essential thing is that the
functions exist.

Indeed, what really matters for the proof of the Bell theorem is the
dependence with respect to the settings $a$ and $b$: \ the function $A$ must
depend on $a$ only, while $B$ must depend on $b$ only.\ Locality expressed
mathematically in terms of $a$ and $b$ is the crucial ingredient.\ For
instance we could, if we wished, assume that the result $A$ of one measurement
is also function of fluctuating random variables attached to the other
apparatus, which introduces a non-local process; but this does not create any
mathematical problem for the proof (as long as these variables are not
affected by setting $b$). On the other hand, if $A$ becomes a function of $a$
and $b$ (and/or the same for $B$), it is easy to see that the situation is
radically changed: in the reasoning of \S \ \ref{proof} we must now associate
8 numbers to each pair (since there are two results to specify for each of the
4 different combinations of settings), instead of 4, so that the proof
miserably collapses. Appendix I gives another concrete illustration showing
that it is locality, not determinism, which is at stake; see also the appendix
of \cite{FL}).

Needless to say, the independence of $A$ of $b$ does not mean that the result
observed on one side, $A$, is independent of the outcome at the other side,
$B$: one should not confuse setting and outcome dependences!\ It is actually
clear that, in any theory, the correlations would disappear if outcome
dependence was totally excluded.\ We should also mention that the setting
dependence is subject to some constraints, if the theory is to remain
compatible with relativity.\ If, for instance, the probability of observation
of the results on one side, which is a sum of probabilities over the various
possible outcomes on the other side, was still a function of the other
setting, one would run into incompatibility; this is because one could use the
device to send signals without any fundamental delay, thus violating the
constraints of relativity.\ \ See refs. \cite{Jarrett} and
\cite{Ballentine-Jarrett} for a discussion in terms of ``strong locality'' and
``predictive completeness'' (or ``parameter independence'' and of ``outcome
independence'' in ref.\cite{Shimony-events}). Appendix IV discusses how the
general formalism of quantum mechanics manages to ensure compatibility with relativity.

An interesting generalization of the Bell theorem, where time replaces the
settings, has been proposed by Franson \cite{Franson} and implemented in
experiments for an observation of a violation of the Bell inequalities (see
for instance \cite{Brendel}); another generalization shows that a violation of
the Bell inequalities is not limited to a few quantum states (singlet for
instance), but includes all states that are not products \cite{Gisin}
\cite{Popescu-Rohrlich}. For a general discussion of the conceptual impact of
a violation of the inequalities, we refer to the book collecting Bell's
articles \cite{Bell-livre}.

We wish to conclude this section by emphasizing that the Bell theorem is much
more general than many people think.\ All potential authors on the subject
should think twice and remember this carefully before taking their pen and
sending a manuscript to a physics journal: every year a large number of them
is submitted, with the purpose of introducing ``new'' ways to escape the
constraints of the Bell theorem, and to ``explain'' why the experiments have
provided results that are in contradiction with the inequalities.\ According
to them, the non-local correlations would originate from some new sort of
statistics, or from perturbations created by cosmic rays, gas collisions with
fluctuating impact parameters, etc. The imagination is the only limit of the
variety of the processes that can be invoked, but we know from the beginning
that all these attempts are doomed to failure.\ The situation is analogous to
the attempts of past centuries to invent ``perpetuum mobile'' devices: even if
some of these inventions were extremely clever, and if it is sometimes
difficult to find the exact reason why they can not work, it remains true that
the law of energy conservation allows us to know at once that they cannot.\ In
the same way, some of these statistical ``Bell beating schemes'' may be
extremely clever, but we know that the theorem is a very general theorem in
statistics: in all situations that can be accommodated by the mathematics of
the $\lambda$'s and the $A$ and $B$ functions (and there are many!), it is
impossible to escape the inequalities.\ No, non-local correlations can not be
explained cheaply; yes, a violation of the inequalities is therefore a very,
very, rare situation.\ In fact, until now, it has never been observed, except
of course in experiments designed precisely for this purpose.\ In other words,
if we wanted to build automata including arbitrarily complex mechanical
systems and computers, we could never mimic the results predicted by quantum
mechanics (at least for remote measurements); this will remain impossible
forever, or at least until completely different computers working on purely
quantum principles are built\footnote{In terms of the Mendel parable: an
observation of a violation of the Bell inequalities would imply that something
inside both peas (maybe a pair of DNA molecules?) remains in a coherent
quantum superposition, without decoherence, even if the distance between the
peas is large.}.

\subsection{Hardy's impossibilities}

\label{hardy}

Another scheme of the same conceptual type was introduced recently by Hardy
\cite{Hardy}; it also considers two particles but it is nevertheless
completely different since it involves, instead of mathematical constraints on
correlation rates, the very possibility of occurrence for some type of events
- see also \cite{Mermin3} for a general discussion of this interesting
contradiction. As in \S \ \ref{proof}, we assume that the first particle may
undergo two kinds of measurements, characterized by two values $a$ and
$a^{^{\prime}}$ of the first setting; if we reason as in the second half of
\S \ \ref{proof}, within the frame of local realism, we can call $A$ and
$A^{^{\prime}}$ the corresponding results.\ Similar measurements can be
performed on the second particle, and we call $B$ and $B^{^{\prime}}$ the results.

Let us now consider three types of situations:

(i) settings without prime: we assume that the result $A=1$, $B=1$ is
sometimes obtained.

(ii) one prime only: we assume that the ``double one'' is impossible, in other
words that one never gets $A=1$, $B^{^{\prime}}=1$, and never $A^{^{\prime}%
}=1$, $B=1$ either.

(iii) double prime settings: we assume that ``double minus one'' is
impossible, in other words that $A^{^{\prime}}=-1$, $B^{^{\prime}}=-1$ is
never observed.

A closer inspection shows that these three assumptions are in fact
incompatible.\ To see why, let us for instance consider the logical scheme of
figure 2, where the upper part corresponds to the possibility opened by
statement (i); statement (ii) then implies that, if $A=1$, one necessarily has
$B^{^{\prime}}=-1$, which explains the first diagonal in the figure; the
second diagonal follows by symmetry.\ Then we see that all events
corresponding to the results $A=B=1$ also necessarily correspond to
$A^{^{\prime}}=B^{^{\prime}}=-1$, so that a contradiction with statement (iii)
appears: the three propositions are in fact incompatible.\ A way to express it
is to say that the ``sometimes'' of (i) is contradictory with the ``never'' of
proposition (iii). \begin{figure}[ptb]
\begin{center}
\includegraphics[width=8cm]{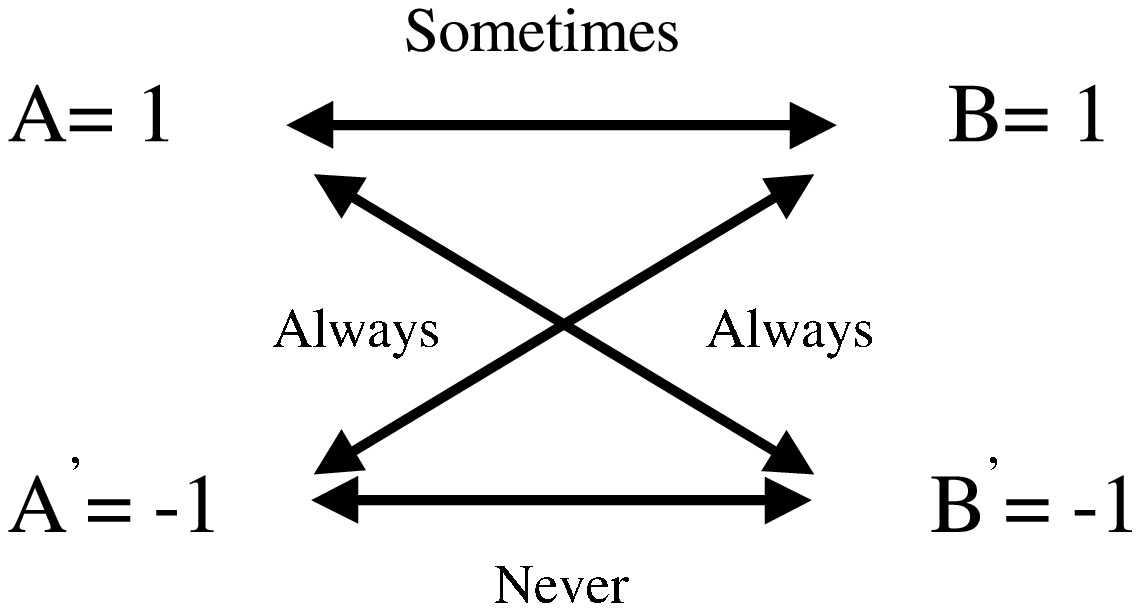}\newline
\end{center}
\par
\label{fig 2}\end{figure}

But it turns out that quantum mechanics does allow a simultaneous realization
of all three propositions! To see how, let us for instance consider a two-spin
state vector of the form:%

\begin{equation}
\mid\Psi>=\alpha\mid+,->+\beta\mid-,+>+\gamma\mid+,+> \label{h1}%
\end{equation}
where the $\mid\pm,\pm>$ refer to eigenstates of $A^{^{\prime}}$ and
$B^{^{\prime}}$ (NB: axis $Oz$ is chosen as the direction of measurement
associated with primed operators). From the beginning, the absence of any
$\mid\Psi>$ component on $\mid-,->$ ensures that proposition (iii) is
true.\ As for the measurements without prime, we assume that they are both
performed along a direction in the plane $xOz$ that makes an angle $2\theta$
with $Oz$; the eigenstate with eigenvalue $+1$ associated in the single-spin
state is then merely:
\begin{equation}
\cos\theta\mid+>+\sin\theta\mid-> \label{h2}%
\end{equation}
The first state excluded by proposition (ii) (diagonal in figure 2) is then
the two-spin state:
\begin{equation}
\cos\theta\mid+,+>+\sin\theta\mid+,-> \label{h3}%
\end{equation}
while the second is:
\begin{equation}
\cos\theta\mid+,+>+\sin\theta\mid-,+> \label{h4}%
\end{equation}
so that the two exclusion conditions are equivalent to the conditions:
\begin{equation}
\alpha\sin\theta+\gamma\cos\theta=\beta\sin\theta+\gamma\cos\theta=0
\label{h5}%
\end{equation}
or, within a proportionality coefficient:
\begin{equation}
\alpha=\beta=-\gamma\cot\theta\label{h6}%
\end{equation}

This arbitrary coefficient may be used to write $\mid\Psi>$ in the form:
\begin{equation}
\mid\Psi>=-\cos\theta\left(  \mid+,->+\mid-,+>\right)  +\sin\theta\mid+,+>
\label{h7}%
\end{equation}
The last thing to do is to get the scalar product of this ket by that where
the two spins are in the state (\ref{h2}); we get the result:
\begin{equation}
-\sin\theta\cos^{2}\theta\label{h8}%
\end{equation}
The final step is to divide this result by the square of the norm of ket
(\ref{h7}) in order to obtain the probability of the process considered in
(iii); this is a straightforward calculation (see appendix II), but here we
just need to point out that the probability is not zero; the precise value of
its $\theta$ maximum found in appendix II is about $9\%$.\ This proves that
the pair of results considered in proposition (i) can sometimes be obtained
together with (ii) and (iii): indeed, in $9\%$ of the cases, the predictions
of quantum mechanics are in complete contradiction with those of a local
realist reasoning.

An interesting aspect of the above propositions is that they can be
generalized to an arbitrary number of measurements \cite{Hardy2}; it turns out
that this allows a significant increase of the percentage of ``impossible
events'' (impossible within local realism) predicted by quantum mechanics -
from $9\%$ to almost $50\%$!\ The generalization involves a chain, which keeps
the two first lines (i) and (ii) unchanged, and iterates the second in a
recurrent way, by assuming that:

(iii) for measurements of the type ($a^{^{\prime}}$, $b^{^{\prime\prime}}$) or
($a^{^{\prime\prime}},$ $b^{^{\prime}}$), one never gets opposite
results\footnote{In fact, the reasoning just requires that the pair $-1$, $+1$
is never obtained, and does not require any statement about $+1$, $-1$.}.

(iv) for measurements of the type ($a^{^{\prime\prime}}$, $b^{^{^{\prime
\prime\prime}}}$) or ($a^{^{^{\prime\prime\prime}}},$ $b^{^{\prime\prime}}$),
one never gets opposite results.

etc..

(n) finally, for measurement of the type ($a^{n}\,$,$\,b^{n}$), one never gets
$-1$ and $-1$.

The incompatibility proof is very similar to that given above; it is
summarized in figure 3. In both cases, the way to resolve the contradiction is
the same as for the Bell theorem: in quantum mechanics, it is not correct to
reason on all 4 quantities $A$, $A^{^{\prime}}$, $B$ and $B^{^{\prime}}$for a
given pair of spins, even as quantities that are unknown and that could be
determined in a future experiment. This is simply because, with a given pair,
it is obviously impossible to design an experiment that will measure all of
them: they are incompatible.\ \ If we insisted on introducing similar
quantities to reproduce the results of quantum mechanics, we would have to
consider 8 quantities instead of 4 (see second paragraph of \S \
\ref{general}).\ For a discussion of non-local effects with other states, see
\cite{Goldstein2}. \begin{figure}[ptb]
\begin{center}
\includegraphics[width=7cm]{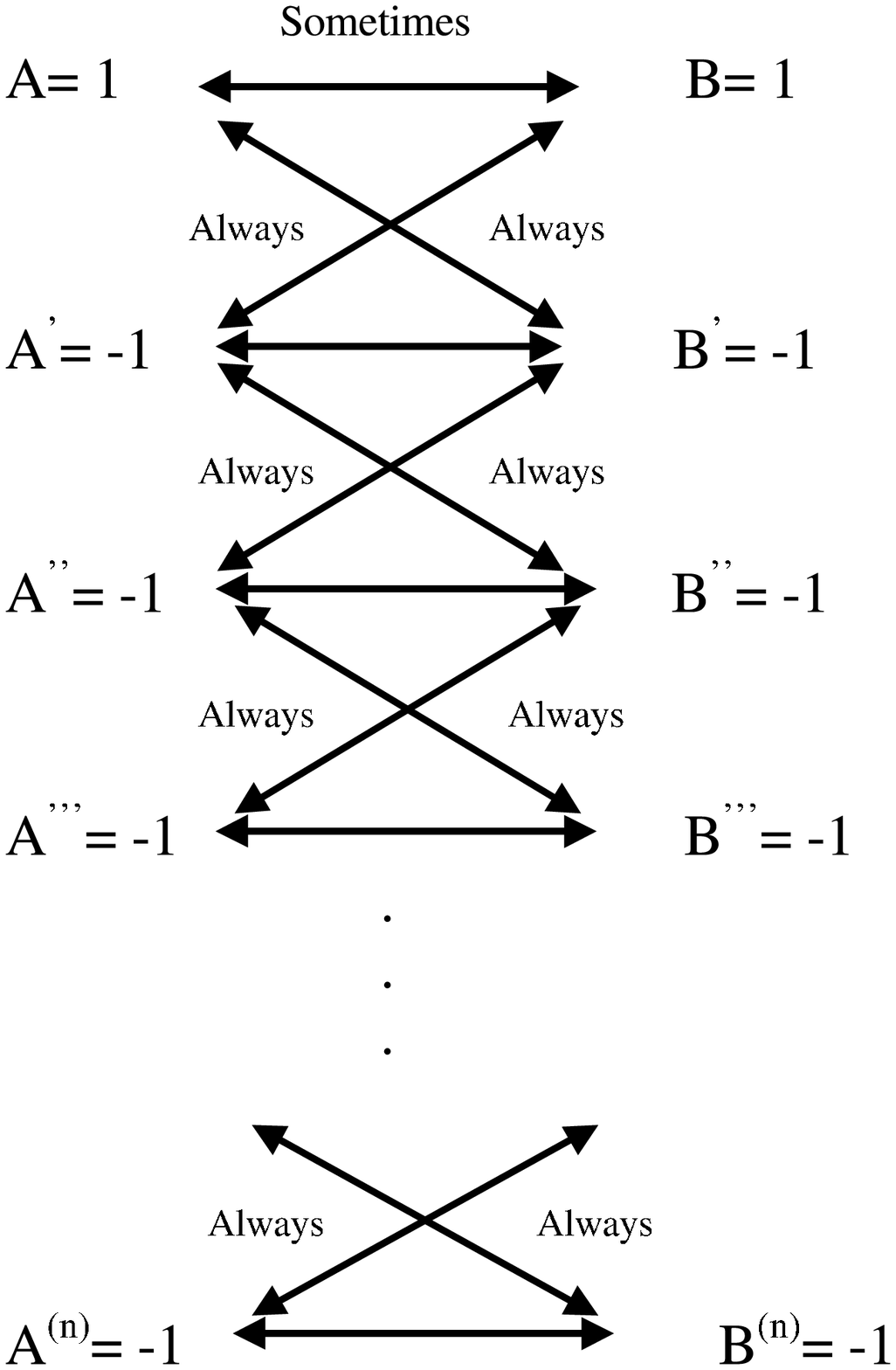}\newline
\end{center}
\caption{{}}%
\label{fig 3}%
\end{figure}

\subsection{GHZ equality}

\label{ghz}

For many years, everyone thought that Bell had basically exhausted the subject
by considering all really interesting situations, in other words that two-spin
systems provided the most spectacular quantum violations of local realism. It
therefore came as a surprise to many when in 1989 Greenberger, Horne and
Zeilinger (GHZ) showed that this is not true: as soon as one considers systems
containing more than two correlated particles, even more dramatic violations
of local realism become possible in quantum mechanics, and even without
involving inequalities. Here, we limit ourselves to the discussion of three
particle systems, as in the original articles \cite{GHZ} \cite{GHZ-bis}, but
generalization to $N$ particles are possible; see for instance
\S \ \ \ref{complete-ent} or \cite{FL2}. While \cite{GHZ-bis} discussed the
properties of three correlated photons, each emitted through two pinholes and
impinging beam splitters, we will follow ref.\ \cite{Mermin5} and consider a
system of three $1/2$ spins (external variables play no role here); we assume
that the system is described by the quantum state:%

\begin{equation}
\mid\Psi>=\frac{1}{\sqrt{2}}\left[  \mid+,+,+>-\mid-,-,->\right]  \label{7}%
\end{equation}
where the $\mid\pm>$ states are the eigenstates of the spins along the $Oz$
axis of an orthonormal frame $Oxyz$. We now calculate the quantum
probabilities of measurements of the spins $\sigma_{1,2,3}$ \ of the three
particles, either along direction $Ox$, or along direction $Oy$ which is
perpendicular.\ More precisely, we assume that what is measured is not
individual spin components, but only the product of three of these components,
for instance $\sigma_{1y}\times\sigma_{2y}\times\sigma_{3x}$.\ A
straightforward calculation (see appendix III) shows that:
\begin{equation}%
\begin{array}
[c]{ll}%
\mathcal{P}(\sigma_{1y}\times\sigma_{2y}\times\sigma_{3x}=1) & =+1\\
\mathcal{P}(\sigma_{1x}\times\sigma_{2y}\times\sigma_{3y}=1) & =+1\\
\mathcal{P}(\sigma_{1y}\times\sigma_{2x}\times\sigma_{3y}=1) & =+1
\end{array}
\label{8}%
\end{equation}
In fact, the state vector written in (\ref{7}) turns out to be a common
eigenstate to all three operator products, so that each of them takes a value
$+1$ that is known before the measurement\footnote{But, if the product is
fixed, each of the individual components still fluctuates with a 100\%
amplitude, between results $+1$ and $-1$.}. Now, if we consider the product of
three spin components along $Ox$, it is easy to check (appendix III) that the
same state vector is also an eigenstate of the product operator $\sigma
_{1x}\times\sigma_{2x}\times\sigma_{3x}$, but now with eigenvalue $-1$, so
that:
\begin{equation}
\mathcal{P}(\sigma_{1x}\times\sigma_{2x}\times\sigma_{3x}=-1)=1 \label{9}%
\end{equation}
This time the result is $-1$, with probability $1$, that is with certainty.

Let us now investigate the predictions of a local realist EPR type point of
view in this kind of situation.\ Since the quantum calculation is so
straightforward, it may seem useless: indeed, no one before GHZ suspected that
anything interesting could occur in such a simple case, where the initial
state is an eigenstate of all observables considered, so that the results are
perfectly certain.\ But, actually, we will see that a complete contradiction
emerges from this analysis! The local realist reasoning is a simple
generalization of that given in \ref{proof}; we call $A_{x,y}$ the results
that the first spin will give for a measurement, either along $Ox$, or $Oy$;
similar letters $B$ and $C$ are used for the measurement on the two other
spins. From the three equalities written in (\ref{8}) we then get:
\begin{equation}%
\begin{array}
[c]{ll}%
A_{y}B_{y}C_{x} & =1\\
A_{x}B_{y}C_{y} & =1\\
A_{y}B_{x}C_{y} & =1
\end{array}
\label{10}%
\end{equation}
Now, if we assume that the observations of the three spins are performed in
three remote regions of space, locality implies that the values measured for
each spin should be independent of the type of observation performed on the
two other spins.\ This means that the same values of $A$, $B$ and $C$ can be
used again for the experiment where the three $Ox$ components are measured:
the result is merely the product $A_{x}B_{x}C_{x}$. But, since the squares
$A_{y}^{2}$ etc. are always equal to $+1$, we can obtain this result by
multiplying all three lines of equation (\ref{10}), which provides:
\begin{equation}
A_{x}B_{x}C_{x}=+1 \label{11}%
\end{equation}
But equality (\ref{9}) predicts the opposite sign!

Here we obtain a contradiction that looks even more dramatic than for the Bell
inequalities: the two predictions do not differ by some significant fraction
(about $40\%$), they are just completely opposite. In addition, all
fluctuations are eliminated since all of the results (the products of the
three components) are perfectly known before measurement: the $100\%$
contradiction is obtained with $100\%$ certainty! Unfortunately, this does not
mean that, experimentally, tests of the GHZ equality are easy.\ Three
particles are involved, which must be put in state (\ref{7}), surely a
non-trivial task; moreover one has to design apparatuses that measure the
product of three spin components.\ To our knowledge, no experiment analogous
to the Bell inequality experiments has been performed on the GHZ equality yet,
at least with macroscopic distances; only microscopic analogues have been
observed, in NMR experiments \cite{NMR-GHZ} - for recent proposals, see for
instance \cite{NMR-GHZ-2} \cite{NMR-GHZ-3}. Nevertheless constant progress in
the techniques of quantum electronics is taking place, and GHZ entanglement
has already been observed \cite{GHZ-1} \cite{GHZ-1bis}, so that one gets the
impression that a full experiment is not too far away in the future.

In a GHZ situation, how precisely is the conflict between the reasoning above
and quantum mechanics resolved? There are different stages at which this
reasoning can be put into question.\ First, we have assumed locality, which
here takes the form of non-contextuality (see \S \ \ref{BKS}): each of the
results is supposed to be independent of the nature of the measurements that
are performed on the others, because they take place in remote regions of
space.\ Clearly, there is no special reason why this should necessarily be
true within quantum mechanics. Second, we have also made assumptions
concerning the nature of the ``elements of reality'' attached to the
particles.\ In this respect, it is interesting to note that the situation is
very different from the EPR-Bell or Hardy cases: Bohr could not have replied
that different elements of reality should be attached to different
experimental setups!\ In the GHZ argument, it turns out that all four quantum
operators corresponding to the measurements commute, so that there is in
principle no impossibility of measuring all of them with a single setup. But
the local realist reasoning also assumes that a measurement of the product of
three operators is equivalent to a separate measurement of each of them, which
attributes to them separate elements of reality. In the formalism of quantum
mechanics, the question is more subtle.\ \ It turns out that the measurement
of a single product of commuting operators is indeed equivalent to the
measurement of each of them; but this is no longer the case for several
product operators, as precisely illustrated by those introduced above:
clearly, all 6 spin component operators appearing in the formulas do not
commute with each other.\ It is therefore impossible to design a single
experimental setup to have access to all 6 quantities $A_{x,y}$, $B_{x,y}$ and
$C_{x,y}$ that we have used in the local realist proof\footnote{The ideal GHZ
experiment would therefore involve only measurements of commuting observables,
i.e. products measured directly without measuring each factor separately.\ In
practice, it is probably easier to measure each factor in the product; if all
four products are needed, this necessarily implies successive measurements of
incompatible observables with different experimental setups; the price to pay,
then, is that loopholes such as the ``biased sample loophole''
(\S \ \ \ref{loopholes}) may be opened again in the interpretation of the
results.}.

When the measurements are imperfect, the GHZ equality can give rise to
inequalities (as in the BCHSH theorem), as discussed in \cite{GHZ-bis} and
\cite{Mermin4}; this latter reference also presents a generalization to an
arbitrary number $N$ of particles; in the same line, ref.\ \cite{FL2} provides
a discussion of the $N$-particle correlation function with varying angles for
the analyzers, which we partially reproduce in \S \ \ref{complete-ent}.

\subsection{Bell-Kochen-Specker; contextuality.}

\label{BKS}

Another theorem was introduced also by Bell \cite{Bell-1} as well as
(independently and very shortly after) by Kochen and Specker
\cite{Kochen-Specker}, hence the name ``BKS theorem'' that is often used for
it. This theorem is not particularly related to locality, as opposed to those
that we have already discussed in the preceding subsections.\ It is actually
related to another notion, called ``contextuality'': an additional variable
attached to a physical system is called ``contextual'' if its value depends,
not only of the physical quantity that it describes, but also of the other
physical quantities that can be measured at the same time on the same system
(in quantum mechanics they correspond to commuting observables). If, on the
other hand, its value is completely independent of all the other observables
that the experimenter may decide to measure at the same time, the additional
variable is called ''non-contextual''; one can then say that it describes a
property of the physical system only, and not a combined property of the
system and the measurement apparatus; it may have pre-existed in the system
before any measurement.\ The notion of distance is no longer relevant in this
context; for instance, the theorem applies to a single system with no
extension in space.

Let us first consider a spin 1 particle in quantum mechanics, with three
quantum states $\mid-1>$ $\mid0>$ and $\mid+1>$ as a basis of a state space of
the dimension 3.\ The three components $S_{x}$, $S_{y}$, and $S_{z}$, do not
commute (they obey the usual commutation relation for the angular momentum),
but it is easy to show that all the squares of all these three operators do
commute; this is a specific property of angular momentum 1, and can be seen by
an elementary matrix calculation with the usual operators $S_{\pm}%
$.\ Moreover, the sum of these squares is a constant (a c-number) since:%
\begin{equation}
S_{x}^{2}+S_{y}^{2}+S_{z}^{2}=2\hbar^{2} \label{12}%
\end{equation}
It is not against any fundamental principle of quantum mechanics, therefore,
to imagine a triple measurement of the observables $S_{x}^{2}$, $S_{y}^{2}$,
and $S_{z}^{2}$; we know that the sum of the three results will always be $2$
(from now on we drop the factor $\hbar^{2}$, which plays no role in the
discussion). Needless to say, the choice of the three orthogonal directions is
completely arbitrary, and the compatibility is ensured for any choice of this
triad, but not more than one: the measurements for different choices remain
totally incompatible.

In passing, we note that the measurement of the square $S_{x}^{2}$ of one
component cannot merely be seen as a measurement of $S_{x}$ followed by a
squaring calculation made afterwards by the experimentalist! Ignoring
information is not equivalent to not measuring it (we come to this point in
more detail, in terms of interferences and decoherence, at the end of
\S \ \ref{correlation}).\ There is indeed less information in $S_{x}^{2}$ than
in $S_{x}$ itself, since the former has only two eigenvalues ($1$ and $0$),
while the latter has three ($-1$ is also a possible result).\ What is needed
to measure $S_{x}^{2}$ is, for instance, a modified Stern-Gerlach system where
the components of the wave function corresponding to results $\pm1$ are not
separated, or where they are separated but subsequently grouped together in a
way they makes them impossible to distinguish. Generally speaking, in quantum
mechanics, measuring the square of an operator is certainly not the same
physical process as measuring the operator itself!

Now, suppose that we try to attach to each individual spin an EPR element of
reality/additional variable that corresponds to the result of measurement of
$S_{x}^{2}$; by symmetry, we will do the same for the two other components, so
that each spin now gets three additional variables $\lambda$ to which we may
attribute values that determine the possible results: $1$ or $0$.\ The results
are described by functions of these variables, which we note $A_{x,y,z}$:%
\begin{equation}
\text{ \ }\,A_{x}=0\text{\ or }1\text{;\ }\,A_{y}=0\text{\ or }1\text{;}%
\,\ A_{z}=0\text{\ or }1 \label{13}%
\end{equation}
At first sight, this seems to provide a total number of 8 possibilities; but,
if we want to preserve relation (\ref{12}), we have to select among these 8
possibilities only those three for which two $A$'s are one, one is zero.\ As
traditional, for this particular spin we then attribute colors to the three
orthogonal directions $Ox$, $Oy$ and $Oz$: the two directions that get an
$A=1$ are painted in red, the last in blue \cite{Belifante}.

The same operation can obviously be made for all possible choices of the
triplet of directions $Oxyz$. A question which then naturally arises is: for
an arbitrary direction, can one attribute a given color (a given value for
$\,A_{x}$) that remains independent of the context in which it was
defined?\ Indeed, we did not define the value as a property of an $Ox$
direction only, but in the context of two other directions $Oy$ and $Oz$; the
possibility of a context independent coloring is therefore not obvious.\ Can
we for instance fix $Oz$ and rotate $Ox$ and $Oy$ around it, and still keep
the same color for $Oz$? We are now facing an amusing little problem of
geometry that we might call ``ternary coloring of all space
directions''.\ Bell as well as Kochen and Specker showed that this is actually
impossible; for a proof see either the original articles, or the excellent
review \cite{Mermin} given by Mermin.

In the same article, this author shows how the complications of the
geometrical problem may be entirely avoided by going to a space of states of
dimension $4$ instead as $3$.\ He considers two spin 1/2 particles and the
following table of $9$ quantum variables (we use the same notation as in
\S \ref{ghz}):%
\begin{equation}%
\begin{array}
[c]{ccc}%
\sigma_{x}^{1} & \sigma_{x}^{2} & \sigma_{x}^{1}\sigma_{x}^{2}\\
\sigma_{y}^{2} & \sigma_{y}^{1} & \sigma_{y}^{1}\sigma_{y}^{2}\\
\sigma_{x}^{1}\sigma_{y}^{2} & \sigma_{y}^{1}\sigma_{x}^{2} & \sigma_{z}%
^{1}\sigma_{z}^{2}%
\end{array}
\label{14}%
\end{equation}
All operators have eigenvalues $\pm1$.$\;$It is easy to see why all three
operators belonging to the same line, or to the same column, always commute
(the products of two $\sigma$'s that anti-commute are commuting operators,
since the commutation introduces two $-1$ signs, with cancelling
effects).\ Moreover, the products of all three operators is always $+1$,
except the last column for which it is $-1$\ \footnote{This can easily be
checked from the well-known properties of the Pauli matrices; the minus sign
for the third column comes from the product of the two $i$'s, arising from the
relation $\sigma_{x}\sigma_{y}=i\sigma_{z}$; on the other hand, in the third
line one gets $i\times(-i)=1$ because of the change of order of the
operators.}. Here, instead of an infinite number of triplet of directions in
space, we have $6$ groups of three operators, but the same question as above
arises: can we attribute a color to each of the $9$ elements of matrix
(\ref{14}), red for result $+1$ and yellow for result $-1$, in a way that is
consistent with the results of quantum mechanics?\ For this consistency to be
satisfied, all lines and columns should either contain three red cases, or one
red and two yellow, except the last column that will contain one or three
yellow cases (in order to correspond to $-1$ instead of $+1$).

This little matrix coloring problem is much simpler than the geometrical
coloring problem mentioned above: it is obviously impossible to find $9$
numbers with a product that is at the same time equal to $1$, condition on
rows, and $-1$, condition on columns (we note in passing that Mermin's
reasoning is very close to that of \S \ \ref{ghz}, which illustrates how
similar the GHZ theorem and this form of the BKS theorem are).\ Here, as in
the three direction problem, non-contextuality leads us to an impossible
coloring problem.\ For another illustration of the impossibility, see also
\S \ VI of ref. \cite{Mermin} which deals with three 1/2 spins instead of two.

What can we conclude from this contradiction?\ Certainly that the predictions
of quantum mechanics are incompatible with a non-contextual view on the EPR
elements of reality/additional variables, where the result of the measurement
should depend solely of the system measured - see for instance the discussion
given in ref. \cite{AP}.\ But is this a good argument against these elements
of reality, or at least an indication that, if they exist, their properties
are completely unexpected?\ Not really.\ As Bell noted \cite{Bell-1}, ``the
result of an observation may reasonably depend not only on the state of the
system (including hidden/additional variables) but also on the complete
disposition of the apparatus''.\ There is for instance no special conceptual
difficulty in building a theory where additional variables are attributed to
the apparatuses, and where both kinds of additional variables collaborate in
order to determine the observed result. Violations of the Bell theorem by
quantum mechanics are therefore generally considered as much more significant
quantum manifestations than violations of the BKS theorem.\ For a general
discussion of the status of the various ``impossibility theorems'' with
emphasis on the BKS theorems, see ref.\ \cite{Mermin}

\section{Non-locality and entanglement: where are we now?}

In view of the locality theorems as well as their violation by the modern
experimental results, which were not available when the orthodox
interpretation of quantum mechanics was invented, some physicists conclude
triumphantly: ``Bohr was right!'', while others will claim with the same
enthusiasm ``Bohr was wrong!''.\ Both these opinions make sense, depending on
what aspect of the debate one wishes to favor.\ We have already touched the
question at the end of \S \ \ref{transpo}; here, we will just add that,
whether one personally feels closer to the orthodox camp or to local realism,
it remains clear that the line initiated by Einstein and Bell had the decisive
role in the last 50 years.\ In fact, they are the ones who pointed out the
inadequate character of some impossibility theorems, as well as the crucial
importance of the notion of locality in all these discussions.\ This resulted
in much more progress and understanding than the simple re-statement of the
orthodox position. For instance, even now, the introduction of the reduction
of the state vector is sometimes ``explained'' by invoking the ``unavoidable
perturbations that the measurement apparatus brings to the measured system'' -
see for instance the traditional discussion of the Heisenberg microscope which
still appears in textbooks!\ But, precisely, the EPR-Bell argument shows us
that this is only a cheap explanation: in fact, the quantum description of a
particle can be modified without any mechanical perturbation acting on it,
provided the particle in question was previously correlated with another
particle.\ So, a trivial effect such as a classical recoil effect in a
photon-electron collision cannot be the real explanation of the nature of the
wave packet reduction! It is much more fundamentally quantum and may involve
non-local effects.

Another lesson is that, even if quantum mechanics and relativity are not
incompatible, they do not fit very well together: the notion of events in
relativity, which are supposed to be point-like in space-time, or the idea of
causality, are still basic notions, but not as universal as one could have
thought before the Bell theorem. Indeed, quantum mechanics teaches us to take
these notions ``with a little grain of salt''.\ Still another aspect is
related to the incredible progress that experiments have made in the $20^{th}$
century, whether or not stimulated by fundamental quantum mechanics.\ One gets
the impression that this progress is such that it will allow us to have access
to objects at all kinds of scale, ranging from the macroscopic to the
microscopic.\ Therefore, while at Bohr's time one could argue that the precise
definition of the border line between the macroscopic world of measurement
apparatuses was not crucial, or even academic, the question may become of real
importance; it may, perhaps, even give rise to experiments one day.\ All these
changes, together, give the impression that the final stage of the theory is
not necessarily reached and that conceptual revolutions are still possible.

In this section, we give a brief discussion of some issues that are related to
quantum non-locality and entanglement, with some emphasis on those that are
presently, or may soon be, the motivation of experiments (\S \ \ref{contra} is
an exception, since it is purely theoretical).\ Going into details would
certainly bring us beyond the scope of this article, so that we will limit
ourselves to a few subjects, which we see as particularly relevant, even if
our choice may be somewhat arbitrary.\ Our main purpose is just to show that,
even if theoretically it is really difficult to add anything to what the
founding fathers of quantum mechanics have already said long ago, it still
remains possible to do interesting physics in the field of fundamental quantum
mechanics! Even if we treat the subject somewhat superficially, the hope is
that the reader will be motivated to get more precise information from the references.

\subsection{Loopholes, conspiracies}

\label{loopholes}

One sometimes hears that the experiments that have been performed so far are
not perfectly convincing, and that no one should claim that local realism
\`{a} la Bell has been disproved.\ Strictly speaking, this is true: there are
indeed logical possibilities, traditionally called ``loopholes'', which are
still open for those who wish to restore local realism.\ One can for instance
deny the existence of any real conflict between the experimental results and
the Bell inequalities.\ First of all, of course, one can always invoke trivial
errors, such as very unlikely statistical fluctuations, to explain why all
experiments seem to ``mimic'' quantum mechanics so well; for instance some
authors have introduced ad hoc fluctuations of the background noise of
photomultipliers, which would magically correct the results in a way that
would give the impression of exact agreement with quantum mechanics.\ But the
number and variety of Bell type experiments supporting quantum mechanics with
excellent accuracy is now large; in view of the results, very few physicists
seem to take this explanation very seriously.

Then one could also think of more complicated scenarios: for instance, some
local unknown physical variables may couple together in a way that will give
the (false) impression of non-local results, while the mechanism behind them
remains local.\ One possibility is that the polarization analyzers might,
somehow, select a subclass of pairs which depend on their settings; then, for
each choice $(a$, $b)$, only a small fraction of all emitted pairs would be
detected; one could then assume that, when the orientation of the analyzers
are changed by a few degrees, all the pairs that were detected before are
eliminated, and replaced with a completely different category of physical
systems with arbitrary properties.\ In this situation, everything becomes
possible: one can ascribe to each category of pairs whatever ad hoc physical
properties are needed to reproduce any result, including those of quantum
mechanics, while remaining in a perfectly local context.

Indeed, in the derivation of the Bell inequalities, one assumes the existence
of ensemble averages over a non-biased, well defined, ensemble of pairs, which
are completely independent of the settings $a$ and $b$.\ Various proofs of the
Bell inequalities are possible, but in many of them one explicitly writes the
averages with an integral containing a probability distribution $\varrho
(\lambda)$; this function mathematically defines the ensemble on which these
averages are taken. The non-biasing assumption is equivalent to assuming that
$\rho$ is independent of $a$ and $b$; on the other hand, it is easy to
convince oneself that the proof of the Bell inequalities is no longer possible
if $\rho$ becomes a function of $a$ and $b$. In terms of the reasoning of
\S \ \ref{proof}, where no function $\rho$ was introduced, what we have
assumed is that the four numbers $A$, $A^{^{\prime}}$, $B$ and $B^{^{\prime}}$
are all attached to the same pair; if $M$ was built from more numbers, such as
numbers associated to different pairs, the algebra would clearly no longer
hold, and the rest of the proof of the inequality would immediately collapse..

Of course, no problem occurs if every emitted pair\ is detected and provides
two results $\pm1$, one on each side, whatever the choice of $a$ and $b$ (and
even if this choice is made after the emission of the pair).\ It then makes
sense to obtain the ensemble average $<M>$ from successive measurements of
four average values $<AB>$, $<AB^{\prime}>$, etc..\ But, if many pairs are
undetected, one can not be completely sure that the detection efficiency
remains independent of the settings $a$ and $b$; if it is not, the four
averages may in principle correspond to different sub-ensembles, and there is
no special reason why their combination by sum and difference should not
exceed the limit of $2$ given by the Bell theorem\footnote{Another intuitive
way to understand why experiments where most pairs go undetected are useless
for a violation of the inequality is the following: if one associates $0$ to
the absence of result, the occurrence of many zeros in the results will bring
the correlations rates closer to zero and the combination will never exceed
$2$.}.\ The important point is not necessarily to capture all pairs, since one
could in theory redefine the ensemble in relation with detection; but what is
essential, for any perfectly convincing experiment on the violation of the
Bell inequalities, is to make sure that the sample of counted events is
completely independent of their settings $a$ and $b$ (unbiased sample).\ This,
in practice, implies some sort of selection (or detection) that is completely
independent of the settings, which is certainly not the case in any experiment
that detect only the particles that have crossed the analyzers.

An ideal situation would be provided by a device with a triggering button that
could be used by an experimentalist, who could at will launch a pair of
particles (with certainty); if the pair in question was always analyzed and
detected, with 100\% efficiency, the loophole would definitely be closed! When
discussing thought experiments, Bell introduced in some of his talks the
notion of ``preliminary detectors'' \cite{Bell-oral}, devices which he
sketched as cylinders through which any pair of particles would have to
propagate before reaching both ends of the experiment (where the $a$ and $b$
dependent measurement apparatuses sit); the idea was that the preliminary
detectors should signal the presence of pairs and that, later, the
corresponding pairs would always be detected at both ends.\ The role of these
cylinders was therefore to make the definition of the sample perfectly
precise, even if initially the pairs were emitted by the source in all
directions.\ Such class of systems, which allow a definition of an ensemble
that is indeed totally independent of $a$ and $b$, are sometimes called an
``event ready detectors''.\ See also reference \cite{Bell-EGAS} where Bell
imagines a combination of veto and go detectors associated with the first
detected particles in a ternary emission, precisely for the purpose of better
sample definition.

Needless to say, in practice, the situation is very different. First, one
should realize that, in all experiments performed until now, most pairs are
simply missed by the detectors.\ There are several reasons for this situation:
in photon experiments, the particles are emitted in all directions, while the
analyzers collect only a small solid angle and, therefore, only a tiny
fraction of the pairs (this was especially true in the initial experiments
using photon cascades; in more modern experiments \cite{parametric}, the use
of parametric photon conversion processes introduces a strong correlation
between the direction of propagation of the photons and a much better
collection efficiency, but it still remains low).\ Moreover, the transmission
of the analyzers is less than one (it is actually less than 1/2 if ordinary
photon polarization filters are used, but experiments have also been performed
with birefringent two-channel analyzers \cite{Aspect}, which are not limited
to 50\% efficiency). Finally, the quantum efficiency of particle detectors
(photomultipliers for photons) is not 100\% either, so that pairs of particles
are lost at the last stage too. There is no independent way to determine the
sample of detected pairs, except of course the detection process itself, which
is obviously $a$ and $b$ dependent; as a consequence, all experimental results
become useful only if they are interpreted within a ``no-biasing'' assumption,
considering that the settings of the analyzers does not bias the statistics of
events. On the other hand, we should also mention that there is no known
reason why such a sample biasing should take place in the experiments, and
that the possibility remains speculative. For proposals of ``loophole free''
experiments\footnote{A perfect correlation between the detections on each side
(in an ideal experiment with parametric generation of photons for instance)
would provide another possible scheme for a loophole free experiment - this,
of course, implies that two channel detectors with a 100\% efficiency are used
on both ends of the experiment. In itself, the fact that any click at one side
is always correlated with a click at the other, independently of the settings
$a$ and $b$, is not sufficient to exclude a setting dependence of the ensemble
of detected pairs.\ But, if one assumes locality at this stage also, a simple
reasoning shows that a perfect detection correlation is sufficient to ensure
the independence: how could a particle on one side ``know'' that it belongs to
the right sub-ensemble for the other particle to be detected, without knowing
the other setting? In other words, locality arguments may be used, not only
for the results of the apparatuses (the functions $A$ and $B$), but also in
order to specify the ensemble of observed pairs (the distribution function
$\rho$).\ Under these conditions, the observation (in some future experiment)
of a violation of the Bell inequalities with a perfect detection correlation
would be sufficient to exclude local theories, and therefore to close the
loophole.}, see refs. \cite{Kwiat} and \cite{Fry-loop}; actually, there now
seems to be a reasonable hope that this loophole will be closed by the
experiments within the next few years.

Other loopholes are also possible: even if experiments were done with 100\%
efficiency, one could also invoke some possibilities for local processes to
artificially reproduce quantum mechanics.\ One of them is usually called the
``conspiracy of the polarizers'' (actually, ``conspiracy of the analyzers''
would be more appropriate; the word polarizer refers to the experiments
performed with photons, where the spin orientation of the particles is
measured with polarizing filters; but there is nothing specific of photons in
the scenario, which can easily be transposed to massive spin 1/2 particles) -
or also ``communication loophole''. The idea is the following: assume that, by
some unknown process, each analyzer could become sensitive to the orientation
of the other analyzer; it would then acquire a response function which depends
on the other setting and the function $A$ could acquire a dependence on both
$a$ and $b$. The only way to beat this process would be to choose the settings
$a$ and $b$ at the very last moment, and to build an experiment with a large
distance between the two analyzers so that no information can propagate (at
the speed of light) between the two of them.\ A first step in this direction
was done by Aspect and coll. in 1982 \cite{Aspect-temps}, but more recent
experiments have beautifully succeeded in excluding this possibility in an
especially convincing way \cite{Zeilinger-temps}. So there no longer exist a
real conspiracy loophole; quantum mechanics seems to still work well under
these more severe time dependent conditions.

Along a similar line is what is sometimes called the ``fatalistic loophole''
(or also ``superdeterminism'').\ The idea is to put into question an implicit
assumption of the reasoning that leads to the Bell theorem: the completely
arbitrary choice of the settings $a$ and $b$ by the experimenters. Usually,
$a$ and $b$ are indeed considered as free variables: their values that are not
the consequence of any preliminary event that took place in the past, but
those of a free human choice.\ On the other hand, it is true that there is
always some an overlap between the past cones of two events, in this case the
choice of the settings.\ It is therefore always possible in theory to assume
that they have a common cause; $a$ and $b$ are then no longer free parameters,
but variables that can fluctuate (in particular, if this cause itself
fluctuates) with all kinds of correlations.\ In this case,\ it is easy to see
that the proof of the Bell theorem is no longer possible \footnote{For
instance, in the proof that makes uses of a probability density $\rho
(\lambda)$, if one assumes that $a$ and $b$ become two functions $a(\lambda)$
and $b(\lambda)$, it makes no sense to compare the average values for
different fixed values of $a$ and $b$.}, so that any contradiction with
locality is avoided.\ What is then denied is the notion of free will of the
experimenters, whose decisions are actually predetermined, without them being
aware of this fact; expressed more technically, one excludes from the theory
the notion of arbitrary external parameters, which usually define the
experimental conditions.\ This price being paid, one could in theory build an
interpretation of quantum mechanics that would remain at the same time
realistic, local and (super)deterministic, and would include a sort of
physical theory of human decision.\ This is, of course, a very unusual point
of view, and the notion of arbitrary external parameters is generally
accepted; in the words of Bell \cite{Bell-once-only}: ``A respectable class of
theories, including quantum theory as it is practised, have free external
variables in addition to those internal to and conditioned by the
theory....They are invoked to represent the experimental conditions.\ They
also provide a point of leverage for free willed experimenters,
...''.\ Needless to say, the fatalist attitude in science is even more subject
to the difficulties of orthodox quantum mechanics concerning the impossibility
to develop a theory without observers, etc..

We could not conclude honestly this section on loopholes without mentioning
that, while most specialists acknowledge their existence, they do not take
them too seriously because of their ``ad hoc'' character.\ Indeed, one should
keep in mind that the explanations in question remain artificial, inasmuch
they do not rest on any precise theory: no-one has the slightest idea of the
physical processes involved in the conspiracy, or of how pair selection would
occur in a way that is sufficiently complex to perfectly reproduce quantum
mechanics. By what kind of mysterious process would experiments mimic quantum
mechanics so perfectly at low collection efficiencies, and cease to do so at
some threshold of efficiency? Bell himself was probably the one who should
have most liked to see that his inequalities could indeed be used as a logical
tool to find the limits of quantum mechanics; nevertheless, he found these
explanation too unaesthetic to be really plausible.\ But in any case logic
remains logic: yes, there still remains a slight possibility that, when the
experiments reach a level of efficiency in pair collection where the loophole
becomes closed, the results concerning the correlation rates will
progressively deviate from those of quantum mechanics to reach values
compatible with local realism. Who knows?

\subsection{Locality, contrafactuality}

\label{contra}

One can find in the literature various attitudes concerning the exact relation
between quantum mechanics and locality.\ Some authors consider that the
non-local character of quantum mechanics is a well-known fact, while for
others quantum non-locality is an artefact created by the introduction into
quantum mechanics of notions which are foreign to it (typically the EPR
elements of reality). Lively discussions to decide whether or not quantum
mechanics in itself is inherently non-local have taken place and are still
active \cite{Stapp} \cite{Stapp-3} \cite{Mermin-6}; see also references
\cite{d'Espagnat} and \cite{Redhead} \cite{CS}. Delicate problems of logic are
involved and we will not discuss the question in more detail here.

What is easier to grasp for the majority of physicists is the notion of
``contrafactuality'' \cite{Stapp-3}.\ A counterfactual reasoning consists in
introducing the results of possible experiments that can be envisaged for the
future as well-defined quantities, and valid mathematical functions to use in
equations, even if they are still unknown - in algebra one writes unknown
quantities in equations all the time. This is very natural: as remarked by
d'Espagnat \cite{BD-1} \cite{BD-2} and by Griffiths \cite{Griffiths-5},
``counterfactuals seem a necessary part of any realistic version of quantum
theory in which properties of microscopic systems are not created by the
measurements''.\ One can also see the EPR reasoning as a justification of the
existence of counterfactuals.\ But it also remains true that, in practice, it
is never possible to realize more than one of these experiments: for a given
pair, one has to choose a single orientation of the analyzers, so that all
other orientations will remain forever in the domain of speculation.\ For
instance, in the reasoning of \S \ \ref{proof}, at least some of the numbers
$A$, $A^{^{\prime}}$, $B$ and $B^{^{\prime}}$ are counterfactuals, and we saw
that using them led us to a contradiction with quantum mechanics through the
Bell inequalities.\ One could conclude that contrafactuality should be put
into question in quantum mechanics; alternatively, one could maintain
counterfactual reasoning, but then the price to pay is the explicit appearance
of non-locality.\ We have already quoted a sentence by Peres \cite{Peres}
which wonderfully summarizes the situation as seen within orthodoxy:
''unperformed experiments have no results''; as Bell once regretfully remarked
\cite{Bell-once-only}: ``it is a great inconvenience that the real world is
given to us once only''!

But, after all, one can also accept contrafactuality as well as explicit
non-locality together, and obtain a perfectly consistent point of view; it
would be a real misunderstanding to consider the Bell theorem as an
impossibility theorem, either for contrafactuality, or for hidden
variables.\ In other words, and despite the fact that the idea is still often
expressed, it is not true that the Bell theorem is a new sort of Von Neumann
theorem.\ The reason is simple: why require that theories with
contrafactuality/additional variables should be explicitly local at all
stages, while it is not required from standard quantum mechanics? Indeed,
neither the wave packet reduction postulate, nor the calculation of
correlation of experimental results in the correlation point of view
(\S \ \ref{correlation}), nor again the expression of the state vector itself,
correspond to mathematically local calculations.\ In other words, even if one
can discuss whether or not quantum mechanics is local or not at a fundamental
level, it is perfectly clear that its formalism is not; it would therefore be
just absurd to request a local formalism from a non-orthodox theory -
especially when the theory in question is built in order to reproduce all
results of quantum mechanics! As an illustration of this point, as seen from
theories with additional variables, we quote Goldstein \cite{Goldstein}: ``in
recent years it has been common to find physicists .... failing to appreciate
that what Bell demonstrated with his theorem was not the impossibility of
Bohmian mechanics, but rather a more radical implication - namely non-locality
- that is intrinsic to quantum theory itself''.

\subsection{``All-or-nothing coherent states''; decoherence}

In this section, we first introduce many particle quantum states which have
particularly remarkable correlation properties; then we discuss more precisely
a phenomenon that we have already introduced above, decoherence, which tends
to reduce their lifetime very efficiently, especially if the number of
correlated particles is large.

\subsubsection{Definition and properties of the states}

\label{complete-ent}

The states that we will call ``all-or-nothing coherent states'' (or
all-or-nothing states for short) could also be called ``many-particle GHZ
states'' since they are generalizations of (\ref{7}) to an arbitrary number
$N$ of particles:%
\begin{equation}
\mid\Psi>=\alpha\mid1:+;2:+;....;N:+>+\beta\mid1:-;2:-;...;N:-> \label{20}%
\end{equation}
where $\alpha$ and $\beta$ are arbitrary complex numbers satisfying $\left|
\alpha\right|  ^{2}+\left|  \beta\right|  ^{2}=1$. In fact, the most
interesting situations generally occur when $\alpha$ and $\beta$ have
comparable modulus, meaning that there are comparable probabilities to find
the system in states where all, or none, of the spins is flipped (hence the
name we use for these states); when $\alpha$ and $\beta$ are both equal to
$1/\sqrt{2}$, these states are sometimes called ``states of maximum
entanglement'' in the literature, but since the measure of entanglement for
more than two particles is not trivial (several different definitions have
actually been proposed in the literature), here we will use the words
``all-or-nothing states'' even in this case.

In order to avoid a frequent cause of confusion, and to better emphasize the
peculiarities of these entangled states, let us first describe a preparation
procedure that would NOT lead to such a state.\ Suppose that $N$ spin 1/2
particles oriented along direction $Ox$ enter a Stern-Gerlach magnet oriented
along direction $Oz$, or more generally that $N$ particles cross a filter
(polarization dependent beam splitter, Stern-Gerlach analyzer, etc.) while
they are initially in a state which is a coherent superposition of the
eigenstates of this filter, with coefficients $\alpha_{1}$ and $\beta_{1}$.
The effect of the filter on the group of particles is to put them into a state
which is a product of coherent superpositions of the two outputs of the
filter, namely:%
\begin{equation}%
\begin{array}
[c]{cl}%
\mid\Psi>= & \left[  \alpha\mid1:+>+\beta\mid\overset{}{1}:->\right]
\otimes\left[  \alpha\mid2:+>+\beta\mid\overset{}{2}:->\right]  \otimes\\
& \otimes....\otimes\left[  \alpha\mid N:+>+\beta\mid\overset{}{N}:->\right]
\end{array}
\label{21}%
\end{equation}
The point we wish to make is that this state is totally different from
(\ref{20}), since it contains many components of the state vector where some
of the spins are up, some down.\ In (\ref{21}), each particle is in a coherent
superposition of the two spin states, a situation somewhat analogous to a
Bose-Einstein condensate where all particles are in the same coherent state -
for instance two states located on either sides of a potential barrier as in
the Josephson effect. By contrast, in (\ref{20}) all spins, or none, are
flipped from one component to the other\footnote{In a all-or-nothing coherent
state, all spins\ are not necessarily up in the first component of the state
vector, while they are down in the second; what matters is that every spin
changes component from one component to the other and reaches an orthogonal
state (the quantization axis of every spin is not even necessarily the same).}
so that the coherence is essentially a $N$-body coherence only; it involves
entanglement and is much more subtle than in (\ref{21}). For instance, one can
show \cite{FL2} that the coherence in question appears only ``at the last
moment'', when all particles are taken into account: as long as one considers
any subsystem of particles, even $N-1$, it exhibits no special property and
the spins are correlated in an elementary way (as they would be in a classical
magnet); it is only when the last particle is included that quantum coherence
appears and introduces effects which are completely non-classical.

Assume for instance that:%
\begin{equation}
\alpha=1/\sqrt{2}\,\,\,\,\,\,\,\,\,;\,\,\,\,\,\,\,\,\,\,\beta=e^{i\varphi
}/\sqrt{2} \label{21bis}%
\end{equation}
and that a measurement is performed of a component of each spin that belongs
to the $Oxy$ plane and is defined by its angle $\theta_{1}$ with $Ox$ for the
first particle, $\theta_{2}$ for the second,...$\theta_{N}$ for the last. It
is an easy exercise of standard quantum mechanics to show that the product of
all results has the following average value:
\begin{equation}
E(\theta_{1},\theta_{2},....\theta_{N})=\cos(\theta_{1}+\theta_{2}%
+....\theta_{N}-\varphi) \label{211}%
\end{equation}
(we take the usual convention where the results of each measurement is $\pm
1$). For instance, each time the sum $\theta_{1}+\theta_{2}+....\theta
_{N}-\varphi$ is equal to an integer even multiple of $\pi$, the average is
$1$, indicating that the result is certain and free from any fluctuation (in
the same way, an odd multiple of $\pi$ would give a certain value, $-1$).
Indeed, the result of the quantum calculation may look extremely simple and
trivial; but it turns out that, once more, it is totally impossible to
reproduce the oscillations contained in (\ref{211}) within local realism.\ In
the case $N=2$, this is of course merely the consequence of the usual Bell
theorem; as soon as $N$ becomes $3$ or takes a larger value, the contradiction
becomes even more dramatic.\ Actually, if one assumes that a local
probabilistic theory reproduces (\ref{211}) only for some sets of particular
value of the angles $\theta$'s (those for which the result is certain), one
can show \cite{FL2} that the theory in question necessarily predicts that $E$
is independent of all $\theta$'s.\ The average keeps a perfectly constant
value $+1$! Indeed, the very existence of the oscillation predicted by
(\ref{211}) can be seen as a purely quantum non-local effect (as soon as
$N\geq2$).

This is, by far, not the only remarkable property of all-or-nothing coherent
states.\ For instance, it can be shown that they lead to exponential
violations of the limits put by local realistic theories \cite{Mermin4}; it
has also been pointed out \cite{Bollinger} that these states, when relation
(\ref{21bis}) is fulfilled (they are then called ``maximally correlated
states'' in \cite{Bollinger}), have interesting properties in terms of
spectroscopic measurements: the frequency uncertainty of measurements
decreases as $1/N$ for a given measurement time, and not as $1/\sqrt{N}$ as a
naive reasoning would suggest. This is of course a more favorable situation,
and the quantum correlation of these states may turn out to be, one day, the
source of improved accuracy on frequency measurements. How to create such
states with massive particles such as atoms, and not with photons as usual,
was demonstrated experimentally by Hagley et al.\ in 1997 \cite{Hagley} in the
case $N=2$.\ We have already mentioned in \S \ \ref{ghz} that entanglement
with $N=3$ was reported in refs. \cite{GHZ-1} and \cite{GHZ-1bis}.\ Proposals
for methods to generalize to larger values of $N$ with ions in a trap were put
forward by Molmer et al. \cite{Molmer}; the idea exploits the motion
dependence of resonance frequencies for a system of several ions in the same
trap, as well as on some partially destructive interference effects.\ The
scheme was successfully put into practice in a very recent experiment by
Sackett et al. \cite{Sackett} where ``all-or-nothing states'' were created for
$N=2$ as well as $N=4$ ions in a trap.

\subsubsection{Decoherence}

\label{decoherence}

We have defined in \S \ref{Neumann} decoherence as the initial part of the
phenomenon associated with the Von Neumann infinite regress: coherent
superpositions tend to constantly propagate towards the environment, they
involve more and more complex correlations with it, so that they become
rapidly completely impossible to detect in practice. To see more precisely how
this happens, let us for instance consider again state (\ref{20}); we now
assume that the single particle states $\mid+>$ and $\mid->$ are convenient
notations for two states where a particle has different locations in space
(instead of referring only to opposite spin directions): this will happen for
instance if the particles cross a Stern-Gerlach analyzer which correlates the
spin directions to the positions of the particles. Under these conditions, it
is easy to see that the coherence contained in the state vector becomes
extremely fragile to any interaction with environment.\ To see why, let us
assume that an elementary particle (photon for instance), initially in state
$\mid\mathbf{k}_{0}>$, interacts with the particles.\ It will then scatter
into a quantum state that is completely different, depending on where the
scattering event took place: if the scattering atoms are in the first state
$\mid+>$, the photon is scattered by the atoms into state $\mid\mathbf{k}%
_{+}>$; on the other hand, if it interacts with atoms in state $\mid->$, it is
scattered into state $\mid\mathbf{k}_{-}>$ \footnote{We could also have
assumed that the photon is focussed so that it can interact only with one sort
of atoms, but is not scattered by the other, without changing the conclusion
of this discussion.}. As soon as the new particle becomes correlated with the
atoms, the only state vector that can be used to describe the system must
incorporate this new particle as well, and becomes:%

\begin{equation}%
\begin{array}
[c]{cl}%
\mid\Psi^{^{\prime}}>= & \alpha\mid1:+;2:+;....;N:+>\otimes\mid\mathbf{k}%
_{+}>+\\
& +\beta\mid1:-;2:-;...N:->\otimes\mid\mathbf{k}_{-}>
\end{array}
\label{23}%
\end{equation}

Assume now that we are interested only in the system of $N$ atoms; the reason
might be, for instance, that the scattered photon is impossible (or very
difficult) to detect (e.g. it may be a far-infrared photon).\ It is then
useful to calculate the partial trace over this photon in order to obtain the
density operator which describes the atoms only.\ A straightforward
calculation shows that this partial trace can be written, in the basis of the
two states $\mid+,+,+,..>$ and $\mid-,-,-,..>:$%
\begin{equation}
\rho=\left(
\begin{array}
[c]{cc}%
\mid\alpha\mid^{2} & \alpha\beta^{\ast}<\mathbf{k}_{-}\mid\mathbf{k}_{+}>\\
\alpha^{\ast}\beta<\mathbf{k}_{+}\mid\mathbf{k}_{-}> & \mid\beta\mid^{2}%
\end{array}
\right)  \label{24}%
\end{equation}
(for the sake of simplicity we assume that the states $\mid\mathbf{k}_{\pm}>$
are normalized). We see in this formula that, if the scalar product
$<\mathbf{k}_{-}\mid\mathbf{k}_{+}>$ was equal to one, the density matrix of
the atoms would not be affected at all by the scattering of the single
photon.\ But this would assume that the photon is scattered exactly into the
same state, independently of the spatial location of the scatterers! This
cannot be true if the distance between the locations is much larger than the
photon wavelength.\ Actually, it is much more realistic to assume that this
scalar product is close to zero, which means that the off-diagonal element of
(\ref{24}), in turn, becomes almost zero.\ We then conclude that the
scattering of even a single particle destroys the coherence between atomic
states, as soon as they are located at different places.

The situation becomes even worse when more and more photons (assumed to be all
in the same initial state $\mid\mathbf{k}_{0}>$) are scattered, since one then
has to replace (\ref{23}) by the state:%
\begin{equation}%
\begin{array}
[c]{cl}%
\mid\Psi^{^{\prime}}>= & \alpha\mid1:+;2:+;....;N:+>\otimes\mid\mathbf{k}%
_{+}>\mid\mathbf{k}_{+}^{^{\prime}}>\mid\mathbf{k}_{+}^{^{^{\prime\prime}}%
}>..+\\
& +\beta\mid1:-;2:-;...N:->\otimes\mid\mathbf{k}_{-}>\mid\mathbf{k}%
_{-}^{^{\prime}}>\mid\mathbf{k}_{-}^{^{^{\prime\prime}}}>..
\end{array}
\label{25}%
\end{equation}
with obvious notation (the states with $n$ primes correspond to the $n-1$ th.
scattered photon);\ the same calculation as above then provides the following
value for $\rho$:%
\begin{equation}
\left(
\begin{array}
[c]{cc}%
\mid\alpha\mid^{2} & \alpha\beta^{\ast}<\mathbf{k}_{-}\mid\mathbf{k}%
_{+}><\mathbf{k}_{-}^{^{\prime}}\mid\mathbf{k}_{+}^{^{\prime}}>...\\
\alpha^{\ast}\beta<\mathbf{k}_{+}\mid\mathbf{k}_{-}><\mathbf{k}_{+}^{^{\prime
}}\mid\mathbf{k}_{-}^{^{\prime}}>... & \mid\beta\mid^{2}%
\end{array}
\right)  \label{26}%
\end{equation}
Since we now have, in the off-diagonal elements, the product of all single
scalar product $<\mathbf{k}_{-}\mid\mathbf{k}_{+}>$, it is clear that these
elements are even more negligible than when a single photon is scattered.
Actually, as soon as the two states $\mid\mathbf{k}_{+}>$ and $\mid
\mathbf{k}_{-}>$ are not strictly identical, they tend exponentially to zero
with the number of scattering events.

This is a completely general property: objects (especially macroscopic
objects) have a strong tendency to leave a trace in the environment by
correlating themselves with any elementary particle which passes by; in the
process, they lose their own coherence, which regresses into a coherence
involving the environment and more and more complex correlations with it (the
scattered photon, in turn, may correlate with other particles); soon it
becomes practically impossible to detect. The phenomenon is unavoidable,
unless the scattering properties of both states symbolized by $\mid+>$ and
$\mid->$ are exactly the same, which excludes any significant spatial
separation between the states.\ In particular, it is impossible to imagine
that a cat, whether dead or alive, will scatter photons exactly in the same
way, otherwise we could not even see the difference! This shows how fragile
coherent superpositions of macroscopic objects are, as soon as they involve
states that can be seen as distinct.

We are now in a position where we can come back in more detail to some
questions that we already discussed in passing in this text, and which are
related to decoherence and/or the Schr\"{o}dinger cat. The first question
relates to the conceptual status of the phenomenon of decoherence.\ Some
authors invoke this phenomenon as a kind of ``explanation'' of the postulate
of wave packet reduction: when the superposition of the initial system becomes
incoherent, are we not in presence of a statistical mixture that resembles the
description of a classical object with well defined (but ignored) properties?
On this point, we do not have much to add to what was already said in
\S \ \ref{diff}: this explanation is unsatisfactory because the purpose of the
postulate of wave packet reduction is not to explain decoherence, which can
already be understood from the Schr\"{o}dinger equation, but the uniqueness of
the result of the measurement - in fact, the effect of the wave packet
reduction is sometimes to put back \ the measured sub-system into a pure
state, which is the perfect opposite of a statistical mixture, so that the
real question is to understand how the (re)emergence of a pure state should be
possible \cite{London-Bauer}.\ Indeed, in common life, as well as in
laboratories, one never observes superposition of results; we observe that
Nature seems to operate in such a way that a single result always emerges from
a single experiment; this will never be explained by the Schr\"{o}dinger
equation, since all that it can do is to endlessly extend its ramifications
into the environment, without ever selecting one of them only.

Another way to say the same thing is to emphasize the logical structure of the
question.\ The starting point is the necessity for some kind of limit of the
validity of the linear Schr\"{o}dinger equation, the initial reason being that
a linear equation can never predict the emergence of a single result in an
experiment.\ The difficulty is where and how to create this
border.\ Logically, it is then clear that this problem will never be solved by
invoking any process that is entirely contained in the linear Schr\"{o}dinger
equation, such as decoherence or any other similar linear process; common
sense says that, if one stays in the middle of a country one never reaches its
borders. Actually, no one seriously doubts that a typical measurement process
will involve decoherence at some initial stage, but the real question is what
happens after.

Pressed to this point, some physicists reply that one can always assume that,
at some later stage, the superposition resolves into one of its branches only;
this is of course true, but this would amount to first throwing a problem out
by the door, and then letting it come back through the window! (see
discussions above, for instance on the status of the state vector and the
necessity to resolve the Wigner friend paradox). A more logical attitude,
which is indeed sometimes proposed as a solution to the problem, is to
consider that the natural complement of decoherence is the Everett
interpretation of quantum mechanics (see \S \ \ref{ever}); indeed, this
provides a consistent interpretation of quantum mechanics, where the emergence
of a single result does not have to be explained, since it is assumed never to
take place (the Schr\"{o}dinger equation then has no limit of validity).\ But,
of course, in this point of view, one has do deal with all the intrinsic
difficulties of the Everett interpretation, which we will discuss later.

Concerning terminology, we have already mentioned in \S \ \ref{cat} that,
during the last few years, it has become rather frequent to read the words
``Schr\"{o}dinger cat'' (SC ) used in the context of states such as (\ref{20})
for small values of $N$ (actually even for a single ion, when $N=1$). This is
a redefinition of the words, since the essential property of the original cat
was to have a macroscopic number of degree of freedom, which is not the case
for a few atoms, ions or photons.\ But let us now assume that someone
succeeded in preparing an all-or-nothing state with a very large value of $N$,
would that be a much better realization of the Schr\"{o}dinger cat as meant by
its inventor?\ To some extent, yes, since the cat can be seen as a symbol of
\ a system of many particles that change their state, when one goes from one
component of the state vector to the other.\ Indeed, it is likely that many of
the atoms of a cat take part in different chemical bonds if the cat is alive
or dead, which puts them in a different quantum state.\ But it seems rather
hard to invent a reason why every atom, every degree of freedom, should
necessarily be in an orthogonal state in each case, while this is the
essential property of ``all-or-nothing states''. In a sense they do too much
for realizing a standard Schr\"{o}dinger cat, and the concepts remain somewhat
different, even for large values of $N$.

From an experimental point of view, decoherence is an interesting physical
phenomenon that is certainly worth studying in itself, as recent experiments
have illustrated \cite{Brune}; a result of these studies and of the related
calculations, among others, is to specify the basis in the space of states
that is relevant to the decoherence process, as a function of the coupling
Hamiltonian, as well as the characteristic time constants that are
associated.\ One can reasonably expect that more experiments on decoherence
will follow this initial breakthrough and provide a more detailed
understanding of many aspects of the phenomenon. Nevertheless decoherence is
not to be confused with the measurement process itself; it is just the process
which takes place just before: during decoherence, the off-diagonal elements
of the density matrix\footnote{\ The formalism of the density operator, or
matrix, is elegant and compact, but precisely because it is compact it
sometimes partially hides the physical origin of the mathematical terms. The
density matrix allows one to treat in the same way classical probabilities,
arising from non fundamental uncertainties and imperfect knowledge of a
physical system, and purely quantum probabilities which are more fundamental
and have nothing to do with any particular observer.\ But mathematical
analogies should not obscure conceptual difficulties!} vanish (decoherence),
while in a second step all diagonal elements but one should vanish (emergence
of a single result).

\subsection{Quantum cryptography, teleportation}

\label{crypto}

In the two subsections below, we discuss two concepts that have some
similarity, quantum cryptography and teleportation.

\subsubsection{Sharing cryptographic keys by quantum measurements}

A subject that is relatively new and related to the EPR correlations, is
quantum cryptography \cite{Ekert} \cite{Bennett2}. The basic idea is to design
a perfectly safe system for the transmission at a distance of \ a
cryptographic key - the word refers to a random series of numbers 0 or 1,
which are used to code, and subsequently decode, a message.\ In a first step,
the two remote correspondents A (traditionally called Alice) and B
(traditionally called Bob) share this key; they then use it to code, or to
decode, all the messages that they exchange; if the key is perfectly random,
it becomes completely impossible for anyone who does not know it to decode any
message, even if it is sent publicly.\ But if, during the initial stage of key
exchange, someone can eavesdrop the message (the actor in question is
traditionally called Eve), in the future he/she will be able to decode all
messages sent with this key.\ Exchanging keys is therefore a really crucial
step in cryptography.\ The usual strategy is to take classical methods to keep
the secret: storage in a safe, transport of the keys by \ secure means, etc.;
it is always difficult to assess its safety, which depends on many human factors.

On the other hand, quantum sharing of keys relies on fundamental physical
laws, which are impossible to break: however clever and inventive spies may
be, they will not be able to violate the laws of quantum mechanics! The basic
idea is that Alice and Bob will create their common key by making quantum
measurements on particles in an EPR correlated\ state; in this way they can
generate series of random numbers that can be subsequently used as a secret
communication.\ What happens if Eve tries to intercept the photons, for
example by coupling some elaborate optical device to the optical fiber where
the photons propagate between Alice and Bob, and then making measurements?

If she wants to operate unnoticed, she clearly cannot just absorb the photons
in the measurement; this would change the correlation properties observed by
Alice and Bob.\ The next idea is to try to ``clone'' photons, in order to make
several identical copies of the initial photon; she could then use a few of
them to measure their properties, and re-send the last of them on the line so
that no-one will notice anything.\ But, it turns out that ''quantum cloning''
is fundamentally impossible: within the rules of quantum mechanics, there is
absolutely no way in which several particles can be put into the same
arbitrary and unknown state $\mid\varphi>$ as one given particle
\cite{Wooters-Zurek} \cite{Dieks} - see also \cite{Gisin-Massar} for a
discussion of multiple cloning. In appendix IV we discuss why, if state
cloning were possible, it would be possible to apply it to each particle of an
EPR pair of correlated particles; then the multiple realization of the states
could be used to transmit information on the orientations $a$ and $b$ used by
the experimenters.\ Since such a scheme would not be subject to any minimum
time delay, it could also transmit messages at superluminal velocities, and be
in contradiction with relativity.\ Fortunately for the consistency of physics,
the contradiction is avoided by the fact that clonig of single systems is
impossible in quantum mechanics!

So, whatever Eve does to get some information will automatically change the
properties of the photons at both ends of the optical fibre, and thus be
detectable by Alice and Bob, if they carefully compare their data and their
correlation rates.\ Of course, they do not necessarily have a way to prevent
Eve's eavesdropping, but at least they know what data can be used as a
perfectly safe key.\ There is actually a whole variety of schemes for quantum
cryptography, some based on the use of EPR correlated particles, others not
\cite{Bennett2}; but this subject is beyond our scope here.

\subsubsection{Teleporting a quantum state}

The notion of quantum teleportation \cite{Bennett1} is also related to quantum
non-locality; the idea is to take advantage of the correlations between two
entangled particles, which initially are for instance in state (\ref{20}) (for
$N=2$), in order to reproduce at a distance any arbitrary spin state of a
third particle. The scenario is the following: initially, two entangled
particles propagate towards two remote regions of space; one of them reaches
the laboratory of the first actor, Alice, while the second reaches that of the
second actor, Bob; a third particle in quantum state $\mid\varphi>$ is then
provided to Alice in her laboratory; the final purpose of the all the scenario
is to put Bob's particle into exactly the same state $\mid\varphi>$, whatever
it is (without, of course, transporting the particle itself). One then says
that sate $\mid\varphi>$ has been teleported.

More precisely, what procedure is followed in teleportation?\ Alice has to
resist the temptation of performing any measurement on the particle in state
$\mid\varphi>$ to be teleported; instead, she performs a ``combined
measurement'' that involves at the same time this particle as well as her
particle from the entangled pair.\ In fact, for the teleportation process to
work, an essential feature of this measurement is that no distinction between
the two particles involved must be established.\ With photons one may for
instance, as in ref. \cite{teleport-exp}, direct the particles onto opposite
sides of the same optical beam splitter, and measure on each side how many
photons are either reflected or transmitted; this device does not allow one to
decide from which initial direction the detected photons came, so that the
condition is fulfilled.\ Then, Alice communicates the result of the
measurement to Bob; this is done by some classical method such as telephone,
e-mail etc., that is by a method that is not instantaneous but submitted to
the limitations related to the finite velocity of light.\ Finally, Bob applies
to his particle an unitary transformation that depends on the classical
information he has received; this operation puts it exactly into the same
state $\mid\varphi>$ as the initial state of the third particle, and realizes
the ``teleportation'' of the state.\ The whole scenario is ``mixed'' because
it involves a combination of transmission of quantum information (through the
entangled state) and classical information (the phone call from Alice to Bob).

Teleportation may look trivial, or magical, depending how one looks at
it.\ Trivial because the possibility of reproducing at a distance a state from
classical information is not in itself a big surprise.\ Suppose for instance
that Alice decided to choose what the teleported state should be, and filtered
the spin (she sends particles through a Stern-Gerlach system until she gets a
$+1$ result\footnote{For filtering a spin state, one obviously needs to use a
non-destructive method for detection after the Stern-Gerlach magnet.\ One
could for instance imagine a laser detection scheme, designed is such a way
that the atom goes through an excited state, and then emits a photon by
returning to the same internal ground state (closed optical pumping cycle -
this is possible for well chosen atomic transition and laser polarization).});
she could then ask Bob by telephone to align his Stern-Gerlach filter in the
same direction, and to repeat the experiment until he also observes a +1
result.\ This might be called a trivial scenario, based only on the
transmission of classical information.\ But teleportation does much more than
this! First, the state that is transported is not necessarily chosen by Alice,
but can be completely arbitrary. Second, a message with only binary classical
information, such as the result of the combined experiment made by Alice in
the teleportation scheme, is certainly not sufficient information to
reconstruct a quantum state; in fact a quantum state depends on continuous
parameters, while results of experiments correspond to discrete information
only. Somehow, in the teleportation process, binary information has turned
into continuous information! The latter, in classical information theory,
would correspond to an infinite number of bits (in the trivial scenario above,
sending the complete information on the state with perfect accuracy would
require an infinite time).

\ Let us come back in more detail to these two differences between
teleportation and what we called the trivial scenario. Concerning the
arbitrary character of the state, of course Alice may also, if she wishes,
teleport a known state.\ For this, beforehand, she could for instance perform
a Stern-Gerlach experiment on the third particle in order to filter its spin
state.\ The remarkable point, nevertheless, is that teleportation works
exactly as well is she is given a spin in a completely unknown state, by a
third partner for instance; in this case, it would be totally impossible for
her to know what quantum state has been sent just from the result of the
combined experiment.\ A natural question then arises: if she knows nothing
about the state, is it not possible to improve the transmission efficiency by
asking her to try and determine the state in a first step, making some trivial
single-particle measurement?\ The answer to the question is no, and for a very
general reason: it is impossible to determine the unknown quantum state of a
single particle (even if one accepts only an a posteriori determination of a
perturbed state); one quantum measurement clearly does not provide sufficient
information to reconstruct the whole state; but several measurements do not
provide more information, since the first measurement has already changed the
spin state of the particle. In fact, acquiring the complete information on an
unknown spin-1/2 state would require from Alice to start from an infinite
number of particles that have been initially prepared into this same state,
and to perform measurements on them; this is, once more, because the
information given by each measurement is discrete while the quantum state has
continuous complex coefficients. Alice cannot either clone the state of the
single particle that she is given, in order to make several copies of it and
measure them (see preceding section and appendix IV).\ So, whatever attempt
Alice makes to determine the state before teleportation will not help the process.

Concerning the amount of transmitted information, what Bob receives has two
components: classical information sent by Alice, with a content that is
completely ``uncontrolled'', since it is not decided by her, but just
describes the random result of an experiment; quantum information contained in
the teleported state itself (what we will call a ``q-bit'' in the next
section) and can possibly be controlled.\ We know that neither Bob nor Alice
can determine the teleported state from a single occurrence, but also that
Alice can prepare the state to be teleported by a spin filtering operation in
a direction that she decides arbitrarily; Bob then receives some controlled
information as well.\ For instance, if the teleportation is repeated many
times, by successive measurements on the teleported particles Bob will be able
to determine its quantum state with arbitrary accuracy, including the
direction that was chosen by Alice; he therefore receives a real message from
her (for a discussion of the optimum strategy that Bob should apply, see ref.
\cite{Massar}).

If one wishes to describe teleportation things in a sensational way, one could
explain that, even before Bob receives any classical information, he has
already received ``almost all the information'' on the quantum state, in fact
all the controllable information since the classical message does not have
this property; this ``information'' has come to him instantaneously, exactly
at the time when Alice performed her combined experiment, without any minimum
delay that is proportional to the distance covered. The rest of the
information, which is the ``difference'' between a continuous ``information''
and a discrete one, comes only later and is, of course, subject to the minimum
delay associated with relativity.\ But this is based on an intuitive notion of
``difference between quantum/controllable and classical/non-controllable
information'' that we have not precisely defined; needless to say, this should
not be taken as a real violation of the basic principles of relativity!

Finally, has really something been transported in the teleportation scheme, or
just information?\ Not everyone agrees on the answer to this question, but
this may be just a discussion on words, so that we will not dwell further on
the subject.\ What is perfectly clear in any case is that the essence of the
teleportation process is completely different from any scenario of classical
communication between human beings.\ The relation between quantum
teleportation and Bell-type non-locality experiments is discussed in
\cite{Popescu}; see also \cite{Sudbury} as well as \cite{Physics-Today} for a
review of recent results.

\subsection{Quantum computing and information}

Another recent development is quantum computing \cite{C-H-Bennett}
\cite{Vicenzo} \cite{les2}.\ Since this is still a rather new field of
research, our purpose here cannot be to give a general overview, while new
results are constantly appearing in the literature.\ We will therefore
slightly change our style in the present section, and limit ourselves to an
introduction of the major ideas, at the level of general quantum mechanics and
without any detail; we will provide references for the interested reader who
wishes to learn more.

\ The general idea of quantum computing \cite{Deutsch} is to base numerical
calculations, not on classical ``bits'', which can be only in two discrete
states (corresponding to 0 and 1 in usual binary notation), but on quantum
bits, or ``q-bits'', that is on quantum systems that have access to a
two-dimensional space of states; this means that q-bits can not only be in one
of the two states $\mid0>$ and $\mid1>$, but also in any linear superposition
of them.\ It is clear that a continuum of states is a much ``larger'' ensemble
than two discrete states only; in fact, for classical bits, the dimension of
the state space increases linearly with the number of bits (for instance, the
state of 3 classical bits defines a vector with 3 components, each equal to
$0$ or $1$); for q-bits, the dimension increases exponentially (this is a
property of the tensor product of spaces; for instance, for three q-bits the
dimension of space is $2^{3}=8$).\ If one assumes that a significant number of
q-bits is available, one gets access to a space state with an enormous
``size'', where many kinds of interference effects can take place.\ Now, if
one could somehow make all branches of the state vector ``work in parallel''
to perform independent calculations, it is clear that one could perform much
faster calculations, at least in theory.\ This ``quantum parallelism'' opens
up many possibilities; for instance, the notion of unique computational
complexity of a given mathematical problem, which puts limits on the
efficiency of classical computers, no longer applies in the same way.\ Indeed,
it has been pointed out \cite{Shor} that the factorization of large numbers
into prime factors may become faster than by classical methods, and by
enormous factors. Similar enhancements of the speed of computation are
expected in the simulation of many-particle quantum systems \cite{Abrams}%
.\ For some other problems the gain in speed is only polynomial in theory,
still for some others there is no gain at all.

Fundamentally, there are many differences between classical and quantum
bits.\ While classical bits have two reference states that are fixed once and
for all, q-bits can use any orthogonal basis in their space of
states.\ Classical bits can be copied at will and ad infinitum, while the
no-cloning theorem mentioned in the preceding section (see also appendix IV)
applies to q-bits.\ On the other hand, classical bits can be transmitted only
into the forward direction of light cones, while the use of entanglement and
teleportation removes this limitation for q-bits. But we have to remember that
there is more distance between quantum q-bits and information than there is
for their classical bits: in order to transmit and receive useable information
from q-bits, one has to specify what kind of measurements are made with them
(this is related to the flexibility on space state basis mentioned
above).\ Actually, as all human beings, Alice and Bob can communicate only at
a classical level, by adjusting the macroscopic settings of their measurement
apparatuses and observing the red and green light flashes associated with the
results of measurements.\ Paraphrasing Bohr (see the end of \S \ \ref{status}%
), we could say that ''there is no such concept as quantum information;
information is inherently classical, but may be transmitted through quantum
q-bits''; nevertheless, the whole field is now sometimes called ``quantum
information theory''. For an early proposal of a practical scheme of a quantum
computer with cold trapped ions, see ref. \cite{Cirac-Zoller}.

Decoherence is the big enemy of quantum computation, for a simple reason: it
constantly tends to destroy the useful coherent superpositions; this sadly
reduces the full quantum information to its classical, boolean, component made
of diagonal matrix elements only.\ It is now actually perfectly clear that a
``crude'' quantum computer based on the naive use of non-redundant q-bits will
never work, at least with more than a very small number of them; it has been
remarked that this kind of quantum computer would simply be a sort or
resurgence of the old analog computers (errors in quantum information form a
continuum), in an especially fragile version!\ But it has also been pointed
out that an appropriate use of quantum redundancy may allow one to design
efficient error correcting schemes \cite{Shor2} \cite{Steane}; decoherence can
be corrected by using a system containing more q-bits, and by projecting its
state into some subspaces in which the correct information about the
significant q-bit survives without error \cite{Preskill}; the theoretical
schemes involve collective measurements of several q-bits, which give access
to some combined information on all them, but none on a single q-bit.\ It
turns out that it is theoretically possible to ``purify'' quantum states by
combining several systems in perturbed entangled states and applying to them
local operations, in order to extract a smaller number of systems in
non-perturbed states \cite{Benn}; one sometimes also speaks of ``quantum
distillation'' in this context.\ This scheme applies in various situations,
including quantum computation as well as communication or cryptography
\cite{Benn-PRA}.\ Similarly the notion of ``quantum repeaters'' \cite{Briegel}
has been introduced recently in order to correct for the effect of
imperfections and noise in quantum communication. Another very different
approach to quantum computation has been proposed, based on a semiclassical
concept where q-bits are still used, but communicate only through classical
macroscopic signals, which are used to determine the type of measurement
performed on the next q-bit \cite{Griffiths-comput}; this kind of computer
should be much less sensitive to decoherence.

Generally\ speaking, whether or not it will be possible one day to beat
decoherence in a sufficiently large system for practical quantum computing
still remains to be seen. Moreover, although the factorization into prime
numbers is an important question (in particular for cryptography), as well as
the many-body quantum problem, it would be nice to apply the principles of
quantum computation to a broader scope of problems!\ The question as to
whether of not quantum computation will become a practical tool one day
remains open to debate \cite{JMR} \cite{les2}, but in any case this is an
exciting new field of research.

\section{Various interpretations}

\label{other}

In section \ref{historical} we have already mentioned some of the other,
``unorthodox'' interpretations of quantum mechanics that have been proposed,
some of them long ago and almost in parallel with the ``orthodox'' Copenhagen
interpretation.\ Our purpose was then to discuss, in a historical context, why
they are now generally taken more seriously by physicists than they were in
the middle of the 20th. century, but not to give any detail; this article
would be incomplete without, at least, some introduction to the major
alternative interpretations of quantum mechanics that have been proposed over
the years, and this is the content of the present section.

\ It is clearly out of the question to give here an exhaustive discussion of
all possible interpretations.\ This would even probably be an impossible task!
The reason is that, while one can certainly distinguish big families among the
interpretations, it is also possible to combine them in many ways, with an
almost infinite number of nuances.\ Even the Copenhagen interpretation itself
is certainly not a monolithic construction; it can be seen from different
points of view and can be declined in various forms.\ An extreme case was
already mentioned in \S \ \ref{Wigner}: what is sometimes called the ``Wigner
interpretation'' of quantum mechanics, probably because of the title and
conclusion of ref. \cite{Wigner-friend} - \ but views along similar lines were
already discussed by London and Bauer in 1939 \cite{London-Bauer}. In this
interpretation, the origin of the state vector reduction should be related to
consciousness.\ For instance, London and Bauer emphasize that state vector
reduction restores a pure state from a statistical mixture of the measured
sub-system (see \S \ \ref{decoherence}), and ``the essential role played by
the consciousness of the observer in this transition from a mixture to a pure
state''; they then explain this special role by the faculty of introspection
of conscious observers.\ Others prefer to invoke ``special properties'' of the
electrical currents which correspond to perception in a human brain, but how
seriously this explanation is put forward is not always entirely clear.\ In
fact, Wigner may have seen the introduction of an influence of consciousness
just as an extreme case (exactly as the Schr\"{o}dinger cat was introduced by
Schr\"{o}dinger), just for illustrating the necessity of a non-linear step in
order to predict definite results.\ In any event, the merit of the idea is
also to show how the logic of the notion of measurement in the Copenhagen
interpretation can be pushed to its limits: indeed, how is it possible to
ascribe such special properties to the operation of measurement without
considering that the human mind also has very special properties?

For obvious reasons of space, here we limit ourselves to a sketchy description
of the major families of interpretations.\ We actually start with what we
might call a ``minimal interpretation'', a sort of common ground that the vast
majority of physicists will consider as a safe starting point.\ We will then
proceed to discuss various families of interpretations: additional variables,
non-linear evolution of the state vector, consistent histories, Everett
interpretation. All of them tend to change the status of the postulate of the
wave packet reduction; some interpretations incorporate it into the normal
Schr\"{o}dinger evolution, others make it a consequence of another physical
process that is considered as more fundamental, still others use a formalism
where the reduction is hidden or even never takes place. But the general
purpose always remains the same: to solve the problems and questions that are
associated with the coexistence of two postulates for the evolution of the
state vector.

\subsection{Common ground; ``correlation interpretation''}

\label{correlation}

The method of calculation that we discuss in this section belongs to standard
mechanics; it is actually common to almost all interpretations of quantum
mechanics and, as a calculation, very few physicists would probably put it
into question.\ On the other hand, when the calculation is seen as an
interpretation, it may be considered by some as too technical, and not
sufficiently complete conceptually, to be really called an interpretation. But
others may feel differently, and we will nevertheless call it this way; we
will actually use the words ``correlation interpretation'', since all the
emphasis is put on the correlations between successive results of experiments.

The point of view in question starts from a simple remark: the Schr\"{o}dinger
equation alone, more precisely its transposition to the ``Heisenberg point of
view'', allows a relatively straightforward calculation of the probability
associated with any sequence of measurements, performed at different
times.\ To see how, let us assume that a measurement\footnote{Here, we assume
that all measurements are ideal; if non-ideal measurements are condidered, a
more elaborate treatment is needed.} of a physical quantity associated with
operator $M$ is performed at time $t_{1}$, and call $m$ the possible results;
this is followed by another measurement of observable $N$ at time $t_{2}$,
with possible results $n$, etc.\ Initially, we assume that the system is
described by a pure state $\mid\Psi(t_{0})>$, but below we generalize to a
density operator $\rho(t_{0})$. According to the Schr\"{o}dinger equation, the
state vector evolves between time $t_{0}$ and time $t_{1}$ from $\mid
\Psi(t_{0})>$ to $\mid\Psi(t_{1})>$; let us then expand this new state into
its components corresponding to the various results that can be obtained at
time $t_{1}$:%
\begin{equation}
\mid\Psi(t_{1})>=\sum_{m}\mid\Psi_{m}(t_{1})> \label{dec-1}%
\end{equation}
where$\mid\Psi_{m}(t_{1})>$ is obtained by applying to $\mid\Psi(t_{1})>$ the
projector $P_{M}(m)$ on the subspace corresponding to result $m$:
\begin{equation}
\mid\Psi_{m}(t_{1})>=P_{M}(m)\mid\Psi(t_{1})> \label{dec-2}%
\end{equation}
Now, just after the first measurement, we can ``chop'' the state vector into
different ``slices'', which are each of the terms contained in the sum of
(\ref{dec-1}).\ In the future, these terms will never give rise to
interference effects, since they correspond to different measurement results;
actually, each component becomes correlated to an orthogonal state of the
environment (the pointer of the measurement apparatus for instance) and a full
decoherence will ensure that any interference effect is cancelled.

Each ``slice'' $\mid\Psi_{m}(t_{1})>$ of $\mid\Psi(t_{1})>$ can then be
considered as independent from the others, and taken as a new initial state of
the system under study.\ From time $t_{1}$ to time $t_{2}$, the state in
question will then evolve under the effect of the Schr\"{o}dinger equation and
become a state $\mid\Psi_{m}(t_{2})>$.\ For the second measurement, the
procedure repeats itself; we ``slice'' again this new state according to:%
\begin{equation}
\mid\Psi_{m}(t_{2})>=\sum_{n}\mid\Psi_{m,n}(t_{2})> \label{dec-3}%
\end{equation}
where $\mid\Psi_{m,n}(t_{2})>$ is obtained by the action of the projector
$P_{N}(n)$ $\ $on the subspace corresponding to result $n$:%
\begin{equation}
\mid\Psi_{m,n}(t_{2})>=P_{N}(n)\mid\Psi_{m}(t_{2})> \label{dec-4}%
\end{equation}
The evolution of each $\mid\Psi_{m}(t_{2})>$ will now be considered
independently and, if a third measurement is performed at a later time $t_{3}%
$, generate one more decomposition, and so on.\ It is easy to
check\footnote{This can be done for instance by successive applications of the
postulate of the wave packet reduction and the evaluation of conditional
probabilities.\ Note that we have not restored the norm of any intermediate
state vector to 1, as opposed to what one usually does with the wave packet
reduction; this takes care of intermediate probabilities and explains the
simplicity of result (\ref{dec-5}).} that the probability of any given
sequence of measurements $m$, $n$, $p$, etc. is nothing but by the square of
the norm of the final state vector:%
\begin{equation}
\mathcal{P}(m,t_{1};n,t_{2};p,t_{3};..)=\mid<\Psi_{m,n,p,..q}(t_{q})\mid
\Psi_{m,n,p,..q}(t_{q})>\mid^{2} \label{dec-5}%
\end{equation}

Let us now describe the initial state of the system through a density operator
$\rho(t_{0})$;\ it turns out that the same result can be written in a compact
way, according to a formula that is sometimes called Wigner formula
\cite{Wigner1} \cite{Wigner-formule}.\ For this purpose, we consider the
time-dependent version, in the Heisenberg point of view\footnote{Let
$U(t,t_{0})$ be the unitary operator associated with the evolution of the
state vector between time $t_{0}$ and time $t_{1}$, in the Schr\"{o}dinger
point of view.\ If $P$ is any operator, one can obtain its transform
$\widehat{P}(t)$ in the ``Heisenberg point of view'' by the unitary
transformation: $\widehat{P}(t)=U^{\dagger}(t,t_{0})PU(t,t_{0})$, where
$U^{\dagger}(t,t_{0})$ is the Hermitian conjugate of $U(t,t_{0})$; the new
operator depends in general of time $t$, even if this is not the case for the
initial operator.}, of all projectors: $\widehat{P}_{M}(m;t)$ corresponds to
$P_{M}(m)$, $\widehat{P}_{N}(n;t)$ to $P_{N}(n)$, etc. One can then show that
the probability for obtaining result $m$ followed by result $n$ is given
by\footnote{Using circular permutation under the trace, one can in fact
suppress one of the extreme projectors $\widehat{P}_{N}(n;t_{2})$ in formula
(\ref{Wigner-proba}), but not the others.}:
\begin{equation}
\mathcal{P}(m,t_{1};n,t_{2})=Tr\left\{  \widehat{P}_{N}(n;t_{2})\widehat
{P}_{M}(m;t_{1})\rho(t_{0})\widehat{P}_{M}(m;t_{1})\widehat{P}_{N}%
(n;t_{2})\right\}  \label{Wigner-proba}%
\end{equation}
(generalizing this formula to more than two measurements, with additional
projectors, is straightforward).

Equation (\ref{Wigner-proba}) can be seen as a consequence of \ the wave
packet reduction postulate of quantum mechanics, since we obtained it in this
way.\ But it is also possible to take it as a starting point, as a postulate
in itself: it then provides the probability of any sequence of measurements,
in a perfectly unambiguous way, without resorting, either to the wave packet
reduction, or even to the Schr\"{o}dinger equation itself.\ The latter is
actually contained in the Heisenberg evolution of projection operators, but it
remains true that a direct calculation of the evolution of $\mid\Psi>$ is not
really necessary.\ As for the wave packet reduction, it is also contained in a
way in the trace operation of (\ref{Wigner-proba}), but even less
explicitly.\ If one just uses formula (\ref{Wigner-proba}), no conflict of
postulates takes place, no discontinuous jump of any mathematical quantity;
why not then give up entirely the other postulates and just use this single
formula for all predictions of results?

This is indeed the best solution for some physicists: if one accepts the idea
that the purpose of physics is only to correlate the preparation of a physical
system, contained mathematically in $\rho(t_{0})$, with all possible sequence
of results of measurements (by providing their probabilities), it is true that
nothing more than (\ref{Wigner-proba}) is needed. Why then worry about which
sequence is realized in a particular experiment? It is sufficient to assume
that the behavior of physical systems is fundamentally indeterministic, and
that there is no need in physics to do more than just giving rules for the
calculation of probabilities.\ The ``correlation interpretation'' is therefore
a perfectly consistent attitude; on the other hand, it is completely opposed
to the line of the EPR reasoning, since it shows no interest whatsoever in
questions related to physical reality as something ``in itself''. Questions
such as: ``how should the physical system be described when one first
measurement has already been performed, but before the second measurement is
decided'' should be dismissed as meaningless. Needless to say, the notion of
the EPR elements of reality becomes completely irrelevant, at least to
physics, a logical situation which automatically solves all potential problems
related to Bell, GHZ and Hardy type considerations.\ The same is true of the
emergence of a single result in a single experiment; in a sense, the
Schr\"{o}dinger cat paradox is eliminated by putting it outside of the scope
of physics, because no paradox can be expressed in terms of correlations.\ An
interesting feature of this point of view is that the boundary between the
measured system and the environment of the measuring devices is flexible; an
advantage of this flexibility is that the method is well suited for successive
approximations in the treatment of a measurement process, for instance the
tracks left by a particle in a bubble chamber as discussed by Bell
\cite{Bell-speakable}.

In practice, most physicists who favor the correlation interpretation do not
feel the need for making it very explicit.\ Nevertheless, this is not always
the case; see for instance the article by Mermin \cite{Mermin-corr}, which
starts from the statement: ``Throughout this essay, I shall treat correlations
and probabilities as primitive concepts''.\ In a similar context, see also a
recent ``opinion'' in Physics Today by Fuchs and Peres \cite{Fuchs-Peres} who
emphasize ``the internal consistency of the theory without
interpretation''.\ On the other hand, the correlation interpretation is seen
by some physicists as minimalistic because it leaves aside, as irrelevant, a
few questions that they find important; an example is the notion of physical
reality, seen as an entity that should be independent of measurements
performed by human beings.\ Nevertheless, as we have already mentioned, the
interpretation can easily be supplemented by others that are more
specific.\ In fact, experience shows that defenders of the correlation point
of view, when pressed hard in a discussion to describe their point of view
with more accuracy, often express themselves in terms that come very close to
the Everett interpretation (see \S \ \ref{ever}); in fact, they may sometimes
be proponents of this interpretation without realizing it!

Let us finally mention in passing that formula (\ref{Wigner-proba}) may be the
starting point for many interesting discussions, whether or not it is
considered as basic in the interpretation, or just as a convenient
formula.\ Suppose for instance that the first measurement is associated with a
degenerate eigenvalue of an operator, in other words that $\widehat{P}%
_{M}(m;t_{1})$ is a projector over a subspace of more than one dimension:%
\begin{equation}
\widehat{P}_{M}(m;t_{1})=\sum_{i=1}^{n}\mid\varphi_{i}><\varphi_{i}%
\mid\label{proj}%
\end{equation}
(for the sake of simplicity we assume that $t_{1}=t_{0}$, so that no time
dependence appears in this expression).\ Inserting this expression into
(\ref{Wigner-proba}) immediately shows the appearance of interference terms
(or crossed terms) $i\neq j$ between the contribution of the various
$\mid\varphi_{i}>$.\ Assume, on the other hand, that more information was
actually obtained in the first measurement, so that the value of $i$ was also
determined, but that this information was lost, or forgotten; the experimenter
ignores which of two or more $i$ results was obtained.\ Then, what should be
calculated is the sum of the probabilities associated with each possible
result, that is a single sum over $i$ from which all crossed term $i\neq j$
have disappeared.\ In the first case, interference terms arise because one has
to add probability amplitudes; in the second, they do not because one has to
add the probabilities themselves (exclusive events).\ The contrast between
these two situations may be understood in terms of decoherence: in the first
case, all states of the system correlate to the same state of the measuring
apparatus, which plays here the role of the environment; they do not in the
second case, so that by partial trace all interference effect vanish.\ This
remark is useful in the discussion of the close relation between the so called
``Zeno paradox in quantum mechanics'' \cite{Zeno} and decoherence; it is also
basic in the definition of consistency conditions in the point of view of
decoherent histories, to which we will come back later (\S \ \ref{histories}).

\subsection{Additional variables}

\label{additional}

We now leave the range of considerations that are more or less common to all
interpretations; from now on, we will introduce in the discussion some
elements that clearly do not belong to the orthodox interpretation.\ We begin
with the theories with additional variables, as the De Broglie theory of the
pilot wave \cite{De-Broglie}; the work of Bohm is also known as a major
reference in the subject \cite{Bohm} \cite{Bohm2}; see also the almost
contemporary work of Wiener and Siegel \cite{Wiener-Siegel}.\ More generally,
with or without explicit reference to additional variables, one can find a
number of authors who support the idea that the quantum state vector should be
used only for the description of statistical ensembles, not of single events,
- see for instance \cite{Ballentine} \cite{Pearle-3}.

We have already emphasized that the EPR theorem itself can be seen as an
argument in favor of the existence of additional variables (we will come back
later to the impact of the Bell and BKS theorems).\ Theories with such
variables can be built mathematically in order to reproduce exactly all
predictions of orthodox quantum mechanics; if they give exactly the same
probabilities for all possible measurements, it is clear that there is no hope
to disprove experimentally orthodox quantum mechanics in favor of these
theories, or the opposite.\ In this sense they are not completely new
theories, but rather variations on a known theory. They nevertheless have a
real conceptual interest: they can restore not only realism, but also
determinism (this is a possibility but not a necessity: one can also build
theories with additional variables that remain fundamentally non-deterministic).

\subsubsection{General framework}

None of the usual ingredients of orthodox quantum mechanics disappears in
theories with additional variables.\ In a sense, they are even reinforced,
since the wave function loses its subtle status (see \S \ \ref{status}), and
becomes an ordinary field with two components (the real part and the imaginary
part of the wave function - for simplicity, we assume here that the particle
is spinless); these components are for instance similar to the electric and
magnetic components of the electromagnetic field\footnote{The components of
the electromagnetic field are vectors while, here, we are dealing with scalar
fields; but this is unessential.}.\ The Schr\"{o}dinger equation itself
remains strictly unchanged. But a completely new ingredient is also
introduced: in addition to its wave function field, each particle gets an
additional variable $\lambda$, which evolves in time according to a new
equation.\ The evolution of $\lambda$ is actually coupled to the real field,
through a sort of ``quantum velocity term''\footnote{In Bohm's initial work, a
Newton law for the particle acceleration was written in terms of a ``quantum
potential''.\ Subsequent versions of Bohmian mechanics discarded the quantum
potential in favor of a quantum velocity term providing directly a
contribution to the velocity. Both points of view are nevertheless
consistent.\ An unexpected feature of the quantum velocity term is that it
depends only on the gradient of the phase of the wave function, not on its
modulus.\ Therefore, vanishingly small wave functions may have a finite
influence on the position of the particles, which can be seen as a sort of
non-local effect.} that depends on the wave function; but, conversely, there
is no retroaction of the additional variables onto the wave function. From the
beginning, the theory therefore introduces a marked asymmetry between the two
mathematical objects that are used to describe a particle; we will see later
that they also have very different physical properties.

\ For anyone who is not familiar with the concept, additional variables may
look somewhat mysterious; this may explain why they are often called
``hidden'', but this is only a consequence of our much better familiarity with
ordinary quantum mechanics! In fact, these variables are less abstract than
the wave functions, which in these theories becomes a sort of auxiliary field,
even if perfectly real.\ The additional variables are directly ``seen'' in a
measurement, while the state vector remains invisible; it actually plays a
rather indirect role, through its effect on the additional variables.\ Let us
take the example of a particle creating a track in a bubble chamber: on the
photograph we directly see the recording of the successive values of an
additional variable, which is actually nothing but.. the position of the
particle! Who has ever taken a photograph of a wave function?

For a single particle, the additional variable $\lambda$ may therefore also be
denoted as $\mathbf{R}$ since it describes its position; for a many particle
system, $\lambda$ is nothing but a symbol for the set of positions
$\mathbf{R}_{1}$, $\mathbf{R}_{2}$ etc. of all the particles.\ The theory
postulates an initial random distribution of these variables that depends on
the initial wave function $\Psi(\mathbf{r}_{1},\mathbf{r}_{2},...)$ and
reproduces exactly the initial distribution of probability for position
measurements; using hydrodynamic versions of the Schr\"{o}dinger equation
\cite{Madelung}, one can easily show that the evolution under the effect of
the ``quantum velocity term'' ensures that this property continues to be true
for any time.\ This provides a close contact with all the predictions of
quantum mechanics; it ensures, for instance, that under the effect of the
quantum velocity term the position of particles will not move independently of
the wave function, but always remain inside it.

At this point, it becomes natural to restore determinism, and to assume that
the results of measurements merely reveal the initial pre-existing value of
$\lambda$, chosen among all possible values in the initial probability
distribution.\ This assumption solves many difficulties, all those related to
the Schr\"{o}dinger cat paradox for instance: depending on the exact initial
position of a many-dimension variable $\lambda,$ which belongs to an enormous
configuration space (including the variables associated with the radioactive
nucleus as well as all variables associated with the cat), the cat remains
alive or dies.\ But restoring determinism is not compulsory, and
non-deterministic versions of additional variables can easily be designed.\ In
any case, the theory will be equivalent to ordinary quantum mechanics; for
instance, decoherence will act exactly in the same way, and make it impossible
in practice to observe interferences with macroscopic objects in very
different states.

To summarize, we have in this context a description of physical reality at two
different levels:

(i) one corresponding to the elements associated with the state vector, which
can be influenced directly in experiments, since its evolution depends on a
Hamiltonian that can be controlled, for instance by applying fields; this
level alone is not sufficient to give a complete description of a physical system.

(ii) another corresponding to the additional variables, which cannot be
manipulated directly (see appendix V), but obey evolution equations containing
the state vector.

The two levels together are necessary and sufficient for a complete
description or reality.\ There is no retroaction of the additional variables
onto the state vector, which creates an unusual situation in physics (usually,
when two physical quantities are coupled, they mutually influence each
other).\ Amusingly, we are now contemplating another sort of duality, which
distinguishes between direct action on physical systems (or preparation) and
results of observations performed on them (results of measurements).

A similar line of thought has been developed by\ Nelson \cite{Nelson}, who
introduces stochastic motions of point particles in such a way that their
statistical behavior reproduces exactly the predictions of the Schr\"{o}dinger
equation.\ The difference is that the evolution of the wave function is not
given by a postulate, but actually derived from other postulates that are
considered more fundamental.\ This leads to a natural derivation of the
Schr\"{o}dinger equation; the formalism is built to lead exactly to the same
predictions as orthodox quantum mechanics, so that its interest is mostly
conceptual. For the discussion of statistical mixtures in this context, see
ref. \cite{Werner}.

\subsubsection{Bohmian trajectories}

As soon as particles regain a position, they also get a trajectory, so that it
becomes natural to study their properties in various situations; actually one
then gets a variety of unexpected results.\ Even for a single particle in free
space, because of the effects of its wave function on the evolution of its
position, it turns out that the trajectories are not necessarily simple
straight lines \cite{trajectories}; in interference experiments, particles may
actually follow curved trajectories even in regions of space where they are
free, an unusual effect indeed\footnote{Another unusual effect takes place for
a particle with spin: the spin direction associated with the position of the
particle may sometimes spontaneously flip its direction, without any external
coupling \cite{Englert}.}! But this feature is in fact indispensable for the
statistics of the positions to reproduce the usual predictions of quantum
mechanics (interference fringes) \cite{Bell-Bohm}.\ Bell studied these
questions \cite{Bell-speakable} and showed, for instance, that the observation
of successive positions of a particle allows one to reconstruct a trajectory
that remains physically acceptable.

For systems of two particles or more, the situation becomes even more
interesting. Since the Schr\"{o}dinger equation remains unchanged, the wave
functions continues to propagate in the configuration space, while on the
other hand the positions propagate in ordinary three dimensional space.\ The
effects of non-locality become especially apparent through the ``quantum
velocity term'', since the velocity has to be evaluated at a point of
configuration space that depends on the positions of both particles; the
result for the velocity of particle $1$ may then depend explicitly on the
position of particle $2$.\ Consider for instance an EPRB experiment of the
type described in \S \ \ref{proof} and the evolution of the positions of the
two particles when they are far apart.\ If particle $1$ is sent through a
Stern-Gerlach analyzer oriented along direction $a$, the evolution of its
Bohmian position will obviously be affected in a way that depends on $a$ (we
remarked above that the positions have to follow the quantum wave functions;
in this case, it has the choice between two separating wave packets).\ But
this will also change the position $(\mathbf{R}_{1},\mathbf{R}_{2})$ of the
point representing the system in the six dimension configuration space, and
therefore change the quantum velocity term for particle $2$, \ in a way that
depends explicitly on $a$.\ No wonder if such a theory has no difficulty in
reproducing the non-local features of quantum mechanics!\ The advantage of
introducing additional variables is, in a sense, to emphasize the effects of
non-locality, which often remain relatively hidden in the orthodox formalism
(one more reason not to call these variables ``hidden''!).\ Bell for instance
wrote ``it is a merit of the Broglie-Bohm interpretation to bring this
(non-locality) out so explicitly that it can not be ignored'' - in fact,
historically, he came to his famous inequalities precisely through this channel.

An interesting illustration of this fact can be found in the study of Bohmian
trajectories in a two-particle interference experiment, or in a similar case
studied in reference \cite{Englert}.\ The authors of this reference study a
situation which involves an interference experiment supplemented by
electromagnetic cavities, which can store the energy of photons and be used as
a ``Welcher Weg'' device (a device that tells the experimenter which hole the
particle went through in an interference experiment).\ A second particle
probes the state of the field inside the cavity, and when leaving it takes a
trajectory that depends on this field. These authors show that, in some
events, a particle can leave a photon in a cavity and influence a second
particle, while the trajectory of the latter never crosses the cavity; from
this they conclude that the Bohmian trajectories are ``surrealistic''.\ Of
course, considering that trajectories are surrealistic or not is somewhat a
matter of taste.\ What is clear, however, is that a firm believer in the
Bohmian interpretation will not consider this thought experiment as a valid
argument against this interpretation - at best he/she will see it as a valid
attack against some truncated form of Bohmian theory.\ One should not mix up
orthodox and Bohmian theories, but always keep in mind that, in the latter
theory, the wave function has a totally different character: it becomes a real
classical field, as real as a laser field for instance.\ As expressed by Bell
\cite{Bell-no-one}: ``No one can understand this theory until he is willing to
think of $\Psi$ as a real objective field rather than just a probability
amplitude''.\ Therefore, a ``particle'' always involves a combination of both
a position and the associated field, which can not be dissociated; there is no
reason whatsoever why the latter could not also influence its surrounding.\ It
would thus be a mistake to assume that influences should take place in the
vicinity of the trajectory only.

In this context, the way out of the paradox is then simple: just to say that
the real field associated to the first particle\footnote{One sometimes
introduces the notion of the ``empty part of the wave function'' to
characterize the wave packet which does not contain a trajectory of the
particle, for instance in one arm of a Mach Zehnder interferometer.\ In the
present case, this empty part would deposit something (a photon?) in the
cavity that, later, would influence the trajectory of a second particle - in
other words we would have an indirect influence of the empty part on the
second particle.} interacted with the electromagnetic field in the cavity,
leaving a photon in it; later this photon acted on the trajectory of the
second particle.\ In other words, the effect is a crossed field-trajectory
effect, and in these terms it is even perfectly local!\ One could even add
that, even if for some reason one decided to just consider the trajectories of
the two particles, the fact that they can influence each other, even if they
never come close to each other creates no problem in itself; it is just an
illustration of the explicit character of non-locality in the Bohm theory -
see the quotation by Bell above, as well as the discussion of this thought
experiment by Griffiths \cite{Griffiths-Bohm}. So, we simply have one more
example of the fact that quantum \ phenomena are indeed local in configuration
space, but not necessarily in ordinary space.

This thought experiment nevertheless raises interesting questions, such as: if
in this example a particle can influence events outside of its own Bohmian
trajectory, what is then the physical meaning of this trajectory in
general?\ Suppose that, in a cloud chamber for instance, a particle could
leave a track that does not coincide at all with the trajectory of the Bohmian
position; in what sense then could this variable be called ``position''? For
the moment, that this strange situation can indeed occur has not been shown
(the example treated in \cite{Englert} is very special and presumably not a
good model for a cloud chamber), but the question clearly requests more
precise investigation.\ Another difficulty of theories with additional
variables is also illustrated by this thought experiment: the necessity for
including fields (in this case the photons in the cavities).\ Quantum
mechanics is used to describe a large variety of fields, from the usual
electromagnetic field (quantum electrodynamics) to quarks for instance, and
this is truly essential for a physical description of the world; at least for
the moment, the complete description of all these fields has not been
developed within theories with additional variables, although attempts in this
direction have been made.

\subsection{Modified (non-linear) Schr\"{o}dinger dynamics}

\label{stoch}

Another way to resolve the coexistence problem between the two postulates of
quantum mechanics is to change the Schr\"{o}dinger equation itself: one
assumes that the equation of evolution of the wave function contains, in
addition to the usual Hamiltonian terms, non-linear (and possibly stochastic)
terms, which will also affect the state vector \cite{Bohm-Bub-1} \cite{Pearle}
\cite{Ghirardi} \cite{stochastic} \cite{Diosi}.\ These terms may be designed
so that their effects remain extremely small in all situations involving
microscopic objects only (atoms, molecules, etc.); this will immediately
ensure that all the enormous amount of successful predictions of quantum
mechanics is capitalized.\ On the other hand, for macroscopic superpositions
involving for instance pointers of measurement apparatuses, the new terms may
mimic the effects of wave packet reduction, by selecting one branch of the
superposition and cancelling all the others. Clearly, one should avoid both
extremes: either perturb the Schr\"{o}dinger equation too much, and make
interference effects disappear while they are still needed (for instance,
possible recombination of the two beams at the exit of a Stern-Gerlach
magnet); or too little, and not ensure the complete disappearance of
Schr\"{o}dinger cats! This result is obtained if the perturbation term becomes
efficient when (but not before) any microscopic system becomes strongly
correlated to a macroscopic environment, which ensures that significant
decoherence has already taken place; we then know that the recovery of
interference effects is impossible in practice anyway. If carefully designed,
the process then reproduces the effect of the postulate of the wave function
collapse, which no longer appears as an independent postulate, but as a
consequence of the ``normal'' evolution of the wave function.

\subsubsection{Various forms of the theory}

There are actually various versions of theories with modified Schr\"{o}dinger
dynamics.\ Some versions request the introduction of additional variables into
the theory, while others do not.\ The approach proposed in 1966 by Bohm and
Bub \cite{Bohm-Bub-1} belongs to the first category, since these authors
incorporate in their theory additional variables previously considered by
Wiener and Siegel \cite{Wiener-Siegel}; these variables are actually contained
in a ``dual vector'', similar to the usual state vector $\mid\Psi>$, but
obeying an entirely different equation of motion -in fact, both vectors evolve
with coupled equations.\ What is then obtained is a sort of combination of
theories with additional variables and modified dynamics. For some ``normal''
distribution of the new variables, the prediction of usual quantum mechanics
are recovered; but it is also possible to assume the existence ``dispersion
free'' distributions that lead to non-orthodox predictions. An example of
models that are free of additional variables is given by the work of Pearle
\cite{Pearle}, published ten years later, in which nothing is added to the
usual conceptual frame of standard quantum mechanics.\ The theory is based on
a modified dynamics for the modulus and phases of the quantum amplitudes,
which get appropriate equations of evolution; the result is that, depending on
the initial values of the phases before a measurement, all probability
amplitudes but one go to zero during a measurement.\ Because, when a
microscopic system in sent towards a macroscopic apparatus, the initial phases
are impossible to control with perfect mathematical accuracy, an apparent
randomness in the results of experiments is predicted; the equations are
designed so that this randomness exactly matches the usual quantum
predictions.\ In both theories, the reduction of the state vector becomes a
dynamical process which, as any dynamical process, has a finite time duration;
for a discussion of this question, see \cite{Pearle-4}, which remarks that the
theory of ref. \cite{Bohm-Bub-1} introduces an infinite time for complete reduction.

Another line of thought was developed from considerations that were initially
not directly related to wave function collapse, but to continuous observations
and measurements in quantum mechanics \cite{BLP} \cite{Barchielli}. This was
the starting point for the work of Ghirardi et al. \cite{Ghirardi}, who
introduce a random and sudden process of ``spontaneous localization'' with an
arbitrary frequency (coupling constant), which resembles the effect of
approximate measurements in quantum mechanics.\ The constant is adjusted so
that, for macroscopic systems (and for them only), the occurrence of
superposition of far-away states is destroyed by the additional process; the
compatibility between the dynamics of microscopic and macroscopic systems is
ensured, as well as the disappearance of macroscopic coherent superpositions
(transformation of coherent superpositions into statistical mixtures).\ This
approach solves problems that were identified in previous work \cite{Pearle},
for instance the ``preferred basis problem'', since the basis is that of
localized states; the relation to the quantum theory of measurement is
examined in detail in \cite{BGRW}.\ In this model, for individual
systems\footnote{For ensemble of systems, the discontinuities are averaged,
and one recovers continuous equations of evolution for the density
operator.\ Since most of the discussion of \cite{Ghirardi} is given in terms
of density operators/matrices, and of the appearance of statistical mixtures
(decoherence), one may get the (incorrect) impression that individual
realizations are not considered in this work; but this is in fact not the case
and ``hitting processes'' are indeed introduced at a fundamental level.} the
localization processes are sudden (they are sometimes called ``hitting
processes''), which makes them completely different from the usual
Schr\"{o}dinger dynamics.\ Nevertheless, later work \cite{Pearle-5} showed
that it is possible to design theories involving only continuous evolution
that retain the attractive features of the model.\ For instance, the
discontinuous Markov processes in Hilbert space reduce, in an appropriate
limit, to a continuous spontaneous localization, which may result in a new
version of non-linear Schr\"{o}dinger dynamics \cite{GPR} called continuous
spontaneous localization (CSL); another achievement of \cite{GPR} is a full
compatibility with the usual notion of identical particles in quantum
mechanics. See also \cite{stochastic} for an earlier version of modified
Schr\"{o}dinger dynamics with very similar equations of evolution.

A similar line was followed by Diosi \cite{Diosi}, who also started initially
from the treatment of continuous measurements \cite{Diosi-2} by the
introduction of stochastic processes (``quantum Wiener processes''
\cite{Wiener-Siegel}) that are added to the usual deterministic
Schr\"{o}dinger dynamics.\ This author then introduced a treatment of the
collapse of the wave function from an universal law of density localization
\cite{Diosi-3}, with a strength that is proportional to the gravitational
constant, resulting in a parameter free unification of micro- and
macro-dynamics. Nevertheless, this approach was found to create severe
problems at short distances by the authors of \cite{GGR}, who then proposed a
modification of the theory that solves them, but at the price of
re-introducing a constant with dimension (a length).

Generally speaking, beyond their fundamental purpose (an unification of all
kinds of physical evolution, including wave function reduction), two general
features of these theories should of be emphasized.\ The first is that new
constants appear, which may in a sense look like ad hoc constants, but
actually have an important conceptual role: they define the limit between the
microscopic and macroscopic world (or between reversible and irreversible
evolution); the corresponding border is no longer ill-defined, as opposed to
the situation for instance in the Copenhagen interpretation.\ The second
(related) feature is that these theories are more predictive.\ They are
actually the only ones which propose a real physical mechanism for the
emergence of a single result in a single experiment, which is of course
attractive from a physical point of view.\ At the same time, and precisely
because they are more predictive, these theories become more vulnerable to
falsification, and one has to carefully design the mechanism in a way that
satisfies many constraints.\ For instance, we have already mentioned that, in
the initial Bohm-Bub theory, a complete collapse of the wave function is never
obtained in any finite time.\ The same feature actually exists in CSL: there
is always what is called a ``tail'' and, even when most of the wave function
goes to the component corresponding to one single outcome of an experiment,
there always remain a tiny component on the others (extremely small and
continuously going down in size).\ The existence of this component is not
considered as problematic by the proponents of the CSL theory, as illustrated
by the contributions of Pearle and Ghirardi in \cite{Shimony-Festschrift}.\ In
the context of possible conflicts with experiments, see also the discussion of
\cite{GGR} concerning incompatibilities of another form of the theory with the
well-known properties of microscopic objects, as well as \cite{BGRW-2} for a
critical discussion of another version of non-linear dynamics. A similar case
is provided by the generalization of quantum mechanics proposed by Weinberg
\cite{Weinberg}, which this author introduced as an illustration of a
non-linearity that is incompatible with available experimental data; see also
\cite{WS} for an application of the same theory to quantum optics and
\cite{Gisin-Weinberg} for a proof of the incompatibility of this theory
relativity, due to the prediction of superluminal communication (the proof is
specific of the Weinberg form of the non-linear theory and does not apply to
the other forms mentioned above).

\subsubsection{Physical predictions}

Whatever specific form of the theory is preferred, similar physical
descriptions are obtained.\ For instance, when a particle crosses a bubble
chamber, the new terms creates the appearance (at a macroscopic level) of a
particle trajectory; they also select one of the wave packets at the
measurement output of a Stern-Gerlach analyzer (and eliminate the other), but
not before these packets become correlated to orthogonal states of the
environment (e.g. detectors). Of course, any process of \ localization of the
wave function tends to operate in the space of positions rather than in the
space of momenta, which reduces to some extent the usual symmetry between
positions and momenta in quantum mechanics. This is actually not a problem,
but a convenient feature: one can easily convince oneself that, in practice,
what is measured in all experiments is basically the positions of particles or
objects (pointers, etc.), while momenta are only indirectly measured.
Generally speaking, it is a different spatial localization that produces wave
packet collapse.

How is an EPRB experiment described in this point of view?\ In the case of
Bohmian trajectories, we emphasized the role of the ``quantum velocity term'',
which has a value defined in configuration space and not in ordinary space;
here, what is essential is the role of the added non-linear localization term
in the Schr\"{o}dinger equation, which also acts in the 6 dimensional
configuration space.\ This term is designed so that, when correlation with the
environment takes place, one of the components in the corresponding basis
(``basis of decoherence'') is selected.\ Nothing special then occurs as long
as particle $1$ propagates within a Stern-Gerlach analyzer, since it is
microscopic and can perfectly well go through superpositions of far away
states; but as soon as it hits a detector at the output of the magnet, the
system develops correlations with the particles contained in the detector, the
amplifier, etc.\ so that a macroscopic level is reached and the localization
term becomes effective.\ Here, we see that it is the $a$ dependence of the
spatial localization (in other words, the basis of decoherence) that
introduces an overall effect on the two-particle state vector; it provides
particle $2$ with, not only a privileged spin-state basis, but also a
reduction of its spin state to one single component (when particle $1$ hits
the detector). Since this point of view emphasizes the role of the detectors
and not of the analyzers, it is clearly closer to the usual interpretation, in
terms of wave packet reduction, than the Bohmian interpretation. Nevertheless
it also puts into light the role of non-locality in an explicit way, as this
interpretation does.

What about the Schr\"{o}dinger cat and similar paradoxes? If the added
non-linear term has all the required properties and mimic the wave packet
reduction, they are easily solved.\ For instance, a broken poison bottle must
have at least some parts that have a different spatial localization (in
configuration space) than an unbroken bottle; otherwise it would have all the
same physical properties.\ It is then clear that the modified dynamics will
resolve the components long before it reaches the cat, so that the emergence
of a single possibility is ensured. For a recent discussion of the effects of
the modified dynamics on ``all or nothing coherent states''
(\S \ \ref{complete-ent}) in the context of quantum optics, and of the effects
on perception in terms of the ``relative state of the brain'' (\S \ref{ever}),
see ref. \cite{Ghirardi-2}.

The program can be seen as a sort of revival of the initial hopes of
Schr\"{o}dinger, where all relevant physics was contained in the wave function
and its progressive evolution (see the end of \S \ref{undulatory}); this is
especially true, of course, of the versions of non-linear dynamics that are
continuous (even if fluctuating extra quantities may be introduced), and not
so much of versions including ``hits'' that are too reminiscent of the wave
packet reduction.\ Here, the state vector directly describes the physical
reality, in contrast with our discussion of \S \ref{status}; we have a new
sort of wave mechanics, where the notion of point particles is given up in
favor of tiny wave packets.\ The theory is different from theories with
additional variables, because the notion of precise position in configuration
space never appears.\ As we have seen, another important difference is that
these theories with modified dynamics are really new theories: they may, in
some circumstances, lead to predictions that differ from those of orthodox
quantum mechanics, so that experimental tests might be possible.\ We should
emphasize that, in this point of view, the wave function can still not be
considered as an ordinary field: it continues to propagate in a high dimension
configuration space instead of the usual three dimension space.

A mild version of these theories is found in a variant where the
Schr\"{o}dinger equation remains exactly the same, but where stochastic terms
are introduced as a purely computational tool, and without any fundamental
purpose, for the calculation of the evolution of a partial trace density
matrix describing a subsystem \cite{Gisin2} \cite{Percival} \cite{Knight}; in
other words, a master equation for a density operator is replaced by an
average over several state vectors submitted to a random perturbation, which
may is some circumstances turn out to save computing time very efficiently.
Another line of thought that can be related to some extent to modified
Schr\"{o}dinger dynamics is the ``transactional interpretation'' of quantum
mechanics \cite{transactional}, where a quantum event is described by the
exchange of advanced and retarded waves; as in modified non-linear
Schr\"{o}dinger dynamics, these waves are then interpreted as real, and
non-locality is made explicit.

.

\subsection{History interpretation}

\label{histories}

The interpretation of ``consistent histories'' is also sometimes called
``decoherent history interpretation'', or just ``history interpretation'' as
we prefer to call it here (because the notion of consistency is essential at
the level of families of histories, rather than at the level of individual
histories).\ It proposes a logical framework that allows the discussion of the
evolution of a closed quantum system, without reference to measurements.\ The
general idea was introduced and developed by Griffiths \cite{Griffiths} but it
has also been used, and sometimes adapted, by other authors \cite{Omnès-2}
\cite{Gell-Mann} \cite{Omnès-3}. Since this interpretation is the most recent
among those that we discuss in this article, we will examine it in somewhat
more detail than the others.\ We will nevertheless remain within the limits of
a non-specialized introduction; the reader interested in more precise
information on the subject should go to the references that are provided - see
also a recent article in Physics Today \cite{G-O} and the references contained.

\subsubsection{Histories, families of histories}

Consider any orthogonal projector $P$ on a subspace $\mathcal{F}$ of the space
of states of a system; it has two eigenvalues, $+1$ corresponding to all the
states belonging to $\mathcal{F}$,$\mathcal{\,}$\ and $0$ corresponding to all
states that are orthogonal to $\mathcal{F}$ (they belong to the supplementary
sub-space, which is associated with the projector $Q=1-P$).\ One can associate
a measurement process with $P$: if the result of the measurement is $+1$, the
state of the system belongs to $\mathcal{F}$; if it is zero, it is orthogonal
to $\mathcal{F}$.\ Assume now that this measurement is made at time $t_{1}$ on
a system that is initially (at time $t_{0}$) described by a density operator
$\rho(t_{0})$; the probability for finding the state of the system in
$\mathcal{F}$ at time $t_{1}$ is then given by formula (\ref{Wigner-proba}),
which in this case simplifies into:%
\begin{equation}
\mathcal{P}(\mathcal{F},t_{1})=\text{Tr}\left\{  \widehat{P}(t_{1})\rho
(t_{0})\widehat{P}(t_{1})\right\}  \label{hist-1}%
\end{equation}
This result can obviously be generalized to several subspaces $\mathcal{F}%
_{1}$, $\mathcal{F}_{2}$, $\mathcal{F}_{3}$ , etc.\ and several measurement
times $t_{1}$, $t_{2}$, $t_{3}$, etc. (we assume $t_{1}<t_{2}<t_{3}<..$).\ The
probability that the state of the system belongs to $\mathcal{F}_{1}$ at time
$t_{1}$, then to $\mathcal{F}_{2}$ at time $t_{2}$, then to $\mathcal{F}_{3}$
at time $t_{3}$, etc. is, according to the Wigner formula:%
\begin{equation}
\mathcal{P}(\mathcal{F}_{1},t_{1};\mathcal{F}_{2},t_{2};\mathcal{F}_{3}%
,t_{3}..)=\text{Tr}\left\{  ..\widehat{P}_{3}(t_{3})\widehat{P}_{2}%
(t_{2})\widehat{P}_{1}(t_{1})\rho(t_{0})\widehat{P}_{1}(t_{1})\widehat{P}%
_{2}(t_{2})\widehat{P}_{3}(t_{3})...\right\}  \label{hist-2}%
\end{equation}
where, as above, the $\widehat{P}_{i}(t_{i})$ are the projectors over
subspaces $\mathcal{F}_{1}$, $\mathcal{F}_{2}$, $\mathcal{F}_{3}$ in the
Heisenberg point of view.\ We can now associate an ``history'' of the system
with this equation: an history $\mathcal{H}$ is defined by a series of
arbitrary times $t_{i}$, each of them associated with an orthogonal projector
$P_{i}$ over any subspace; its probability is given by (\ref{hist-2}) which,
for simplicity, we will write as $\mathcal{P}(\mathcal{H})$. In other words,
an history is the selection of a particular path, or branch, for the state
vector in a Von Neumann chain, defined mathematically by a series of
projectors.\ Needless to say, there is an enormous number of different
histories, which can have all sorts of properties; some of them are accurate
because they contain a large number of times associated with projectors over
small subspaces $\mathcal{F}$'s; others remain very vague because they contain
a few times only with projectors over large subspaces $\mathcal{F}$'s (one can
even decide that $\mathcal{F}$ is the entire states of spaces, so that no
information at all is contained in the history at the corresponding time).

There are in fact so many histories that it useful to group them into
families, or sets, of histories.\ A family is defined again by an arbitrary
series of times $t_{1}$, $t_{2}$, $t_{3}$, .., but now we associate to each of
these times $t_{i}$ an ensemble of orthogonal projectors $P_{i,j}$ that, when
summed, restore the whole initial space of states.\ For each time we then
have, instead of one single projector, a series of orthogonal projectors that
provide a decomposition of the unity operator:%
\begin{equation}
\sum_{j}P_{i,j}=1 \label{hist-3}%
\end{equation}
This gives the system a choice, so to say, among many projectors for each time
$t_{i}$, and therefore a choice among many histories of the same family.\ It
is actually easy to see from (\ref{hist-3}) and (\ref{hist-2}) that the sum of
probabilities of all histories of a given family is equal to one:%
\begin{equation}
\sum_{\text{histories of a family}}\mathcal{P}(\mathcal{H})=1 \label{hist-4}%
\end{equation}
which we interpret as the fact that the system will always follow one, and
only one, of them.

A family can actually also be built from a single history, the simplest way to
incorporate the history into a family is to associate, at each time $t_{i}$
($i=1,2,..,N)$, in addition to the projector $P_{i}$, the supplementary
projector $Q_{i}=1-P_{i}$; the family then contains $2^{N}$ individual
histories. Needless to say, there are many other ways to complement to single
family with ``more accurate'' histories than those containing the $Q$'s; this
can be done by decomposing each $Q$ into many individual projectors, the only
limit being the dimension of the total space of states.

\subsubsection{Consistent families}

All this looks very simple, but in general it is actually too simple to ensure
a satisfactory logical consistency in the reasonings. Having chosen a given
family, it is very natural to also enclose in the family all those histories
that can be built by replacing any pair or projectors, or actually any group
of projectors, by their sum; this is because the sum of two orthogonal
projectors is again a projector (onto a subspace that is the direct sum of the
initial subspaces). The difference introduced by this operation is that, now,
at each time, the events are no longer necessarily exclusive\footnote{For
these non exclusive families, relation (\ref{hist-4}) no longer holds since it
would involve double counting of possibilities.}; the histories incorporate a
hierarchy in their descriptive accuracy, including even cases where the
projector at a given time is just the projector over the whole space of states
(no information at all on the system at this time).

Consider the simplest case where two projectors only, occurring at time
$t_{i}$, have been grouped into one single projector to build a new history.
The two ``parent'' histories correspond to two exclusive possibilities (they
contain orthogonal projectors), so that their probabilities add independently
in the sum (\ref{hist-4}). What about the daughter history?\ It is exclusive
of neither of its parents and, in terms of the physical properties of the
system, it contains less information at time $t_{i}$: the system may have
either of the properties associated to the parents.\ But a general theorem in
probability theory states that the probability associated to an event than can
be realized by either of two exclusive events is the sum of the individual
probabilities; one then expects that the probability of the daughter history
should be the sum of the parent probabilities. On the other hand, inspection
of \ (\ref{hist-2}) shows that this is not necessarily the case; since any
projector, $\widehat{P}_{2}(t_{2})$ for instance, appears twice in the
formula, replacing it by a sum of projectors introduces four terms: two terms
that give the sum of probabilities, as expected, but also two crossed terms
(or ``interference terms'') between the parent histories, so that the
probability of the daughter history is in general different from the sums of
the parent probabilities. These crossed terms are actually very similar to the
right hand side of (\ref{hist-2}), but the trace always contains at some time
$t_{i}$ one projector $\widehat{P}_{i,j}(t_{i})$ on the left of $\rho(t_{0})$
and an orthogonal projector $\widehat{P}_{i,k}(t_{i})$ on the right. This
difficulty was to be expected: we know that quantum mechanics is linear at the
level of probability amplitudes, not probabilities themselves; interferences
occur because the state vector at time $t_{i}$, in the daughter story, may
belong to one of the subspaces associated with the parents, but may also be
any linear combination of such states.\ As a consequence, a linearity
condition for probabilities is not trivial.

One way to restore the additivity of probabilities is to impose the condition:%
\begin{equation}%
\begin{array}
[c]{l}%
\text{Tr}\left\{  ..\widehat{P}_{3,j_{3}}(t_{3})\widehat{P}_{2,j_{2}}%
(t_{2})\widehat{P}_{1,j_{1}}(t_{1})\rho(t_{0})\widehat{P}_{1,j_{1}^{^{\prime}%
}}(t_{1})\widehat{P}_{2,j_{2}^{^{\prime}}}(t_{2})\widehat{P}_{3,j_{3}%
^{^{\prime}}}(t_{3})...\right\} \\
\propto\delta_{j_{1},j_{1}^{^{\prime}}}\times\delta_{j_{2},j_{2}^{^{\prime}}%
}\times\delta_{j_{3},j_{3}^{^{\prime}}}\times...
\end{array}
\label{hist-5}%
\end{equation}
Because of the presence of the product of $\delta$'s in the right hand side,
the left hand side of (\ref{hist-5}) vanishes as soon as a least one pair of
the indices $(j_{1},j_{1}^{^{\prime}})$, $(j_{2},j_{2}^{^{\prime}})$,
$(j_{3},j_{3}^{^{\prime}})$, etc. contains different values; if they are all
equal, the trace merely gives the probability $\mathcal{P}(\mathcal{H})$
associated with the particular history of the family.\ What is important for
the rest of the discussion is the notion of consistent family: if condition
(\ref{hist-5}) is fulfilled for all projectors of a given family of histories,
we will say that this family is logically consistent, or consistent for short.
Condition (\ref{hist-5}) is basic in the history interpretation of quantum
mechanics; it is sometimes expressed in a weaker form, as the cancellation of
the real part only of the left hand side; this, as well as other points
related to this condition, is briefly discussed in Appendix VI. We now discuss
how consistent families can be used as an interpretation of quantum mechanics.

\subsubsection{Quantum evolution of an isolated system}

Let us consider an isolated system and suppose that a consistent family of
histories has been chosen to describe it; any consistent family may be
selected but, as soon as the choice is made, it cannot be modified and all the
other families are excluded (we discuss later what happens if one attempts to
describe the same system with more than one family). This unique choice
provides us with a well-defined logical frame, and with a series of possible
histories that are accessible to the system and give information at all
intermediate times $t_{1}$, $t_{2}$, ..\ Which history will actually occur in
a given realization of the physical system is not known in advance: we
postulate the existence of some fundamentally random process of Nature that
selects one single history among all those of the family.\ The corresponding
probability $\mathcal{P}(\mathcal{H})$ is given by the right hand side of
(\ref{hist-2}); since this formula belongs to standard quantum mechanics, this
postulate ensures that the standard predictions of the theory are
automatically recovered.\ For each realization, the system will then possess
at each time $t_{i}$ all physical properties associated to the particular
projectors $P_{i,j}$ that occur in the selected history. This provides a
description of the evolution of its physical properties that can be
significantly more accurate than that given by its state vector; in fact, the
smaller the subspaces associated to the projectors $P_{i,j}$'s, the more
accuracy is gained (obviously, no information is gained if all $P_{i,j}$'s are
projectors over the whole space of states, but this corresponds to a trivial
case of little interest). For instance, if the system is a particle and if the
projector is a projector over some region of space, we will say that the
particle is in this region at the corresponding time, even if the whole
Schr\"{o}dinger wave function extends over a much larger region.\ Or, if a
photon strikes a beam splitter, or enters a Mach-Zehnder interferometer, some
histories of the system may include information on which trajectory is chosen
by the photon, while standard quantum mechanics considers that the particle
takes all of them at the same time. Since histories contain several different
times, one may even attempt to reconstruct an approximate trajectory for the
particle, even in cases where this is completely out of the question in
standard quantum mechanics (for instance for a wave function that is a
spherical wave); but of course one must always check that the projectors that
are introduced for this purpose remain compatible with the consistency of a family.

In general, the physical information contained in the histories is not
necessarily about position only: a projector can also project over a range of
eigenstates of the momentum operator, include mixed information on position
and momentum (subject, of course, to Heisenberg relations, as always in
quantum mechanics), information on spin, etc.. There is actually a huge
flexibility on the choice of projectors; for each choice, the physical
properties that may be ascribed to the system are all those that are shared by
all states of the projection subspace, but not by any orthogonal state. A
frequent choice is to assume that, at a particular time $t_{i}$, all
$P_{i,,j}$'s are the projectors over the eigenstates of some Hermitian
operator $H$: the first operator $P_{i,,j=1}$ is the projector over all the
eigenstates of $H$ corresponding to the eigenvalue $h_{1}$, the second
$P_{i,,j=2}$ the corresponding projector for the eigenvalue $h_{2}$, etc. In
this case, all histories of the family will include an exact information about
the value of the physical quantity associated at time $t_{i}$ to $H$ (for
instance the energy if $H$ is the Hamiltonian). Let us nevertheless caution
the reader once more that we are not free to choose any operator $H_{i}$ at
any time $t_{i}$: in general, there is no reason why the consistency
conditions should be satisfied by a family built in this way.

Using histories, we obtain a description of the properties of the system in
itself, without any reference to measurements, conscious observers, etc.. This
does not mean that measurements are excluded; they can be treated merely as
particular cases, by incorporating the corresponding physical devices in the
system under study.\ Moreover, one attributes properties to the system at
different times; this is in contrast with the orthodox interpretation; where a
measurement does not necessarily reveal any pre-existing property of the
physical system, and projects it into a new state that may be totally
independent of the initial state.\ It is easy to show that the whole formalism
of consistent families is invariant under time reversal, in other words that
it makes no difference between the past and the future (instead of the initial
density operator $\rho(t_{0})$, one may use the final density operator
$\rho(t_{N})$ and still use the same quantum formalism \cite{ABL}) - for more
details, and even an intrinsic definition of consistency that involves no
density operator at all, see \S III of ref. \cite{Griffiths-2}.\ In addition,
one can develop a relation between consistent families and semi-classical
descriptions of a physical system; see ref. \cite{Gell-Mann} for a discussion
of how classical equations can be recovered for a quantum system provided
sufficient coarse graining is included (in order to ensure, not only
decoherence between the various histories of the family, but also what these
authors call ``inertia'' to recover classical predictability). See also chap.
16 of \cite{Omnès-3}\ for a discussion of how classical determinism is
restored, in a weak version that ensures perfect correlations between the
values of quasi-classical observables at different times (or course, there is
no question of fundamental determinism in this context).\ The history point of
view undoubtedly has many attractive features, and seems to be particularly
clear and easy to use, at least as long as one limits oneself to one single
consistent family of histories.

How does the history interpretation deals with the existence of several
consistent families?\ They are all a priori equally valid, but they will
obviously lead to totally different descriptions of the evolution of the same
physical system; this is actually the delicate aspect of the interpretation
(we will come back to it in the next subsection).\ The answer of the history
interpretation to the question is perfectly clear: different consistent
families are to be considered as mutually exclusive (except, of course, in
very particular cases where the two families can be embedded into a single
large consistent family); all families may be used in a logical reasoning, but
never combined together.\ In other words: the physicist is free to choose any
point of view in order to describe the evolution of the system and to ascribe
properties to the system; in a second independent step, another consistent
family may also be chosen in order to develop other logical considerations
within this different frame; but it would be totally meaningless (logically
inconsistent) to combine considerations arising from the two frames. This a
very important fundamental rule that must be constantly kept in mind when one
uses this interpretation. We refer the reader to \cite{Griffiths-2} for a
detailed and systematic discussion of how to reason consistently in the
presence of disparate families, and to \cite{Griffiths-3} for simple examples
of incompatible families of histories (photon hitting a beam splitter, \S II)
and the discussion of quantum incompatibility (\S V); various classical
analogies are offered for this incompatibility, including a two-dimensional
representation of a three-dimensional object by a draftsman, who can choose
many points of view to make a drawing, but can certainly not take several at
the same time - otherwise the projection would become inconsistent.

\subsubsection{Comparison with other interpretations}

In the history interpretation, as we have already seen, there is no need to
invoke conscious observers, measurement apparatuses, etc..; the system has
properties in itself, as in the non-orthodox interpretations that we discussed
before (considering that the correlation interpretation is orthodox).\ A
striking feature of the history interpretation, when compared to the others,
is the enormous flexibility that exists for the selection of the point of view
(family) that can be chosen for describing the system, since all the times
$t_{1}$, $t_{2}$, . are arbitrary (actually their number is also arbitrary)
and, for each of them, many different projectors $P$ may be introduced.\ One
may even wonder if the interpretation is sufficiently specific, and if this
very large number of families of histories is not a problem. This question
will come naturally in a comparison between the history interpretation to the
other interpretations that we have already discussed.

First, what is the exact relation between the history interpretation and the
orthodox theory?\ The relation is certainly very close, but several concepts
are expressed a more precise way.\ For instance, complementarity stands in the
Copenhagen interpretation as a general, almost philosophical, principle.\ In
the history interpretation, it is related to mathematical conditions, such as
consistency conditions; also, every projector can not be more precise that the
projector over a single quantum state $\mid\varphi>$, which is itself
obviously subject to the uncertainty relations because of the very structure
of the space of states.\ Of course, considerations on incompatible measurement
devices may still be made but, as the Bohrian distinction between the
macroscopic and microscopic worlds, they lose some of their fundamental
character. In the same vein, the history interpretation allows a quantum
theory of the universe (compare for instance with quotation v at the end of
\S \ \ref{diff}); we do not have to worry about dividing the universe into
observed systems and observers.\ The bigger difference between the orthodox
and the history interpretations is probably the way they describe the time
evolution of a physical system.\ In the usual interpretation, we have two
different postulates for the evolution of a single entity, the state vector,
which may sometimes create conflicts; in the history interpretation, the
continuous Schr\"{o}dinger evolution and the random evolution of the system
among histories are put at very different levels, so that the conflict is much
less violent.\ Actually, in the history point of view, the Schr\"{o}dinger
evolution plays a role only at the level of the initial definition of
consistent families (through the evolution operators that appear in the
Heisenberg operators) and in the calculation of the probability $\mathcal{P}%
\mathbb{(\mathcal{H})}$; the real time evolution takes place between the times
$t_{i}$ and $t_{i+1}$ and is purely stochastic.\ In a sense, there is a kind
of inversion of priorities, since it is now the non-determinist evolution that
becomes the major source of evolution, while in the orthodox point of view it
is rather the deterministic evolution of an isolated system. Nevertheless, and
despite these differences, the decoherent history interpretation remains very
much in the spirit of the orthodox interpretation; indeed, it has been
described as an ``extension of the Copenhagen interpretation'', or as ``a way
to emphasize the internal logical consistency of the notion of
complementarity''.\ On the other hand, Gell-Mann takes a more general point of
view on the history interpretation which makes the Copenhagen interpretation
just ``a special case of a more general interpretation in terms of the
decoherent histories of the universe.\ The Copenhagen interpretation is too
special to be fundamental, ...''\cite{reactions}.

What about the ``correlation interpretation''?\ In a sense, this minimal
interpretation is contained in both the orthodox interpretation (from which
some elements such as the reduction of the state vector have been removed) and
in the history interpretation.\ Physicists favoring the correlation
interpretation would probably argue that adding a physical discussion in terms
of histories to their mathematical calculation of probabilities does not add
much to their point of view: they are happy with the calculation of
correlations and do not feel the need for making statements on the evolution
of the properties of the system itself. Moreover, they might add that they
wish to insert whatever projectors correspond to a series of measurements in
(\ref{Wigner-proba}), and not worry about consistency conditions: in the
history interpretation, for arbitrary sequences of measurements, one would get
inconsistent families for the isolated physical system, and one has to include
the measurement apparatuses to restore consistency.\ We have already remarked
in \S \ \ref{correlation} that the correlation interpretation allows a large
flexibility concerning the boundary between the measured system and the
environment.\ For these physicists, the history description appears probably
more as an interesting possibility than as a necessity; but there is no
contradiction either.

Are there also similarities with theories with additional variables?\ To some
extent, yes.\ Within a given family, there are many histories corresponding to
the same Schr\"{o}dinger evolution and, for each history, we have seen that
more information on the evolution of physical reality is available than
through the state vector (or wave function) only.\ Under these conditions, the
state vector can be seen as a non-complete description of reality, and one may
even argue that the histories themselves constitute additional variables (but
they would then be family dependent, and therefore not EPR elements of
reality, as we discuss later).\ In a sense, histories provide a kind of
intermediate view between an infinitely precise Bohmian trajectory for a
position and a very delocalized wave function. In the Bohm theory, the wave
function pilots the position of the particles; in the decoherent history
interpretation, the propagation of the wave function pilots rather the
definition of histories (through a consistency condition) as well as a
calculation of probabilities, but not the evolution between times $t_{i}$ and
$t_{i+1}$, which is supposed to be fundamentally random.\ Now, of course, if
one wished, one could make the two sorts of theories even more similar by
assuming the existence of a well defined point in the space of histories; this
point would then be defined as moving in a completely different space from the
Bohm theory: instead of the configuration space, it would move in the space
defined by the family, and thus be defined as family dependent. In this way,
the history interpretation could be made deterministic if, for some reason,
this was considered useful. On many other aspects, the theories with
additional variables are very different from the history interpretation and we
can probably conclude this comparison by stating that they belong to rather
different point of view on quantum mechanics.

Finally, what is the comparison with theories incorporating additional
non-linear terms in the Schr\"{o}dinger evolution?\ In a sense, they
correspond to a completely opposite strategy: they introduce into one single
equation the continuous evolution of the state vector as well as a
\ non-linear deterministic mechanism simulating the wave packet reduction when
needed; the history interpretation puts on different levels the continuous
Schr\"{o}dinger evolution and a fundamentally random selection of history
selection by the system. One might venture to say that the modified non-linear
dynamics approach is an extension of the purely wave program of
Schr\"{o}dinger, while the history interpretation is a modern version of the
ideas put forward by Bohr.\ Another important difference is that a theory with
modified dynamics is not strictly equivalent to usual quantum mechanics, and
could lead to experimental tests, while the history interpretation is built to
reproduce exactly the same predictions in all cases - even if it can sometimes
provide a convenient point of view that allows to grasp its content more
conveniently \cite{Griffiths-comput}.

\subsubsection{A profusion of points of view; discussion}

We finally come back to a discussion of the impact of the profusion of
possible points of view, which are provided by all the families that satisfy
the consistency condition. We have already remarked that there is, by far, no
single way in this interpretation to describe the evolution of properties of a
physical system - for instance all the complementary descriptions of the
Copenhagen interpretation appear at the same level. This is indeed a large
flexibility, much larger than in classical physics, and much larger than in
the Bohmian theory for instance. Is the ``no combination of points of view''
fundamental rule really sufficient to ensure that the theory is completely
satisfactory?\ The answer to this question is not so clear for several
reasons.\ First, for macroscopic systems, one would like an ideal theory to
naturally introduce a restriction to sets corresponding to quasi-classical
histories; unfortunately, the number of consistent sets is in fact much too
large to have this property \cite{Zurek}.\ This is the reason why more
restrictive criteria for mathematically identifying the relevant sets are (or
have been) proposed, but no complete solution or consensus has yet been found;
the detailed physical consequences of consistency conditions are still being
explored, and actually provide an interesting subject of research.\ Moreover,
the paradoxes that we have discussed above are not all solved by the history
interpretation.\ Some of them are, for instance the Wigner friend paradox, to
the extent where no reference to observers is made in this
interpretation.\ But some others are not really solved, and the interpretation
just leads to a reformulation in a different formalism and vocabulary.\ Let us
for instance take the Schr\"{o}dinger cat paradox, which initially arose from
the absence of any ingredient in the Schr\"{o}dinger equation for the
emergence of single macroscopic result - in other words, for excluding
impossible macroscopic superpositions of an isolated, non-observed,
system.\ In the history interpretation, the paradox transposes in terms of
choice of families of histories: the problem is that there is no way to
eliminate the families of histories where the cat is at the same time dead and
alive; actually, most families that are mathematically acceptable through the
consistency condition contain projectors on macroscopic superpositions, and
nevertheless have exactly the same status as the families that do not.\ One
would much prefer to have a ``super-consistency'' rule that would eliminate
these superpositions; this would really solve the problem, but such a rule
does not exist for the moment.\ At this stage, one can then do two things:
either consider that the choice of sensible histories and reasonable points of
view is a matter of good sense - a case in which and one returns to the usual
situation in the traditional interpretation, where the application of the
postulate of wave packet is also left to the good taste of the physicist; or
invoke decoherence and coupling to the external world in order to eliminate
all these unwanted families - a case in which one returns to the usual
situation where, conceptually, it is impossible to ascribe reasonable physical
properties to a closed system without refereeing to the external world and
interactions with it\footnote{For instance, one sometimes invokes the
practical impossibility to build an apparatus that would distinguish between a
macroscopic superposition and the orthogonal superposition; this would justify
the elimination of the corresponding histories from those that should be used
in the description of reality. Such an argument reintroduces the notion of
measurement apparatus and observers in order to select histories, in
contradiction with the initial motivations of this point of view - see
Rosenfeld's citation in \S \ref{diff}. Moreover, this immediately opens again
the door to Wigner friend type paradoxes, etc.}, which opens again the door to
the Wigner friend paradox, etc.

Finally one may note that, in the decoherent history interpretation, there is
no attempt to follow ``in real time'' the evolution of the physical system;
one speaks only of histories that are seen as complete, ``closed in time'',
almost as histories of the past in a sense.\ Basic questions that were
initially at the origin of the introduction of the wave packet postulate, such
as ``how to describe the physical reality of a spin that has already undergone
a first measurement but not yet a second'', are not easily answered.\ In fact,
the consistency condition of the whole history depends on the future choice of
the observable that will be measured, which does not make the discussion
simpler than in the traditional interpretation, maybe even more complicated
since its very logical frame is now under discussion. What about a series of
measurements which may be, or may not be, continued in the future, depending
on a future decision?\ As for the EPR correlation experiments, they can be
re-analyzed within the history interpretation formalism \cite{Griffiths-4}
(see also \cite{Griffiths-5} for a discussion of the Hardy impossibilities and
the notion of ``consistent contrafactuality''); nevertheless, at a fundamental
level, the EPR\ reasoning still has to be dismissed for exactly the same
reason that Bohr invoked already long ago: it introduces the EPR\ notion of
``elements of reality'', or counterfactual arguments, that are not more valid
within the history interpretation than in the Copenhagen interpretation (see
for instance \S V of \cite{Griffiths-4} or the first letter in
\cite{reactions}).\ We are then brought back to almost the same old debate,
with no fundamentally new element. We have nevertheless already remarked that,
like the correlation interpretation, the history interpretation may be
supplemented by other ingredients, such as the Everett
interpretation\footnote{Nevertheless, since the Everett interpretation
completely suppresses from the beginning any specific notion of measurement,
measuring apparatus, etc., the usefulness of completing it with the history
interpretation is not obvious.} or, at the other extreme, EPR or deterministic
ingredients, a case in which the discussion would of course become different.

For a more detailed discussion of this interpretation, see the references
given at the beginning of this section; for a discussion of the relation with
decoherence, the notion of ``preferred (pointer) bases'', and classical
predictability, see \cite{Zurek}; for a critique of the decoherent history
interpretation, see for instance \cite{Dowker}, where it is argued among
others that consistency conditions are not sufficient to predict the
persistence of quasi-classicality, even at large scales in the Universe; see
also \cite{Kent} which claims that they are not sufficient either for a
derivation of the validity of the Copenhagen interpretation in the future; but
see also the reply to this critique by Griffiths in \cite{Griffiths-3}%
.\ Finally, another reference is a recent article in Physics Today
\cite{Goldstein} that contains a discussion of the history interpretation in
terms that stimulated interesting reactions from the proponents of the
interpretation \cite{reactions}.

\subsection{Everett interpretation}

\label{ever}

A now famous point of view is that proposed by Everett, who named it
``relative state interpretation'' - but in its various forms it is sometimes
also called ``many-worlds interpretation'', or ``branching universe
interpretation'' (the word ``branching'' refers here to the state vector of
the universe). In this interpretation, any possible contradiction between the
two evolution postulates is cancelled by a simple but efficient method: the
second postulate is merely suppressed!

In the Everett interpretation \cite{Everett}, the Schr\"{o}dinger equation is
taken even more seriously than in the orthodox interpretation.\ Instead of
trying to explain how successive sequences of well-defined measurement results
are obtained, one merely considers that single results never emerge: all
possibilities are in fact realized at the same time! The Von Neumann chain is
never broken, and its tree is left free to develop its branch ad
infinitum.\ The basic remark of this interpretation is that, for a composite
system of correlated subsystems (observed system, measurement apparatus, and
observer, all considered after a measurement), ``there does not exist anything
like a single state for one subsystem....one can arbitrarily choose a state
for one subsystem and be led to the relative state for the remainder'' - this
is actually just a description of quantum entanglement, a well-known
concept.\ But, now, the novelty is that the observer is considered as a purely
physical system, to be treated within the theory exactly on the same footing
as the rest of the environment.\ It can then be modelled by an automatically
functioning machine, coupled to the recording devices and registering past
sensory data, as well as its own machine configurations.\ This leads Everett
to the idea that ``current sensory data, as well as machine configuration, is
immediately recorded in the memory, so that all the actions of the machine at
a given instant can be considered as functions of the memory contents
only''..; similarly, all relevant experience that the observer keeps from the
past is also contained in this memory.\ Form this Everett concludes that
``there is no single state of the observer; ..with each succeeding observation
(or interaction), the observer state branches into a number of different
states...\ All branches exist simultaneously in the superposition after any
sequence of observations''.\ Under these conditions, the emergence of
well-defined results from experiments is not considered as a reality, but just
as a delusion of the mind of the observer. What the physical system does,
together with the environment, is to constantly ramify its state vector into
all branches corresponding to all measurement results, without ever selecting
one of these branches. The observer is also part of this ramification process,
that nevertheless has properties which prevent him/her to bring to his/her
mind the perception of several of them at the same time.\ Indeed, each
``component of the observer'' remains completely unaware of all the others, as
well as of the state vectors that are associated to them (hence the name
``relative state interpretation''). The delusion of the emergence of a single
result in any experiment then appears as a consequence of the limitations of
the human mind: in fact, the process that we call ``quantum measurement''
never takes place!

How is an EPRB\ experiment seen in this point of view? In the Bohmian
interpretation we emphasized the role of Stern-Gerlach analyzers, in the
non-linear evolution interpretation that of the detectors and decoherence;
here we have to emphasize the role of the correlations with the external world
on the mind of the two human observers. The state vector will actually develop
its Von Neumann chain through the analyzers and the detectors and, at some
point, include these observers whose brain will become part of the
superposition.\ For each choice of the settings $a$ and $b$, four branches of
the state vector will coexist, containing observers whose mind is aware of the
result associated with each branch. So, the choice of $a$ has a distant
influence on the mind of the second observer, through the definition of the
relevant basis for the Von Neumann chain, and non-locality is obtained as a result.

It is sometimes said that ``what is most difficult in the Everett
interpretation is to understand exactly what one does not understand''.
Indeed, it may look simple and attractive at first sight, but turns out to be
as difficult to defend as to attack (see nevertheless \S 3 of ref. \cite{Zeh},
where the author considers the theory as ambiguous because dynamical stability
conditions are not considered). The question is, to some extent, what one
should expect from a physical theory, and what it should explain. Does it have
to explain in detail how we perceive results of experiments, and if so of what
nature should such an explanation be? What is clear, anyway, is that the whole
point of view is exactly opposite to that of the proponents of the additional
variables: the emphasis is put, not on the physical properties of the systems
themselves, but on the effects that they produce on our minds.\ Notions such
as perception (ref.\ \cite{Everett} speaks of ``trajectory of the memory
configuration'') and psychology become part of the debate.\ But it remains
true that the Everett interpretation solves beautifully all difficulties
related to Bohrian dichotomies, and makes the theory at the same time simpler
and more pleasant esthetically. Since the human population of earth is made of
billions of individuals, and presumably since each of them is busy making
quantum measurements all the time without even knowing it, should we see the
state vector of the universe as constantly branching at a really fantastic rate?

\begin{center}
CONCLUSION
\end{center}

Quantum mechanics is, with relativity, the essence of the big conceptual
revolution of the physics of the $20^{th}$ century.\ Now, do we really
understand quantum mechanics? It is probably safe to say that we understand
its machinery pretty well; in other words, we know how to use its formalism to
make predictions in an extremely large number of situations, even in cases
that may be very intricate. Heinrich Hertz, who played such a crucial role in
the understanding of electromagnetic waves in the $19^{th}$ century (Hertzian
waves), remarked that, sometimes, the equations in physics are ``more
intelligent than the person who invented them'' \cite{Hertz}.\ The remark
certainly applies to the equations of quantum mechanics, in particular to the
Schr\"{o}dinger equation, or to the superposition principle: they contain
probably much more substance that any of their inventors thought, for instance
in terms of unexpected types of correlations, entanglement, etc. It is
astonishing to see that, in all known cases, the equations have always
predicted exactly the correct results, even when they looked completely
counter-intuitive. Conceptually, the situation is less clear.\ One major issue
is whether or not the present form theory of quantum mechanics is
complete.\ If it is, it will never be possible in the future to give a more
precise description of the physical properties of a single particle than its
wave function (or of two particles, for instance in an EPR\ type experiment);
this is the position of the proponents of the Copenhagen interpretation.\ If
it is not, future generations may be able to do better and to introduce some
kind of description that is more accurate.

We have shown why the EPR argument is similar to Gregor Mendel's reasoning,
which led him from observations performed between 1854 and 1863 to the
discovery of specific factors, the genes (the word appeared only later, in
1909), which turned out to be associated with microscopic objects hidden
inside the plants that he studied. In both cases, one infers the existence of
microscopic ``elements of reality'' from the results of macroscopic
observations.\ Mendel could derive rules obeyed by the genes, when they
combine in a new generation of plants, but at his time it was totally
impossible to have any precise idea of what they really could be at a
microscopic level (or actually even if they were microscopic objects, or
macroscopic but too small to be seen with the techniques available at that
time).\ It took almost a century before O.T.\ Avery and coll.\ (1944) showed
that the objects in question were contained in DNA molecules; later (1953),
F.\ Crick and J. Watson illustrated how subtle the microscopic structure of
the object actually was, since genes corresponded to subtle arrangement of
nucleic bases hidden inside the double helix of DNA molecules.\ We now know
that, in a sense, rather than simple microscopic objects, the genes are
arrangements of objects, and that all the biological machinery that reads them
is certainly far beyond anything that could be conceived at Mendel's
time.\ Similarly, if quantum mechanics is one day supplemented with additional
variables, these variables will not be some trivial extension of the other
variables that we already have in physics, but variables of a very different
nature.\ But, of course, this is only a possibility, since the histories of
biology and physics are not necessarily parallel! Anyway, the discussion of
additional variables leads to interesting questions, which we have tried to
illustrate in this article by a brief description of several possible
interpretations of quantum mechanics that have been or are still proposed;
some introduce additional variables that indeed have very special properties,
others do not, but in any case the theory contains at some stage features that
are reminiscent of these difficulties.

\ A natural comparison is with special relativity, since neither quantum
mechanics nor relativity is intuitive; indeed, experience shows that both,
initially, require a lot of thought from each of us before they become
intellectually acceptable.\ But the similarity stops here: while it is true
that, the more one thinks about relativity, the more understandable it becomes
(at some point, one even gets the feeling that relativity is actually a
logical necessity!), one can hardly say the same thing about quantum
mechanics.\ Nevertheless, among all intellectual constructions of the human
mind, quantum mechanics may be the most successful of all theories since,
despite all efforts of physicists to find its limits of validity (as they do
for all physical theories), and many sorts of speculation, no one for the
moment has yet been able to obtain clear evidence that they even exist. Future
will tell us if this is the case; surprises are always possible!

\bigskip

\begin{center}
APPENDICES \bigskip\bigskip

I.\ An attempt for constructing a ``separable'' quantum theory
(non-deterministic but local theory)
\end{center}

We come back to the discussion of \S \ \ref{peas} but now give up botany; in
this appendix we consider a physicist who has completely assimilated the rules
of quantum mechanics concerning non-determinism, but who is sceptical about
the essential character of non-locality in this theory (or non-separability;
for a detailed discussion of the meaning of these terms, see for instance
\cite{d'Espagnat} \cite{Bell-speakable}).\ So, this physicist thinks that, if
measurements are performed in remote regions of space, it is more natural to
apply the rules of quantum mechanics separately in these two regions.\ In
other words, in order to calculate the probability of any measurement result,
he/she will apply the rules of quantum mechanics, in a way that is perfectly
correct locally; the method assumes that it is possible to reason separately
in the two regions of space, and therefore ignores the non-separable character
of quantum events (quantum events may actually involve both space regions at
the same time). Let us take an extreme case, where the two measurement take
place in two different galaxies: our physicist would be prepared to apply
quantum mechanics to the scale of a galaxy, but not at an intergalactic scale!

How will he/she then treat the measurement process that takes place in the
first galaxy? It is very natural to assume that the spin that it contains is
described by a state vector (or by a density operator, it makes no difference
for our reasoning here) that may be used to apply the orthodox formula for
obtaining the probabilities of each possible result. If our experimenter is a
good scientist, he/she will realize at once that it is not a good idea to
assume that the two-spin system is described by a tensor product of states (or
of density operators); this would never lead to any correlation between the
results of measurements performed in the two galaxies. Therefore, in order to
introduce correlations, he/she will assume that the states in question (or the
density operators) are random mathematical objects, which fluctuate under the
effect of the conditions of emission of the particles (for instance). The
method is clear: for any possible condition of the emission, one performs an
orthodox quantum calculation in each region of space, and then takes an
average value over the conditions in question. After all, this is nothing but
the universal method for calculating correlations in all the rest of physics!
We note in passing that this approach takes into account the indeterministic
character of quantum mechanics, but introduces a notion of space separability
in the line of the EPR reasoning. Our physicist may for instance assume that
the two measurement events are separated by a space-like interval in the sense
of relativity, so that no causal relation can relate them in any circumstance;
this seem to fully justify an independent calculation of both phenomena.

Even if this is elementary, and for the sake of clarity, let us give the
details of this calculation.\ The fluctuating random variable that introduces
the correlations is called $\lambda$, and the density operator of the first
spin $\rho_{1}(\lambda)$; for a direction of measurement defined by the unit
vector $a$, the eigenstate of the measurement corresponding to result $+1$ is
noted $\mid+/a>$.\ The probability for obtaining result $+$ if the first
measurement is made along direction $a$ is then written as:
\begin{equation}
\mathcal{P}_{+}(a,\lambda)=<+/a\mid\rho_{1}(\lambda)\mid+/a> \label{aa1}%
\end{equation}
In the same way, one writes the probability for the result $-1$ in the form:
\begin{equation}
\mathcal{P}_{-}(a,\lambda)=<-/a\mid\rho_{1}(\lambda)\mid-/a> \label{aa2}%
\end{equation}
If, instead of direction $a$, \ another different direction $a^{^{\prime}}$ is
chosen, the calculations remain the same and lead to two functions
$\mathcal{P}_{\pm}(a^{^{\prime}},\lambda)$. As for measurements performed in
the second region of space, they provide two functions $\mathcal{P}_{\pm
}(b,\lambda)$ and $\mathcal{P}_{\pm}(b,\lambda)$.

We now calculate the number which appears in the Bell theorem (BCHSH
inequality), namely the linear combination, as in (\ref{5}), of four average
values of products of results associated with the couples of orientations
$(a,b)$, $(a,b^{^{\prime}})$, $(a^{^{\prime}},b)$, $(a^{^{\prime}}%
,b^{^{\prime}})$. Since we have assumed that results are always $\pm1$, the
average value depends only on the differences:
\begin{equation}
A(\lambda)=\mathcal{P}_{+}(a,\lambda)-\mathcal{P}_{-}(a,\lambda) \label{aa3}%
\end{equation}
or:
\begin{equation}
A^{^{\prime}}(\lambda)=\mathcal{P}_{+}(a^{^{\prime}},\lambda)-\mathcal{P}%
_{-}(a^{^{\prime}},\lambda) \label{aa4}%
\end{equation}
(with similar notation for the measurements performed in the other region of
space) and can be written as the average value over $\lambda$ of expression:
\begin{equation}
A(\lambda)B(\lambda)+A(\lambda)B^{^{\prime}}(\lambda)-A^{^{\prime}}%
(\lambda)B(\lambda)+A^{^{\prime}}(\lambda)B^{^{\prime}}(\lambda) \label{aa5}%
\end{equation}

We are now almost back to the calculation of section \ref{proof}, with a
little difference nevertheless: the $A$'s and $B$'s are now defined as
probability differences so that their values are not necessarily $\pm1$.\ It
is nonetheless easy to see that they are all between $+1$ and $-1$ , whatever
the value of $\lambda$ is.\ Let us for a moment consider $\lambda$, $A$ and
$A^{^{\prime}}$ as fixed, keeping only $B$ and $B^{^{\prime}}$ as
variables;\ in the space of these variables, expression (\ref{aa5})
corresponds to a plane surface which, at the four corners of the square
$B=\pm1$, $B^{^{\prime}}=\pm1$, takes values $\pm2A$ or $\pm2A^{^{\prime}}$,
which are between $\pm2$; at the center of the square, the plane goes through
the origin. By linear interpolation, it is clear that, within the inside of
the square, the function given by (\ref{aa5}) also remains bounded between
$\pm2$; finally, its average value has the same property.\ Once more we find
that the Bell theorem holds in a large variety of contexts!

Since we know that quantum mechanics as well as experiments violate the Bell
inequality, one may wonder what went wrong in the approach of our physicist;
after all, his/her reasoning is based on the use of the usual formalism of
quantum mechanics.\ What caused the error was the insistence of treating the
measurements as separable events, while orthodox quantum mechanics requires us
to consider the whole two-spin system as a single, non-separable, system; in
this system, no attempt should be made to distinguish subsystems.\ The only
correct reasoning uses only state vectors/density operators that describe this
whole system in one mathematical object.\ This example illustrates how it is
really separability and/or locality which are at stake in a violation of the
Bell inequalities, not determinism.

It is actually instructive, as a point of comparison, to make the calculation
of standard quantum mechanics as similar as possible to the reasoning that led
to the inequality (\ref{aa5}).\ For this purpose, we notice that any density
operator $\rho$ of the whole system belongs to a space that is the tensor
product of the corresponding spaces for individual systems; therefore $\rho$
can always be expanded as:%
\begin{equation}
\rho=\sum_{n,p}c_{n,p}\,\left[  \rho_{n}(1)\otimes\rho_{p}(2)\right]
\label{aa6}%
\end{equation}
From this, one can obtain the probability of obtaining two results $+1$ along
directions $a$ and $b$ as:%
\begin{equation}
\mathcal{P}_{++}(a,b)=\sum_{n,p}c_{n,p}<+/a\mid\rho_{n}\mid+/a><+/b\mid
\rho_{p}\mid+/b> \label{aa7}%
\end{equation}
(probabilities corresponding to the other combinations of results are obtained
in the same way).\ The right hand side of this equation is not completely
different from the sum over $\lambda$ that was used above; actually it is very
similar, since the sum over the indices $n$ and $p$ plays the same role as the
sum over the different values of $\lambda$.\ In fact, if all $c_{n,p}$'s were
real positive numbers, and if all operators $\rho_{n}$ and $\rho_{p}$ were
positive (or semi-positive) operators, nothing would prevent us from doing
exactly the same reasoning again and deriving the Bell inequality; in other
words, any combined system that is a statistical mixture (which implies
positive coefficients) of uncorrelated states satisfies the Bell
inequalities.\ But, in general, the positivity conditions are not fulfilled,
and this is precisely why the quantum mechanical results can violate the
inequalities. \bigskip\bigskip

\begin{center}
II. Maximal probability for a Hardy state.
\end{center}

In this appendix we give more details on the calculations of \S \ \ref{hardy};
the two-particle state corresponding to the measurement considered in (i) is
the tensor product of ket (\ref{h2}) by its correspondent for the second
spin:
\begin{equation}
\cos^{2}\theta\mid+,+>+\sin\theta\cos\theta\left[  \mid+,->+\mid-,+>\right]
+\sin^{2}\theta\mid-,-> \label{c1}%
\end{equation}
which has the following scalar product with ket (\ref{h7}):
\begin{equation}
\cos^{2}\theta\sin\theta-2\sin\theta\cos^{2}\theta=-\sin\theta\cos^{2}%
\theta\label{c2}%
\end{equation}
The requested probability is obtained by dividing the square of this
expression by the square of the norm of the state vector:
\begin{equation}
\mathcal{P}=\frac{\sin^{2}\theta\cos^{4}\theta}{2\cos^{2}\theta+\sin^{2}%
\theta}=\frac{\sin^{2}\theta\left(  1-\sin^{2}\theta\right)  ^{2}}{2-\sin
^{2}\theta} \label{c3}%
\end{equation}
A plot of this function shows that it has a maximum of about 0.09.

\bigskip

\begin{center}
III. Proof of relations (\ref{8}) and (\ref{9}).
\end{center}

Let us start with the ket:%

\begin{equation}
\mid\Psi>\,=\,\mid+,+,+>+\eta\mid-,-,-> \label{a1}%
\end{equation}
where:
\begin{equation}
\eta=\pm1 \label{a2}%
\end{equation}
We wish to calculate the effect of the product operator $\sigma_{1x}%
\sigma_{2y}\sigma_{3y}$ on this ket.\ Since every operator in the product
commutes with the two others, the order in which they are applied is
irrelevant; let us then begin with the operator associated with the first spin:%

\begin{equation}%
\begin{array}
[c]{ll}%
\sigma_{1+}\mid\Psi>= & 2\eta\mid+,-,->\\
\sigma_{1-}\mid\Psi>= & 2\mid-,+,+>
\end{array}
\label{a3}%
\end{equation}
which provides:
\begin{equation}
\sigma_{1x}\mid\Psi>=\mid\Psi^{^{\prime}}>=\eta\mid+,-,->+\mid-,+,+>
\label{a4}%
\end{equation}
For the second spin:
\begin{equation}%
\begin{array}
[c]{ll}%
\sigma_{2+}\mid\Psi^{^{\prime}}>= & 2\eta\mid+,+,->\\
\sigma_{2-}\mid\Psi^{^{\prime}}>= & 2\mid-,-,+>
\end{array}
\label{a5}%
\end{equation}
so that:
\begin{equation}
\sigma_{2y}\mid\Psi^{^{\prime}}>=\mid\Psi^{^{\prime\prime}}>=\frac{1}%
{i}\left(  \eta\mid+,+,->-\mid-,-,+>\right)  \label{a6}%
\end{equation}
Finally, the third spin gives:
\begin{equation}%
\begin{array}
[c]{ll}%
\sigma_{3+}\mid\Psi^{^{\prime\prime}}>= & -2i\eta\mid+,+,+>\\
\sigma_{3-}\mid\Psi^{^{\prime\prime}}>= & +2i\mid-,-,->
\end{array}
\label{a7}%
\end{equation}
which leads to:
\begin{equation}
\sigma_{3y}\mid\Psi^{^{\prime\prime}}>=-\eta\mid+,+,+>-\mid-,-,->=-\eta
\mid\Psi> \label{a8}%
\end{equation}
(since $\eta^{2}=1$). Indeed, we find that $\mid\Psi>$ is an eigenstate of the
product of the three spin operators $\sigma_{1x}\sigma_{2y}\sigma_{3y}$, with
eigenvalue $-\eta$.\ By symmetry, it is obvious that the same is true for the
product operators $\sigma_{1y}\sigma_{2x}\sigma_{3y}$ and $\sigma_{1y}%
\sigma_{2y}\sigma_{3x}$.

Let us now calculate the effect of operator $\sigma_{1x}\sigma_{2x}\sigma
_{3x}$ on $\mid\Psi>$; from (\ref{a5}) we get:
\begin{equation}
\sigma_{2x}\mid\Psi^{^{\prime}}>=\mid\Psi^{^{\prime\prime\prime}}>=\left(
\eta\mid+,+,->+\mid-,-,+>\right)  \label{a9}%
\end{equation}
so that:
\begin{equation}%
\begin{array}
[c]{ll}%
\sigma_{3+}\mid\Psi^{^{\prime\prime\prime}}>= & 2\eta\mid+,+,+>\\
\sigma_{3-}\mid\Psi^{^{\prime\prime\prime}}>= & 2\mid-,-,->
\end{array}
\label{a10}%
\end{equation}
and, finally:
\begin{equation}
\sigma_{3x}\mid\Psi^{^{\prime\prime\prime}}>=\eta\mid+,+,+>+\mid
-,-,->=\eta\mid\Psi> \label{a11}%
\end{equation}
The change of sign between (\ref{a8}) and (\ref{a11}) may easily be understood
in terms of simple properties of the Pauli spin operators (anticommutation and
square equal to one).

\begin{center}
\bigskip\bigskip

\bigskip\bigskip IV.\ Impossibility of superluminal communication and of
cloning quantum states.
\end{center}

In EPR schemes, applying the reduction postulate projects the second particle
instantaneously onto an eigenstate corresponding to the same quantization axis
as the first measurement. If it were possible to determine this state
completely, superluminal communication would become accessible: from this
state, the second experimenter could calculate the direction of the
quantization axis to which it corresponds, and rapidly know what direction was
chosen by the first experimenter\footnote{Note that what is envisaged here is
communication through the choice of the settings of the measurement
apparatuses; this makes sense since the settings are chosen at will by the
experimenters; on the other hand, the results of the experiments are not
controlled, but random, so that they cannot be directly used as signals.},
without any special effect of the distance, for instance even if the
experimenters are in two different remote galaxies.\ This, obviously, could be
used as a sort or telegraph, completely free of any relativistic minimum delay
(proportional to the distance covered).\ The impossibility for superluminal
communications therefore relies on the impossibility of a complete
determination of a quantum state from a single realization of this
state.\ Such a realization allows only one single measurement, which (almost
always) perturbs the state, so that a second measurement on the same state is
not feasible; there is not, and by far, sufficient information in the first
measurement for a full determination of the quantum state - see discussion in
\S \ \ref{crypto}.

\ Now, suppose for a moment that a perfect ``cloning'' of quantum states could
be performed - more precisely the multiple reproduction (with many particles)
of the unknown state of a single particle\footnote{The ``cloning'' operation
is not to be confused with the preparation of a series of particles into a
same known quantum state: this operation can be performed by sending many spin
1/2 half particles through the same Stern-Gerlach magnet, or many photons
through the same polarizing filter.\ What is theoretically impossible is to
perfectly duplicate an initially unknown (and arbitrary) state.}.\ Applying
the cloning process to the second particle of an EPR pair, one could then make
a large number of perfect copies of its state; in a second step, one could
perform a series of measurements on each of these copies, and progressively
determine the state in question with arbitrary accuracy.\ In this way, the
possibility for superluminal communication would be restored!\ But, in
reality, quantum mechanics does not allow either for such a perfect
reproduction of quantum states \cite{Wooters-Zurek} \cite{Dieks}; for
instance, if one envisages using stimulated emission in order to clone the
state of polarization of one single photon into many copies, the presence of
spontaneous emission introduces noise in the process and prevents perfect
copying. A discussion of multi-particle cloning is given in
\cite{Gisin-Massar}.

This, nevertheless, does not completely solve the general question: even
without cloning quantum states, that is only with the information that is
available in one single measurement in each region of space, it is not so
obvious that the instantaneous reduction of the wave packet cannot be used for
superluminal communication.\ After all, it is possible to repeat the
experiment many times, with many independent pairs of correlated particles,
and to try to extract some information from the statistical properties of the
results of all measurements.\ The EPR correlations are very special and
exhibit such completely unexpected properties (e.g. violations of the Bell
inequalities)! Why not imagine that, by using or generalizing EPR schemes
(more than two systems, delocalized systems, etc.), one could invent schemes
where superluminal communication becomes possible?\ Here we show why such
schemes do not exist; we will sketch the general impossibility proof in the
case of two particles (or two regions of space), but the generalization to
more systems in several different regions of space is straightforward.

Suppose that, initially, the two remote observers already possess a collection
of pairs of correlated particles, which have propagated to their remote
galaxies before the experiment starts.\ Each pair is in an arbitrary state of
quantum entanglement, and we describe it with a density operator $\rho$ in a
completely general way.\ The first observer then chooses a setting $a$ or,
more generally, any local observable $O_{A}$ to measure; the second observer
is equally free to choose any local observable $O_{B}$, and may use as many
particles as necessary to measure the frequency of occurrence of each result
(i.e. probabilities); the question is whether the second observer can extract
some information on $O_{A}$ from any statistical property of the observed
results.\ The impossibility proof relies on the fact that all operators
(observables) corresponding to one of the two sub-systems always commute with
all operators corresponding to the other; consequently, for any choice of the
operators, it is always possible to construct a common eigenbasis $\left\{
\mid\varphi_{k}\,,\,\theta_{j}>\right\}  $ in the space of states of the
two-particle system, where the $\mid\varphi_{k}>$'s are the eigenstates of
$O_{A}$ and the $\mid\theta_{j}>$'s are the eigenstates of $O_{B}$.\ We can
then calculate the probability of sequences of measurement where the first
operator obtains result $A_{m}$ (corresponding, if this eigenvalue is
degenerate, to some range $D_{m}$ for the index $k$) and the second result
$B_{n}$ (corresponding to range $D_{n}$ for index $j$).\ But, what we are
interested in is slightly different: the probability that the second observer
will obtain each result $B_{n}$ after a measurement performed by the other
observer, independently of the result $A_{m}$, since there is no way to have
access to this result in the second galaxy; our purpose is to prove that this
probability is independent of the choice of the operator $O_{A}$.

Mathematically, extracting the probabilities concerning the second observer
only amounts to summing over all possible results $A_{m}$, with the
appropriate weights (probabilities); this is a classical problem, which leads
to the notion of ``partial trace'' $\rho_{B}$ over the variables of the
sub-system $A$.\ This operator acts only in the space of states of system $B$
and is defined by its matrix elements:%
\begin{equation}
<\theta_{i}\mid\rho_{B}\mid\theta_{j}>=\sum_{k}<\varphi_{k},\,\theta_{i}%
\mid\rho\mid\varphi_{k}\,,\,\theta_{j}> \label{app4-1}%
\end{equation}
It contains all information that the second experimenter needs for making
predictions, exactly as from any ordinary density operator for an isolated
system; for instance, the probability of observing result $B_{n}$ is simply:%
\begin{equation}
\mathcal{P}(B_{n})=Tr\left\{  \sum_{i\in D_{n}}\mid\theta_{i}><\theta_{i}%
\mid\rho_{B}\right\}  \label{app4-2}%
\end{equation}
Equations (\ref{app4-1}) and (\ref{app4-2}) can be derived in different
ways.\ One can for instance use formula (\ref{Wigner-proba}), if it has been
proved before.\ Otherwise, one can proceed in steps: one first expands $\rho$
in terms of projectors onto its own eigenstates $\mid\Psi_{l}>$, with positive
eigenvalues; one then applies the wave packet reduction postulate to each
$\mid\Psi_{l}>$ separately in order to get the probability of any sequence of
results; one finally performs the sum over $l$ as well as the appropriate sum
over indices $k$ (unknown result) and $j$ (if the observed eigenvalue is
degenerate) in order to obtain the ``reduced probabilities'' - by these words
we mean the probabilities relevant to the second observer, just after the
other has performed a measurement of $O_{A}$, but before it has been possible
to communicate the result to the second by some classical channel. This
calculation provides the above expressions.

From formula (\ref{app4-1}), one might get the impression that the partial
trace depends on the choice of the basis $\left\{  \mid\varphi_{k}>\right\}
$, so that there is some dependence of operator $\rho_{B}$ on the choice of
$O_{A}$. This is a false impression: in fact, a simple algebra shows that the
sum contained in the partial trace is completely independent of the basis
chosen in the traced space of states; it does not even matter if the first
experimenter has performed any experiment of not.\ Therefore, the second
experimenter receives exactly the same information, completely independently
of the decisions made by the first experimenter; no superluminal communication
is possible.

Finally, one could object that it is not indispensable to have one system
located in one region of space, the other in the second region, as we have
assumed until now; each of them could perfectly well be delocalized in a
superposition of states in different locations.\ Does the proof hold in this
case?\ Yes, it does, after some modification.\ In this case, one should now
associate the letters $A$ and $B$, as well as operators $O_{A}$ and $O_{B}$,
not to sub-systems as before, but to measurements performed in each region of
space.\ Each relevant operator can then be put between two projectors onto
states that are localized either in the first (projector $P_{A}$), or the
second (projector $P_{B}$), region of space.\ Since $P_{A}$ and $P_{B}$ are
orthogonal, it is then simple to show that all operators with index $A$
commute with all operators with index $B$ (this is similar, in field theory,
to the commutation of field operators that are outside mutual light cones);
this remains true even if they act in the space of states of the same
particle.\ We are now dealing with a generalization of the notion of partial
trace, which is no longer related to the existence of different sub-systems
(it may actually apply to one particle only), but to two different sets of
operators acting in the same space of states.\ If all operators of one set
commute with all operators of the second set, the notion of partial trace can
indeed be transposed, and it turns out that the final result is independent of
the operator that was chosen in the first set in order to calculate the
trace.\ This allows one to prove that the information available in one region
of space is completely independent of the kind of measurement performed in the
other.\ Indeed, quantum mechanics is not contradictory with relativity!

\bigskip

\begin{center}
V. Manipulating and preparing additional variables

\bigskip
\end{center}

Using the hydrodynamic equations associated with the evolution of the wave
function, in order to guide the evolution of the additional variables
(positions), may look like a very natural idea.\ In other fields of physics,
it is known that the hydrodynamic equations can be obtained by taking averages
of microscopic quantities over positions and velocities of point like
particles; there is some analogy between the guiding term and the force term
in a Landau type kinetic equations, where each particle is subject to an
average force proportional to the gradient of the density for
instance.\ Nevertheless, here we are dealing with a single particle, so that
the guiding term can not be associated with interactions between
particles.\ Moreover, we also know from the beginning that rather unusual
properties must be contained in the guiding equations, at least if the idea is
to exactly reproduce the predictions of usual quantum mechanics: the Bell
theorem states that the additional variables have to evolve non-locally in
ordinary three dimension space (on the other hand, in the configuration space
of the system, they may evolve locally, exactly as for the state vector).\ In
other words, the additional variables must be able to influence each other at
an arbitrary distance in real space. Indeed, in the Bohmian equation of motion
of the additional variables, the velocity of a particle contains an explicit
dependence on its own position, as expected, but also a dependence on the
positions of all the other particles (assuming that the particles are
entangled).\ This is not a problem in itself: as mentioned in the main text,
one can consider that making non-locality completely explicit in the equations
is actually an advantage of Bohmian mechanics.

But one also has to be careful when this non-local term is included in the
equations of motion: since relativity is based on the idea that it is totally
impossible to send a message at a velocity exceeding the velocity of light,
one must avoid features in the theory that would create conflicts with this
principle.\ We must distinguish two cases, depending whether we consider
influences on the additional variables that are direct (one modifies them ``by
hand'', in a completely arbitrary way, as for instance the position of a
billiard ball), or indirect (applying external fields changes the Hamiltonian
of the system, therefore modifies the evolution of the wave function so that,
in turn, the evolution of the additional variables is affected).\ In the
latter case, one can check that the non-local Bohmian term creates no problem:
it cannot be used to transmit instantaneous information through the additional
variables.\ This is a general result, which holds simply because the
statistical predictions of Bohmian theory are equivalent to usual quantum
mechanics, which itself does not allow superluminal communication.\ But assume
for instance that we could manipulate directly the additional variable
attached to a particle belonging to an EPR correlated pair, in a completely
arbitrary way (even at a microscopic scale), and without changing the wave
function; then, the ``quantum velocity term'' acting on the additional
variables of the other particle would instantaneously be affected, and so
would be its subsequent position in space; since that particle may be in
principle at an arbitrary distance, one could use this property to send
messages at a velocity exceeding the velocity of light. The conclusion is that
such manipulations should be considered as impossible: the only possible
source of evolution of the additional variables has to be the wave function
dependent term.

If the additional variables cannot be directly manipulated at a microscopic
scale, can we then somehow filter them in a range of values, as one does for
the state vector when the $Oz$ component is filtered in a Stern-Gerlach
apparatus? Suppose for instance that we could, for a particle in an eigenstate
of the $Oz$ component of its spin, select the values of the additional
variable that will correspond to a result $+1$ in a future measurement of the
$Ox$ component; were such a selection possible with the help of any physical
device, the theory with additional variables would obviously no longer be
completely equivalent to standard quantum mechanics (this is because, within
orthodox theory, if a spin $1/2$ particle is initially selected into the $+1$
spin state by an $Oz$ oriented Stern-Gerlach apparatus, it becomes completely
impossible to make any prediction on the deviation observed later in an $Ox$
oriented Stern-Gerlach apparatus).\ Theories such as that developed in
\cite{Bohm-Bub-1} include this as a possibility; indeed, if it is ever
demonstrated experimentally, there will be very good reasons to abandon
standard quantum theory in favor of theories with additional variables!\ Of
course, we cannot predict the future and conceptual revolutions are always
possible, but for the moment it may seem safer to provide the additional
variable theories with features that make them equivalent to orthodox
theory.\ In this perspective, it becomes necessary to assume that the
additional variables can neither be manipulated directly nor filtered, as
opposed to the state vector. In other words, the additional variables describe
an objective reality, but at a different level from the reality of the field
of the wave function, since only the latter can be influenced directly by
human decisions.

\bigskip

\begin{center}
VI.\ Constructing consistent families of histories

\medskip
\end{center}

This appendix provides a discussion of the consistency condition
(\ref{hist-5}).\ First, we should mention that other conditions have been
proposed and used in the literature; in the initial article on histories
\cite{Griffiths}, a weaker condition involving only the cancellation of the
real part of the left hand side of (\ref{hist-5}) was introduced.\ For
simplicity, here we limit ourselves to the stronger condition (\ref{hist-5}),
which is a sufficient but not necessary condition to the weaker form; it turns
out that, as noted in ref. \cite{Dowker}, it seems more useful in this context
to introduce selectivity than generality in the definition of consistent histories.

At first sight, a natural question that comes to mind is whether or not it is
easy, or even possible at all, to fulfil exactly the large number of
conditions contained in (\ref{hist-5}); actually, it has been proposed by
Gell-Mann and Hartle to give a fundamental role to families that satisfy
consistency conditions in only an approximate way \cite{Gell-Mann}, but here
we leave aside this possibility and consider only exact consistency
conditions.\ Let us assume for instance that the system under study is a
particle propagating in free space; the various projectors may then define
ranges of positions for the particle, playing a role similar to diaphragms or
spatial filters in optics that confine an optical beam in the transverse
direction. Then the consistency condition will appear as similar to a
non-interference condition for the Huyghens wavelets that are radiated by the
inner surface of each diaphragm. But we know that diffraction is unavoidable
in the propagation of light; even if it can be a very small effect when the
wavelength is sufficiently short and the diaphragms sufficiently broad, it is
never strictly zero.\ Can we then satisfy the non-interference conditions
exactly?\ The answer is not obvious.\ It turns out to be yes, but it is
necessary to exploit the enormous flexibility that we have in the choice of
subspaces and projectors in a large space of states, and not to limit
ourselves to projectors over well-defined positions only. To understand why,
we now briefly sketch one possible systematic method to construct consistent
families of histories.

The simplest method is to guide the construction on the structure of
(\ref{hist-5}), and to introduce the eigenstates $\mid\varphi_{n}^{0}>$ of the
density operator $\rho(t_{0})$ (an Hermitian operator can always be
diagonalized); let us then define the operators $\widehat{P}_{1,j_{1}}(t_{1})$
as:%
\begin{equation}
\widehat{P}_{1,n}(t_{1})=\mid\varphi_{n}^{0}><\varphi_{n}^{0}\mid
\label{hist-6}%
\end{equation}
which is equivalent to assuming that their Schr\"{o}dinger counterparts
$P_{1,j}$ are the projectors over the states that have evolved from the
$\mid\varphi_{n}^{0}>$'s from time $t_{0}$ to time $t_{1}$. Because
$\rho(t_{0})$ is of course diagonal in its own basis, this choice already
ensures the presence of a factor $\delta_{j_{1},j_{1}^{^{\prime}}}$ in the
right hand side of (\ref{hist-5}).\ Now, we can also assume that the
$P_{2,j_{2}}$'s are defined as the projectors over the states that have
evolved from the $\mid\varphi_{n}^{0}>$'s from time $t_{0}$ to time $t_{2}$,
so that a relation similar to (\ref{hist-6}) is again obtained; this will
ensure, not only the presence of factors $\delta_{j_{2},j_{2}^{^{\prime}}}$.
in the right hand side of (\ref{hist-5}), but actually also the appearance of
a delta function $\delta_{j_{1},j_{2}}$.\ The procedure can be repeated as
many times as needed, and in this way a consistent family is built.

It is nevertheless a very special family, for several reasons. The first is
that each projector corresponds to a subspace of dimension $1$ only, which
corresponds to histories that are ``maximally accurate''; the second is that
most histories of the family have zero probability: in fact, only those with
$j_{1}=j_{2}=j_{3}=..$ are possible, which means that the only randomness
occurs at time $t_{1}$, and that all subspaces at later times are then
perfectly determined. The description that we obtain is, in a sense, trivial:
initially, the system is in one of the eigenstates that are contained in
$\rho(t_{0})$, and then evolves deterministically from this initial state.

But it is possible to make the family less singular by grouping together, for
each time $t_{i}$, several projectors into one single projector; different
associations of projectors may be used at different times.\ In this way, the
description of the evolution of the state within this family becomes less
accurate, but also less trivial since projectors at different times are no
longer associated pair by pair.\ On the other hand, it is possible to see that
this grouping of projectors has not destroyed the consistent character of the
family; of course, other methods for constructing consistent families are also possible.

\begin{center}
\bigskip
\end{center}

\textit{Acknowledgments}: The first version of this text was written during a
visit to the Institute of Theoretical Physics of the University of California
at Santa Barbara, as a side activity during a session on Bose-Einstein
condensation, making profit of the presence of A.\ Leggett and sharing an
office with W.\ Zurek; the research was supported in part by the National
Science Foundation under grant number PHY94-07194. The final version of the
text was made during a visit to the Lorentz Center of the University of
Leiden, as a side activity during another session on Bose-Einstein
condensation, two years later; Stig Stenholm was kind enough to read the whole
article and to provide useful advice and comments. The intellectual
stimulation of these two visits was wonderful!

The author is also very grateful to William\ Mullin, Philippe Grangier, Jean
Dalibard, S.\ Goldstein, Serge Reynaud and Olivier Darrigol for useful comment
and advice; they stimulated the re-writing of various parts of this text,
sometimes of whole sections, for better clarity. Abner Shimony was kind enough
to carefully read several versions of this manuscript; the author is very
grateful for all useful suggestions that he made at every step.\ Many thanks
are also due to Robert Griffiths for his comments on the section on the
history interpretation, as well as to P.\ Pearle and G.\ Ghirardi concerning
the section on non-linear Schr\"{o}dinger dynamics.\ It is hoped, but not
necessarily true, that all these colleagues will agree with the present text,
or at least most of it; if not, the scientific responsibility should be
considered as entirely that of the present author.\ Finally, an anonymous
referee made two long, careful and interesting reports containing several
especially useful suggestions, which were taken with gratitude.

LKB (Laboratoire Kastler Brossel) is an ``Unit\'{e} Mixte CNRS'', UMR 8552.

\end{document}